\documentclass[aip,amsmath,amssymb, preprint]{revtex4-1}

\usepackage{graphicx}% Include figure files
\usepackage{dcolumn}% Align table columns on decimal point
\usepackage{bm}% bold math
\usepackage[utf8]{inputenc}
\usepackage[T1]{fontenc}
\usepackage{mathptmx}
\usepackage{etoolbox}
\usepackage{xcolor}
\usepackage{mathrsfs}
\usepackage{amsmath}
\usepackage{adjustbox}
\usepackage[version=4]{mhchem}
\usepackage{siunitx}
\usepackage[caption=false]{subfig}  % safer with revtex
\usepackage{url}
\usepackage{tabularx}
\usepackage[section]{placeins} 
 \usepackage{booktabs}

\makeatletter
\def\@email#1#2{%
 \endgroup
 \patchcmd{\titleblock@produce}
  {\frontmatter@RRAPformat}
  {\frontmatter@RRAPformat{\produce@RRAP{*#1\href{mailto:#2}{#2}}}\frontmatter@RRAPformat}
  {}{}
}%
\makeatother
\begin{document}

\preprint{AIP/123-QED}

\title[]{Invariance-embedded Machine Learning Sub-grid-scale Stress Models for Meso-scale Hurricane Boundary Layer Flow Simulation I: Model Development and \textit{a priori} Studies}
% Force line breaks with \\
\author{Md Badrul Hasan}
% \altaffiliation{Email: mdbadrh1@umbc.edu}%Lines break automatically or can be forced with \\
 \email{mdbadrh1@umbc.edu}
\author{Meilin Yu}%
% \altaffiliation{Email:mlyu@umbc.edu}
\affiliation{ 
Department of Mechanical Engineering, University of Maryland, Baltimore County, MD 21250  %\\This line break forced with \textbackslash\textbackslash
}%
\author{Tim Oates}
 %\homepage{http://www.Second.institution.edu/~Charlie.Author.}
%\affiliation{University of Maryland, Baltimore County, Baltimore, MD, 21250, USA}
% \altaffiliation{Email: oates@cs.umbc.edu}
\affiliation{Department of Computer Science and Electrical Engineering, University of Maryland, Baltimore County, MD 21250}

\date{\today}% It is always \today, today,
             %  but any date may be explicitly specified

\begin{abstract}
This study develops invariance-embedded machine learning sub-grid-scale (SGS) stress models admitting turbulence kinetic energy (TKE) backscatter towards more accurate large eddy simulation (LES) of meso-scale turbulent hurricane boundary layer flows. The new machine learning SGS model consists of two parts: a classification model used to distinguish regions with either strong energy cascade or energy backscatter from those with mild TKE transfer and a regression model used to calculate SGS stresses in regions with strong TKE transfer. To ease model implementation in computational fluid dynamics (CFD) solvers, the Smagorinsky model with a signed coefficient $C_s$, where a positive value indicates energy cascade while a negative one indicates energy backscatter, is employed as the carrier of the machine learning model. To improve its robustness and generality, both  physical invariance and geometric invariance features of turbulent flows are embedded into the model input for classification and regression, and the signed Smagorinsky model coefficient is used as the output of the regression model. Different machine-learning methods and input setups have been used to test the classification model's performance. The F1-scores, which measure balanced precision and recall of a model, of the classification models with physical and geometric invariance embedded can be improved by about 17\% over those without considering geometric invariance. Regression models based on ensemble neural networks have demonstrated superior performance in predicting the signed Smagorinsky model coefficient, exceeding that of the dynamic Smagorinsky model in \textit{a priori} tests. 
\end{abstract}

\maketitle

\section{\label{sec:intro}Introduction%\protect\\ The line
%break was forced \lowercase{via} \textbackslash\textbackslash
}
Recent improvements in computing technology have significantly boosted the accuracy of numerical weather models, especially for predicting tropical cyclones (TCs), thus allowing for greater detail and precision in forecasting extreme weather events. Direct numerical simulation (DNS) and large-eddy simulation (LES) are conventional methods for simulating turbulent flows. However, these methods are not computationally feasible for numerical simulations of TCs, which have numerous energy-containing vortices in the eyewall region. For example, traditional sub-grid-scale (SGS) stress models in different applications~\cite{Celik-2005,  matheou2016numerical, Bou-2005} are mainly designed to model turbulence of a typical length scale around meters, designated as ``micro-scale'' phenomena within the domain of geophysical flows. Therefore, their computational cost is prohibitive for a macro-scale weather pattern, such as hurricanes, spanning hundreds of kilometers. Meso-scale non-hydrostatic atmospheric flow simulations~\cite{Giraldo_Restelli_2008_JCP,HasanEtAl2022JAMES} can significantly reduce the computational cost of TCs. However, judicious numerical dissipation, such as localized artificial viscosity~\cite{Yu_et_al_MWR_15}, must be developed to add an adequate amount of artificial diffusion to stabilize numerical simulations. In many applications~\cite{Nadiga-2007,Ballouz_Ouellette_2018,Linkmann_2018,Vela_Martín_2022}, traditional SGS models, which only consider forward-directed turbulence kinetic energy (TKE) cascade, were used to provide numerical dissipation, which, unfortunately, can introduce an overabundance of numerical dissipation on meso-scale hurricane simulations. Acknowledging the importance of energy backscatter in atmospheric and oceanic models has been demonstrated crucial for realistic meso-scale simulations~\cite{JANSEN2014,Grooms-2023, Chang-2023}. 

In SGS closure modeling, machine learning (ML) has introduced innovative approaches to overcome certain turbulence energy transfer challenges. ML's performance in managing multi-dimensional data and using complex non-linear correlations has made it an enticing toolkit~\cite{Brunton_2020, Duraisamy, Gentine-2021, Zanna-2021}. It offers the prospect of uncovering hidden insights from data and delivering superior SGS closures alongside novel perspectives on SGS physics. Guan et al.~\cite{guan_stable_2022} found that using convolutional neural networks (CNNs) with transfer learning can create accurate, stable, and widely applicable LES models admitting energy backscatter for 2D decaying turbulence. They achieved this through \textit{a priori} and \textit{a posteriori} analyses to develop non-local, data-driven SGS models. The development of data-driven SGS closures for canonical geophysical flows, including 2D and quasigeostrophic turbulence as well as broader atmospheric and oceanic circulations, has yielded promising results~\cite{Bolton-2019, Frezat-2022, Cheng-2022, Guillaumin-2021}.
Notably, deep neural networks (NNs) trained on high-fidelity DNS data have facilitated the emergence of stable and precise LES with data-driven SGS closures~\cite{Yuval_2020, Guan_2023}.
 
We notice that using simulation data directly does not show reasonable estimations of energy backscatter, even from \textit{a priori} tests of the model performance~\cite{hasan2023sub}. The use of frame independence showed better performance at turbulence physics prediction based on the flow information in its neighborhood~\cite{Zhou-22}. Decomposing the velocity gradient into its symmetric and anti-symmetric components and embedding invariances~\cite{LING2016}, which are derived from the high fidelity data, also showed a better performance in adding domain knowledge to machine learning models.

SGS stress models in turbulence modeling can be classified into functional~\cite{Meneveau-2000} and structural~\cite{Sagaut2006} models. Functional models primarily use an eddy viscosity approach to approximate the effects of unresolved scales by adding effective viscosity to enhance kinetic energy dissipation. This method is straightforward and effective in reproducing the overall energy transfer between resolved and subgrid scales. However, it often fails to accurately predict the SGS term, leading to inaccuracies in local flow structures~\cite{Vollant-2017}. Structural models, on the other hand, focus on accurately capturing the local SGS stresses and dynamics by reconstructing the smaller scales. While they offer higher accuracy and preserve detailed flow structures, they can be computationally demanding and prone to instability due to incorrect backscatter predictions~\cite{Vreman-1995, Nadiga-2007}. The idea of energy backscatter captures the essence of energy dynamics within TCs, influencing wind speed distribution and the shape of eyewalls and rain bands. 
In fluid dynamics, energy backscatter means a non-linear and often inverse transfer of energy, which can greatly affect turbulence and, in turn, the weather patterns they cause~\cite{Carati-1995, Kerr-1996, Mason_Thomson_1992, Domaradzki-2021, Nabavi_Kim_2024}. To mitigate the adverse effect of inaccurate backscatter modeling, hybrid approaches that combine eddy viscosity components~\cite{Vreman-1995} with structural models can be employed to retain accuracy while damping excessive backscatter. Additionally, adaptive filters or limiting functions can be used to reduce backscatter~\cite{beck_deep_2019, Zhou-2020} selectively, and introducing hyper-viscosity~\cite{Xie-2020, Xie-2020-2} can stabilize the simulation by enhancing dissipation at smaller scales. In a recent study~\cite{backscatter21}, organized kinetic energy backscatter was observed using radar measurements in the hurricane boundary layer. It demonstrates that meso-scale SGS modeling requires non-equilibrium flow effects for accurate dynamics capture. 

In this work, we introduce a novel and interpretable framework that integrates machine learning techniques with physical and geometric invariances to enhance the accuracy of SGS models admitting energy backscatter for meso-scale simulations. A key strength of this approach lies in its explainability. The framework improves predictive performance by embedding domain-specific knowledge into the raw data using tensor invariants and singular values of local flow features. It provides a clear connection to the underlying physics of unresolved turbulence. Unlike traditional ``black-box'' ML models, this framework explicitly incorporates physical constraints and rotational and reflective invariances, ensuring consistency with fundamental turbulence principles. Using these invariances, the model respects the underlying flow structures, enhancing its interpretability and reliability. The research evaluates various NN configurations, including direct feed-forward networks and ensemble-based approaches, demonstrating their ability to stabilize and improve predictions under highly variable flow conditions. The framework improves the prediction of both forward energy cascade and energy backscatter in mesoscale atmospheric conditions resembling hurricanes, as demonstrated in \textit{a priori} tests.

The remainder of the paper is organized as follows. The main machine learning framework, including classification and regression models, is briefly introduced in Section~\ref{Sec:Framework}. Section~\ref{Sec:Data} describes data preparation processes that create suitable inputs for the ML models and briefly introduces the ML configurations employed for classification and regression. Here, we highlight the systematic incorporation of physical and geometric invariances, which ensures that the models align with the physical laws governing turbulent flows. Section~\ref{sec:Res_Dis} presents and compares the results of the classification and regression ML models. Specifically, Section~\ref{sec:Results_class} highlights the ability of classification models to identify critical regions with strong backscatter and forward energy cascade, while Section~\ref{sec:Results_regression} demonstrates the regression models' capability of accurately predicting the signed coefficient \(C_s\) in the Smagorinsky model. The performance comparison between the new ML model and dynamic Smagorinsky model in an \textit{a priori} sense is presented in Section~\ref{Sec:Compare_pro_dyn}. Finally, Section~\ref{Sec:Conclusion} summarizes the findings and discusses 
future works.

\section{Framework for Machine Learning SGS Stress Modeling}\label{Sec:Framework}

This section gives an overview of the invariance-embedded ML SGS stress modeling framework. The Smagorinsky~\cite{Piomelli} model is used as the baseline SGS stress model in this study. It reads
\begin{equation}\label{eq:eddy_visc}
\nu_{\mathrm{t}} = C_{s} \Delta^{2}|\overline{S}_{ij}^{*}|,
\end{equation}
where $\nu_{\mathrm{t}}$ is the turbulence eddy viscosity, $C_s$ is the model constant,  $\Delta$ is the filter size, and $\overline{S}_{ij}^{*}$ is the deviatoric strain rate tensor with its magnitude $|\overline{S}_{ij}^{*}|$ defined as $\sqrt{{2 \overline{S}_{ij}^{*} \hspace{1mm} \overline{S}_{ij}^{*}}}$.
Note that in the classic Smagorinsky model, the coefficient $C_s$ is required to be positive as the eddy viscosity $\nu_{\mathrm{t}}$ is an analogy of the molecular kinematic viscosity. However, many atmospheric and oceanic modeling/simulation studies, such as those reviewed in Section~\ref{sec:intro}, have emphasized the importance of energy backscatter in meso-scale SGS modeling. Therefore, in this study, instead of treating $\nu_{\mathrm{t}}$ as the eddy viscosity, we redefine it as the turbulence momentum diffusion/anti-diffusion coefficient and assume that it follows the format of the original Smagorinsky model, i.e., Eq.~\eqref{eq:eddy_visc}. As a result, $\nu_{\mathrm{t}}$ can be positive or negative, and $C_s$ in the Smagorinsky model becomes a signed coefficient. Although anti-diffusion can cause numerical instability, it has been reported by Guan et al.~\cite{guan_stable_2022} that accurate modeling of turbulence energy backscatter can ensure stable \textit{a posteriori} LES of turbulence simulations. 

As will be reported in Section~\ref{Sec:Compare_pro_dyn},  the dynamic Smagorinsky model (DSM)~\cite{germano1991dynamic} can reasonably predict the variation trend of the signed Smagorinsky coefficient $C_s$. However, it is very challenging for DSM to predict the true values of $C_s$ accurately. This violates the accuracy requirement posed on energy backscatter modeling in Ref.~\onlinecite{guan_stable_2022}. Therefore, we develop an invariance-embedded ML SGS stress modeling framework in this work, which can accurately predict the signed Smagorinsky coefficient $C_s$. Specifically, this framework consists of a classification model that can distinguish $C_s$ with large and small absolute values with a threshold of 0.001 and a regression model that can map localized flow field to signed $C_s$ in a point-to-point sense, thus facilitating practical ML SGS model implementation for LES in \textit{a posteriori} tests. 

\begin{figure}[h]
    \centering    
    \subfloat{\includegraphics[width=\linewidth]{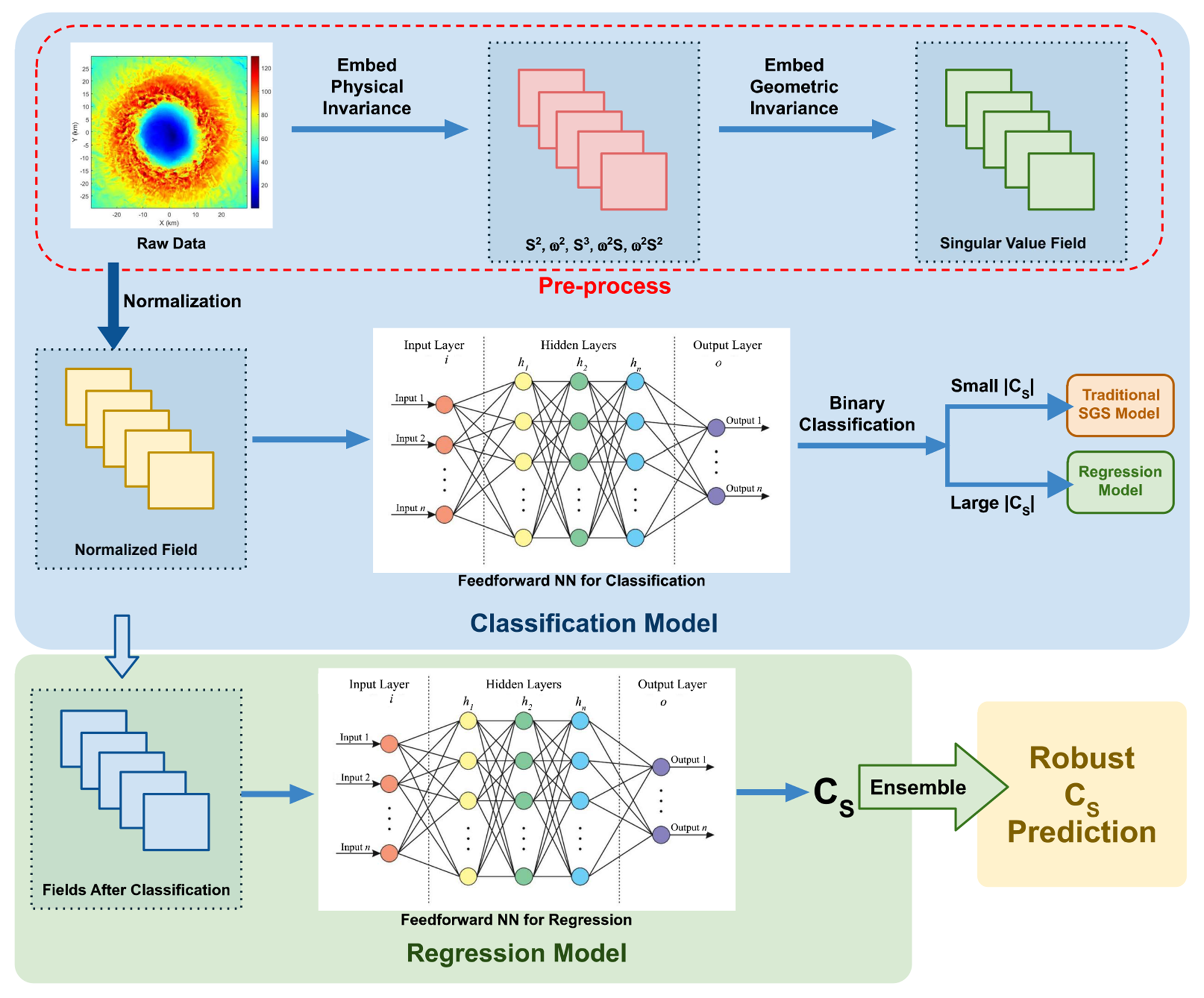}}
    \caption{An illustration of the framework for enhanced sub-grid-scale stress modeling with invariance-embedded machine learning classification and regression models.}
    \label{fig:Framework_1}
\end{figure}

As shown in Figure~\ref{fig:Framework_1},
physical and geometric invariances are extracted from the raw data to embed domain-specific knowledge into the ML model. Specifically in the pre-processing stage, physical invariances, such as tensor invariants from the combination of the strain rate tensor and vorticity tensor, are extracted from the raw outputs of high-fidelity simulation data~\cite{Ren-2020} of a hurricane-like vortex (see Sect.~\ref{subsubsec:Physics_embed}). Geometric invariances are then encapsulated via singular value decomposition (SVD) of the normalized local physical invariance fields, producing corresponding singular value fields with essential structural information about the flow (see Sect.~\ref{subsubsec:Geo_embed}). As the last step of the data pre-processing procedure, the singular value fields are normalized, which transforms to ensure consistency and comparability across the dataset.

The normalized singular value fields are input into an ML classifier, such as the feedforward NN for binary classification. This classifier determines whether the local signed SGS model coefficient (\(C_s\)) should be categorized as small or large based on the learning from the physical and geometric properties of the data. The output from this classification model subsequently informs a regression model based on a feedforward NN structure. This regression model fine-tunes the signed SGS model coefficient (\(C_s\)) predictions, which are critical for accurate TKE transfer modeling. In the final phase, the outputs from the regression models are combined using an ensemble approach, producing a reliable prediction of the SGS model constants. Next, we explain the technical details of data preparation and ML model creation.

\section{Data Preparation and Machine Learning Model Creation}\label{Sec:Data}
The datasets used in this study originated from the high-fidelity simulations by Ren et al.~\cite{Ren-2020}, featuring a flow field resolution of approximately $62$ m. The LES was conducted with the Weather Research and Forecasting Model (WRF)~\cite{WRF-4}, facilitating an accurate presentation of small-scale turbulence patterns crucial to our study. These simulations accounted for varying sea surface temperatures (SSTs), i.e., 26°C, 27°C, 28°C, and 29°C, with the understanding that SSTs above 26°C play a pivotal role in the intensification of hurricanes. Such temperature variations significantly influence a hurricane's strength and structure, directly impacting potential damages along coastal areas.
For an in-depth understanding of these simulations, readers are referred to the work of Ren et al.~\cite{Ren-2020} and Rotunno et al.~\cite{Rotunno-09}.

\begin{figure}[htbp]
    \centering
    \includegraphics[width=0.49\linewidth]{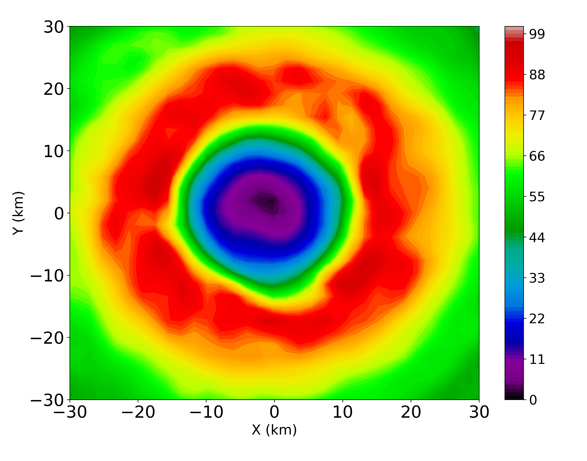}
    \includegraphics[width=0.49\linewidth]{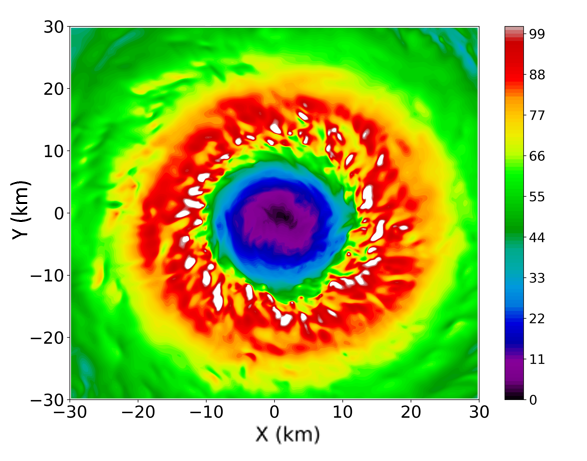}
    
    \vspace{1ex}
    \begin{minipage}{0.49\linewidth}
        \centering \small (a) Resolution at $1666.67$ m
    \end{minipage}%
    \begin{minipage}{0.49\linewidth}
        \centering \small (b) Resolution at $61.72$ m
    \end{minipage}

    \caption{Comparison of instantaneous horizontal wind speed ($m/s$) plots for the hurricane center in a horizontal plane at the height of $303$ m with different grid resolutions.}
    \label{fig:wspd}
\end{figure}

Figure~\ref{fig:wspd} presents the comparison of wind speed fields from simulations with coarse ($1666.67$ m) and fine ($61.72$ m) grid resolutions at the SST of 29°C, highlighting the differences in accurately representing turbulent characteristics at different grid spacings. The larger grid size of about 1.67 km (Figure~\ref{fig:wspd}(a)) depicts the major fluctuations in wind speed, giving a broad picture view of the hurricane's structure without details of the turbulent flow. A narrower grid spacing of 61.72 m (Figure~\ref{fig:wspd}(b)) reveals the effect of turbulent eddies, which can cause local high wind speed that is not captured by the coarse-resolution simulation. 

This study will use the detailed patterns from the high-resolution simulation to figure out important turbulence factors, like eddy viscosity and Smagorinsky model coefficient ($C_{s}$), the basis for our machine/deep learning model. In turn, this model is meant to make storm predictions more accurate. The procedures and methodologies used to refine this data for NN training are further explained in later sub-sections.

\subsection{Filtering Fine-resolution LES data}\label{subsec:filter}
Since only velocity gradient features are used in turbulence modeling with the Smagorinsky model, velocity components in three dimensions $(u, v, w)$ were smoothed with a Gaussian filter. In the Gaussian filter, the weight assigned to each point in a one-dimensional series is expressed as: 
$$
\hat{w}_{ \pm i}(\Delta, n)=\exp \left(\frac{-(i \Delta / n)^2}{2 \sigma^2}\right) ,
$$
normalized to ensure the sum of all weights equals one:
$$
w_{ \pm i}(\Delta, n)=\frac{\hat{w}_{ \pm i}(\Delta, n)}{\sum \hat{w}_{ \pm i}(\Delta, n)}.
$$
Here, $n$ is the number of points in the filter window, $\sigma$ is the standard deviation of the Gaussian distribution, which controls the spread of the Gaussian curve, and $\Delta$ represents the distance between consecutive points in the data series and sets the spacing in the filter application. In this study, $\Delta$ is set as $1$ km.

\begin{figure*}[htbp]
    \centering
    \includegraphics[width=0.32\linewidth]{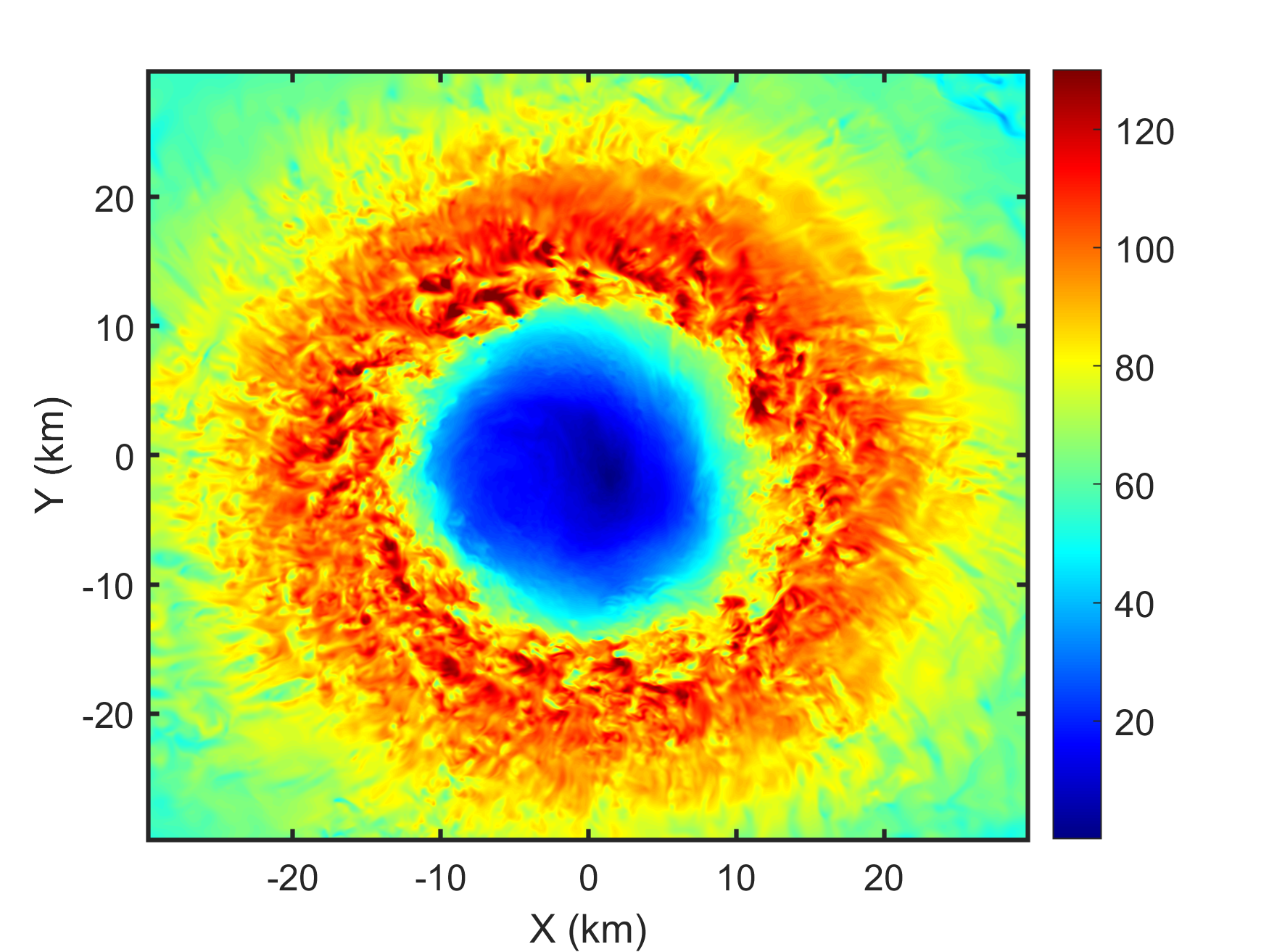}
    \includegraphics[width=0.32\linewidth]{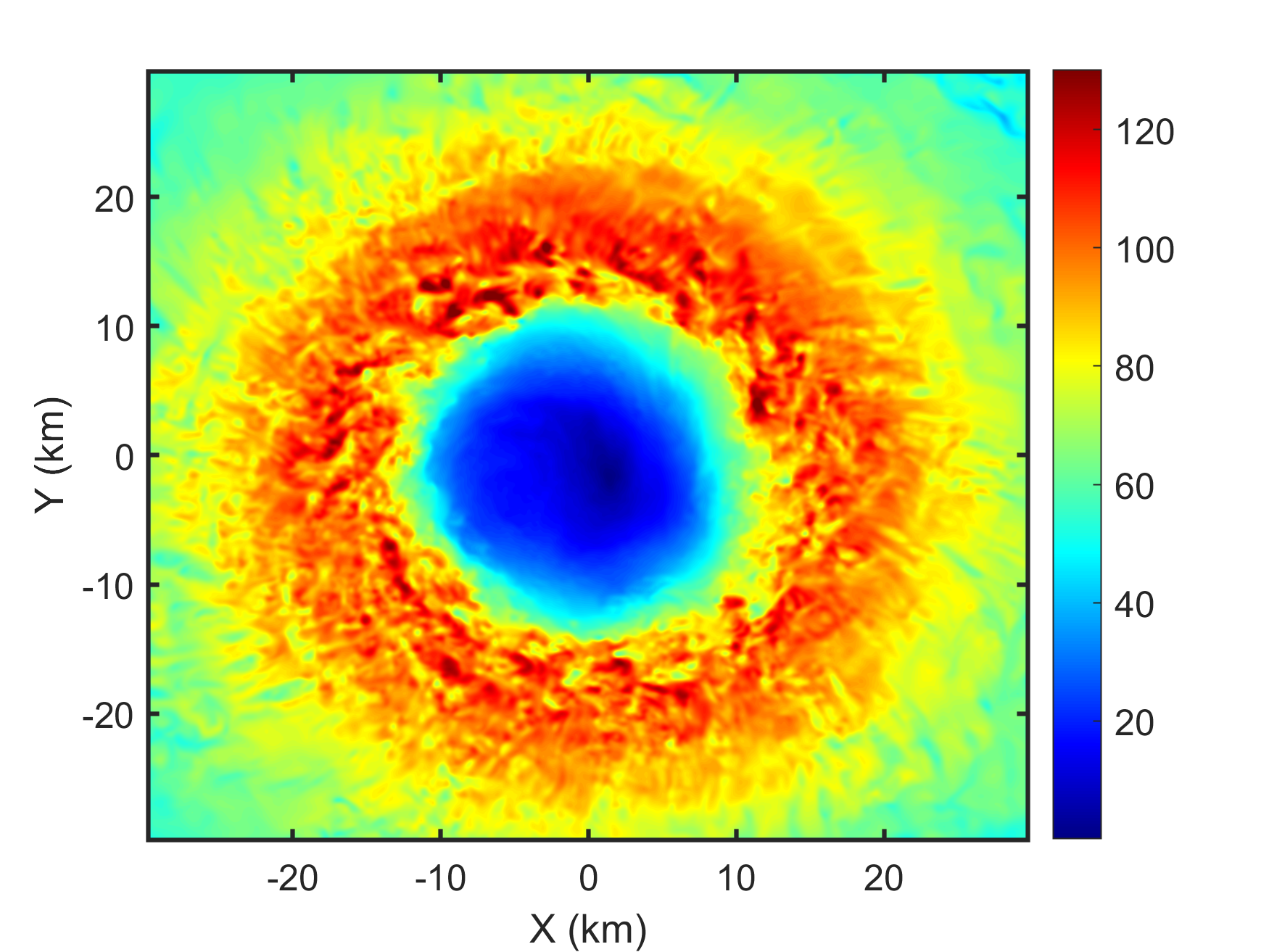}
    \includegraphics[width=0.32\linewidth]{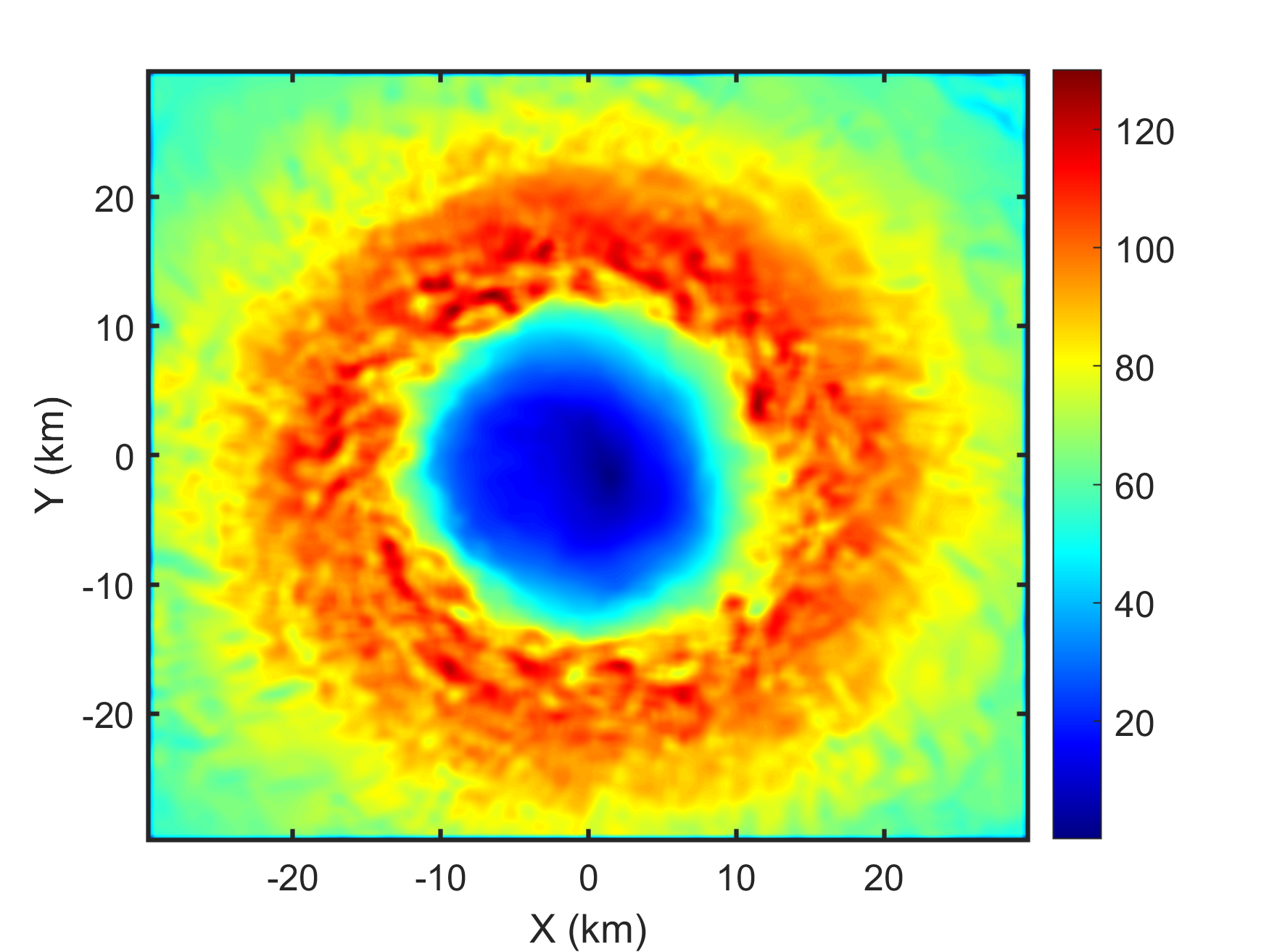}

    \vspace{1ex}
    \begin{minipage}[t]{0.32\linewidth}
        \centering \small(a) Simulation Data
    \end{minipage}%
    \hfill
    \begin{minipage}[t]{0.32\linewidth}
        \centering \small(b) $\sigma = 2$
    \end{minipage}%
    \hfill
    \begin{minipage}[t]{0.32\linewidth}
        \centering \small(c) $\sigma = 20$
    \end{minipage}

    \caption{Horizontal wind speed ($m/s$) field and the corresponding fields after using a $\Delta = 1$ km Gaussian filter with different standard deviations at the 303 m vertical level.}
    \label{fig:wind_speed_fields}
\end{figure*}

Figure~\ref{fig:wind_speed_fields}(a) shows a hurricane simulation's horizontal wind speed field at an SST of 29°C. After applying a Gaussian filter with a standard deviation ($\sigma$) of 2 (Figure~\ref{fig:wind_speed_fields}(b)), smaller-scale features are reduced, highlighting the primary structures such as the hurricane's eye and eyewall. Increasing $\sigma$ to 20 (Figure~\ref{fig:wind_speed_fields}(c)) smooths the field further, emphasizing larger, more organized patterns while diminishing smaller turbulence.

\begin{figure}[h]
    \centering
    \subfloat{\includegraphics[width=0.8\linewidth]{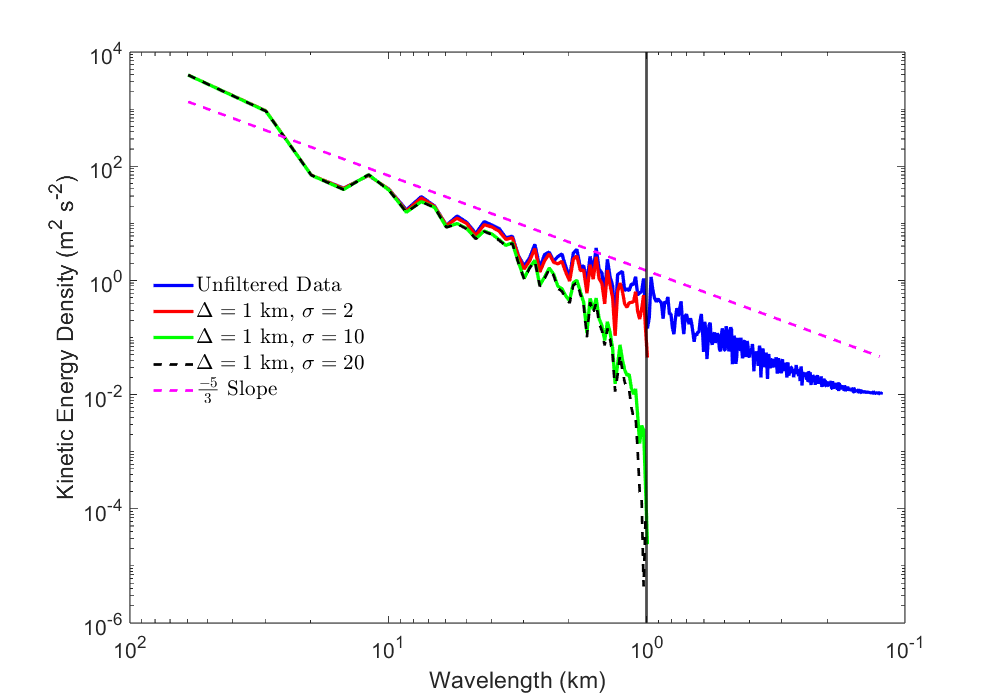}}
 %   \subfloat[ $2$ $km$]{\includegraphics[width=0.5\linewidth]{kes_2km.png}}
    \caption{Kinetic energy spectra after using $1$ km Gaussian filter at $~303$ m vertical level. }
    \label{fig:kes}
\end{figure}
Figure~\ref{fig:kes} presents the kinetic energy spectra from the simulation data and those after filtering, illustrating energy distribution across scales. The unfiltered spectrum (blue line) follows a $-5/3$ slope, consistent with Kolmogorov's~\cite{Kolmogorov-1991} inertial subrange theory. Post-filtering with different $\sigma$ values reduces energy at smaller scales, highlighting the filter's dampening effect, particularly pronounced at higher $\sigma$ values~\cite{pope_2000}. All turbulence features presented in the following (sub)sections are calculated from the original and filtered flow data.

\subsection{Smagorinsky Model Coefficient Evaluation}\label{subsec:Calc}
The production of TKE in turbulent flows can be quantitatively linked to the interaction between the Reynolds stresses and the mean strain rate, a relationship captured by the definition of turbulence production ($\wp^{\text{def}}$)~\cite{durbin-2011}:
\begin{equation}
\wp^{\text{def}} = -\left(\overline{u_i u_j} - \overline{u}_i \overline{u}_j\right) \frac{\partial \overline{u}_i}{\partial x_j} = \tau_{ij} \overline{S}_{ij}.
\label{eq:prod_def}
\end{equation}
Herein, the operator $\overline{(\cdot)}$ represents a general filtering or averaging operation, $u_i$ is the flow velocity, $\tau_{ij}$ is the Reynolds stress tensor, and $\overline{S}_{ij}$ is the strain rate tensor of the filtered or averaged flow field. Eq.~\eqref{eq:prod_def} connects the Reynolds stress tensor with the mean strain rate tensor, which shows how fast TKE is produced due to the mean velocity gradients.

Alternatively, the production of TKE can be modeled through the concept of eddy viscosity ($v_T$), a construct within the Boussinesq hypothesis that allows for the approximation of the Reynolds stresses in terms of the mean strain rate~\cite{wilcox2008turbulence}. This approximation, as given by Eq.~\eqref{eq:prod_eddy}, assumes a proportionality between the turbulent stress and the mean strain rate, transferring energy from the larger to smaller scales:

\begin{equation}
\wp = \left( 2 \nu_{T} \overline{S}_{ij}^{*} + \frac{\tau_{kk}}{3} \delta_{ij} \right) \overline{S}_{ij} =
 2 \nu_{T} \overline{S}_{ij}^{*} \overline{S}_{ij} + \frac{\tau_{kk}}{3} \overline{S}_{ii}. 
\label{eq:prod_eddy}
\end{equation}
Herein, $\overline{S}_{ij}^{*} = \frac{1}{2} (\frac{\partial \overline{u}_i}{\partial x_j} + \frac{\partial \overline{u}_j}{\partial x_i} - \frac{2}{3} \frac{\partial \overline{u}_k}{\partial x_k} \delta_{ij})$ is the deviatoric strain rate tensor and the Smagorinsky eddy viscosity model $\nu_T={C_{s}}\Delta^2\left|\overline{S}_{ij}^{*}\right|$. Note that for a strictly incompressible flow, we have $\overline{S}_{ij}^{*} = \overline{S}_{ij}$ as the divergence of the filtered or averaged flow velocity field $ \partial \overline{u}_k / \partial x_k$ is zero when the partial differentiation operator and the filtering/averaging operator are commutable. By combined Eq.~\eqref{eq:eddy_visc}--\eqref{eq:prod_eddy}, the Smagorinsky model coefficient $C_s$ can be calculated from high-fidelity simulation data and the associated filtered data as

\begin{equation}
{C_{s}}=\frac{\tau_{ij}\overline{S}_{ij}-\frac{\tau_{kk}}{3} \overline{S}_{ii}}{2\Delta^2\left|\overline{S}_{ij}^{*}\right|\overline{S}_{ij}^{*}\overline{S}_{ij}}.
\end{equation}

\begin{figure}[htbp]
    \centering
    \includegraphics[width=0.36\linewidth]{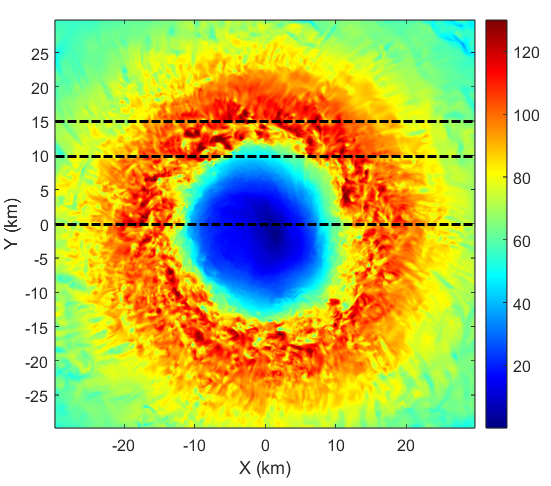}
    \hfill
    \includegraphics[width=0.54\linewidth]{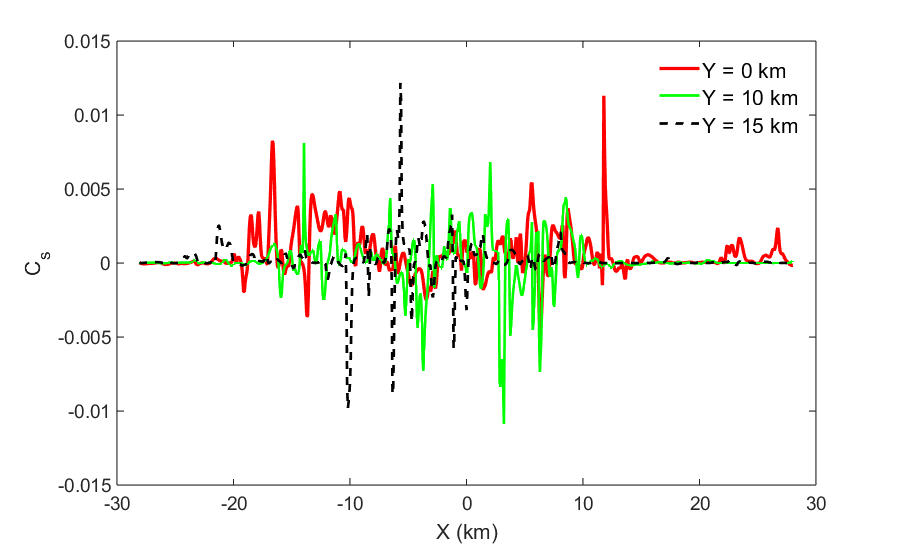}

    \vspace{1ex}
    \begin{minipage}[t]{0.36\linewidth}
        \centering \small (a)
    \end{minipage}%
    \hfill
    \begin{minipage}[t]{0.54\linewidth}
        \centering \small (b)
    \end{minipage}

    \caption{$C_{s}$ values at different positions of $y$ in the $x$-direction at approximately 303 m vertical level.}
    \label{fig:eddy_visc_sign}
\end{figure}

We mention that in the original Smagorinsky model Eq.~\eqref{eq:eddy_visc}, the model coefficient $C_s$ is considered a positive one and a constant value over the whole field. The theoretical value was calculated as 0.027 by Lily \textit{et al.}~\cite{lilly1967} Based on the WRF documentation, $C_{s}$ is a constant of $0.25^2 = 0.0625$. However, in practical applications, $C_s$ can assume both positive and negative values. We presented the $C_s$ distributions at positions of y =  0, 10, and 15 km in Figure~\ref{fig:eddy_visc_sign}. From Figure~\ref{fig:eddy_visc_sign}(b), the large variability of the values of Smagorinsky model coefficient $C_{s}$ can be observed. Note that the y-positions are chosen based on Figure~\ref{fig:eddy_visc_sign}(a), in which the hurricane eye and eye-wall flow structures exhibit large spatial variations.

\subsection{Data Preprocessing and Organization}

\subsubsection{Embedding Physical Invariance} \label{subsubsec:Physics_embed}
Accurately depicting turbulent flows in CFD is essential for precise predictions in engineering and environmental contexts. Recent advancements in data-driven turbulence modeling emphasize selecting physically meaningful and invariant input features to enhance generalizability and interpretability~\cite{LING2016, wang2017physics, Duraisamy, Wu2018}. This study employs tensor invariants derived from the average strain rate tensor $\mathbf{S}$ and the average rotation rate tensor $\mathbf{R}$ as input features for our NN models. These invariants, listed in Table \ref{tab:tensor_invariants}, inherently satisfy rotational and reflectional invariances, providing concise and physically consistent descriptions of the local flow conditions. 

Incorporating these physical invariances ensures that the predictive capability of NN models remains independent of coordinate systems, a crucial advantage for accurately modeling complex mesoscale hurricane flows. Using invariant input features simplifies the NN's learning process by explicitly embedding physical invariance, significantly reducing complexity and improving prediction robustness compared to raw velocity fields or TKE~\cite{hasan2023sub, hasan2025aiaa}. Recent literature underscores the advantage of incorporating such invariances into machine learning frameworks, demonstrating model robustness and predictive accuracy improvements across various turbulent flow scenarios~\cite{tracey2013, Brunton_2020}. By embedding these physical invariances directly, our framework enhances the stability and explainability of NN predictions, thereby making it particularly suitable for complex mesoscale hurricane simulations.
\begin{table}[h]
\centering
\caption{Tensor Invariants Used as Input Features}
\begin{tabular}{ccc}
\hline
\textbf{Invariant} & \quad \textbf{Formula} & \textbf{Description} \\ \hline
$\lambda_1$ & $\operatorname{tr}\left(\mathbf{S}^{2}\right)$ & Trace of the squared strain rate tensor \\ \hline
$\lambda_2$ & $\operatorname{tr}\left(\mathbf{R}^{2}\right)$ & Trace of the squared rotation rate tensor \\ \hline
$\lambda_3$ & $\operatorname{tr}\left(\mathbf{S}^{3}\right)$ & Trace of the cubed strain rate tensor \\ \hline
$\lambda_4$ & $\operatorname{tr}\left(\mathbf{R}^{2} \mathbf{S}\right)$ & Trace of the product of squared rotation rate tensor and strain rate tensor \\ \hline
$\lambda_5$ & $\operatorname{tr}\left(\mathbf{R}^{2} \mathbf{S}^{2}\right)$ &Trace of the product of squared rotation rate tensor and squared strain rate tensor  \\ \hline
\end{tabular}
\label{tab:tensor_invariants}
\end{table}

\subsubsection{Embedding Geometric Invariance} \label{subsubsec:Geo_embed}

\begin{figure*}[htbp]
    \centering
    \includegraphics[width=0.32\linewidth]{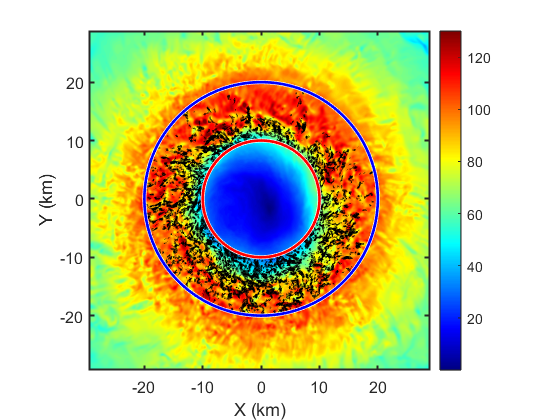}
    \includegraphics[width=0.32\linewidth]{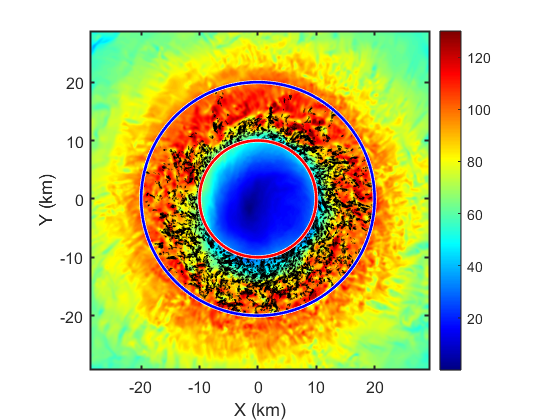}
    \includegraphics[width=0.32\linewidth]{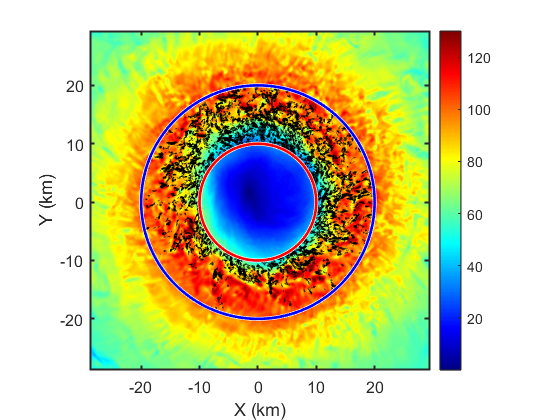}

    \vspace{1ex}
    \begin{minipage}[t]{0.32\linewidth}
        \centering \small (a) Original data
    \end{minipage}%
    \hfill
    \begin{minipage}[t]{0.32\linewidth}
        \centering \small (b) Reflection
    \end{minipage}%
    \hfill
    \begin{minipage}[t]{0.32\linewidth}
        \centering \small (c) Rotation by $\ang{180}$
    \end{minipage}

    \caption{The effect of reflection and rotation by $\ang{180}$ on the actual simulation field.}
    \label{fig:rotation_reflection_inv}
\end{figure*}

The primary focus of this study lies on the eyewall region of a hurricane, which is renowned for its significant contribution to the storm's dynamics and strength fluctuations~\cite{Willoughby1990}. The annular zone within the storm is recognized as a region with concentrated convective activity and powerful angular momentum~\cite{marks2008structure}. In order to isolate this specific location, we establish a defined area bounded by two concentric circles with radii of 10 and 20 km (see Figure~\ref{fig:rotation_reflection_inv}(a)). This effectively marks the eyewall. The black dots in this region are those associated with large absolute $C_s$ values.

\begin{figure}[htbp]
    \centering

    \includegraphics[width=0.40\linewidth]{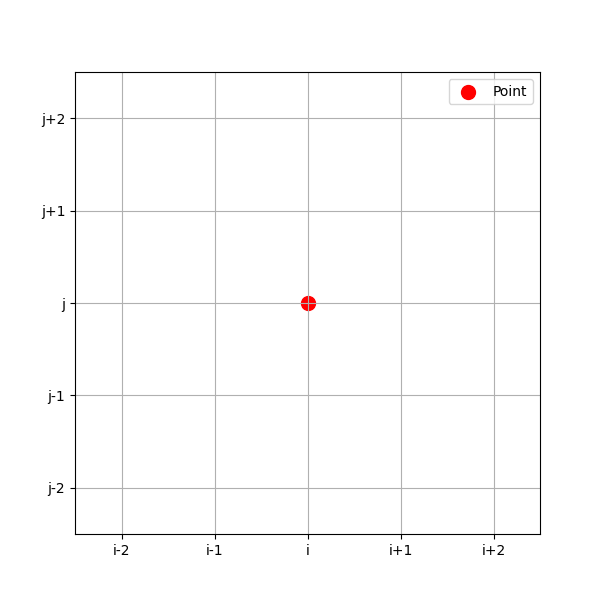}
    \includegraphics[width=0.40\linewidth]{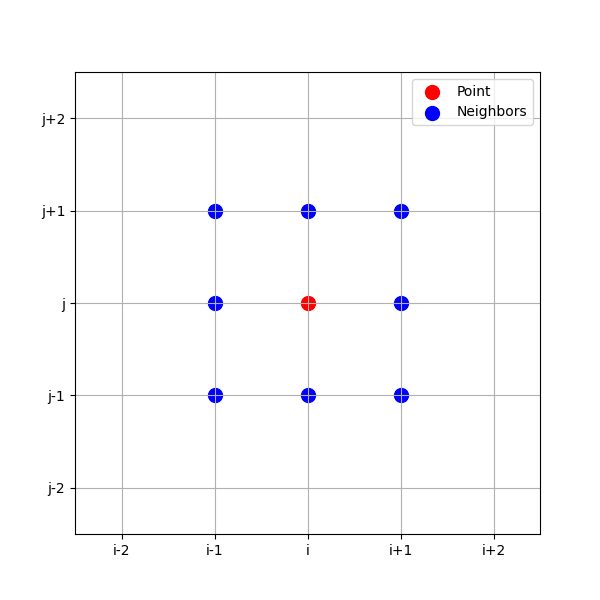}
    \phantomsection \label{subfig:GL}  % optional sublabel reference

    \vspace{0.5ex}
    \begin{minipage}[t]{0.40\linewidth}
        \centering \small (a) Point to point
    \end{minipage}%
    \hspace{0.08\linewidth}
    \begin{minipage}[t]{0.40\linewidth}
        \centering \small (b) Neighboring points to point
    \end{minipage}

    \caption{Different dataset preparation for (a) pointwise learning and (b) group learning.}
    \label{fig:data_1}
\end{figure}

Note that tensor invariants used in Sect.~\ref{subsubsec:Physics_embed} ensure coordinate transformation invariance of pointwise data. It does not present information regarding relative positions (i.e., local pattern) between points. As shown in Figure~\ref{fig:rotation_reflection_inv}(b) and~\ref{fig:rotation_reflection_inv}(c), when a flow pattern is mirrored or rotated, it should be characterized as the same flow feature as the original one by the ML model. We, thus, resort to creating ML classification and regression models with flow features, represented by singular values, of a group of points in the flow field to embed geometric invariance into the model input. The group learning concept and its comparison to the pointwise learning concept are shown in Figure~\ref{fig:data_1}.

Rotation and reflections are a type of orthogonal transformation where the transformation matrix \( Q \) is an orthogonal matrix with a determinant of \(1\) for rotation and \(-1\) for reflection. \( Q \) satisfies the following orthogonality condition:
\[ Q^T Q = Q Q^T = I. \]

Consider the matrix \( A \), and its reflection or rotation is \( QA \), where \( Q \) is the reflection or rotation matrix. To check if the singular values are invariant, consider \( (QA)^T(QA) \):
\[ (QA)^T(QA) = A^TQ^TQA. \]

Since \( Q \) is orthogonal,
%\[ Q^TQ = I \]
we have
\[ (QA)^T(QA) = A^TIA = A^TA. \]

This calculation shows that \( A^TA \) remains unchanged after applying the rotation or reflection transformation. In a geometric sense, both rotation and reflection can be viewed as transformations that alter the orientation or configuration of the matrix \( A \) but do not change the inherent scale lengths measured by the singular values, which are tied to the eigenvalues of \( A^TA \). In the context of hurricanes, where wind directions and flow patterns can vary significantly, maintaining invariance to these transformations ensures that the ML model remains robust and generalizable across different storm scenarios.

In group learning, every point and its neighboring points are assembled to form a matrix. Then, SVD is applied to calculate the singular values. This process not only facilitates a reduction in the dimensionality of the input data but also captures the essential characteristics of the data. By focusing on significant singular values, we retain the most influential features of the turbulence data, reducing noise and less relevant information. In this study, we applied SVD to local $3 \times 3$ or $5 \times 5$ matrices extracted around each point of interest within the hurricane's eyewall. 

\subsection{Machine Learning Methods}\label{Sec:ML}

This subsection briefly introduces machine learning models, including those for classification and regression, used in this study.

\subsubsection{Classification Methods}\label{subsec:Classification}
In the classification task, we focused on predicting whether the absolute value of the local SGS model coefficient (\(C_s\)) should be categorized as small or large. We employed several machine learning methods, including Logistic Regression (LR), Random Forest (RF), Support Vector Machines (SVM), Gradient Boosting (GB), and Neural Network (NN) to evaluate their effectiveness in classifying the \(C_s\) values. The detailed study was reported in Ref.~\onlinecite{hasan2025aiaa}, and the comparison results for the group learning with geometric invariance embedded are briefly discussed in Appendix~\ref{Sec:Appendix}. Based on the test results, NN was selected for further study due to its robustness and adaptability in capturing complex data patterns across various configurations.

We constructed a feedforward NN using PyTorch~\cite{pytorch}, a prominent deep-learning library. The network’s design featured a linear, fully connected architecture with several configurations to assess their influence on the model’s predictive capabilities. Key hyperparameters were tuned for optimal performance, such as the number of hidden layers, the number of nodes in each layer, and the activation functions. For the training of NN-based classification models, the standard backpropagation and gradient descent were used to minimize the Binary Cross-Entropy (BCE) Loss~\cite{Good-1952} defined as:
\begin{equation}
    \text{BCE Loss} = -\frac{1}{N} \sum_{i=1}^{N} \left[ y_i \log(\hat{y}_i) + (1 - y_i) \log(1 - \hat{y}_i) \right],
    \label{eq:BCE_Loss}
\end{equation}
where \( N \) is the number of samples, \( y_i \) is the true label for the \(i\)-th sample (0 or 1), and \( \hat{y}_i \) is the predicted probability for the \(i\)-th sample, which is the output of the sigmoid function applied to the model's output. The model's output is passed through a sigmoid function to produce probabilities, which are then compared to the true labels using the ``BCE Loss'' function.

Training NN involves optimizing weights and biases to minimize a cost function using backpropagation and gradient descent~\cite{rumelhart1986}. The number of hidden layers, nodes per layer, and learning rate significantly impact the model's performance. While larger networks with more layers and nodes can model complex data more effectively, they also risk overfitting, where the model performs well on training data but poorly on unseen data~\cite{hawkins2004overfitting, srivastava}. Proper tuning and regularization are essential to balance model complexity and generalization. Performance test results of different NN setups will be presented in Sect.~\ref{sec:Res_Dis}.

\subsubsection{Regression Methods}\label{subsec:Regression}
The regression task aims to predict the continuous signed values of the Smagorinsky constant ($C_s$) used in turbulence modeling. We employed feedforward NN for this purpose. A dense, fully connected NN was constructed to predict $C_s$ values. The NN training involved minimizing the Mean Squared Error (MSE) between the predicted and true $C_s$ values defined as:
\begin{equation}
\text{MSE} = \frac{1}{n} \sum_{i=1}^{n} (y_i - \hat{y}_i)^2,
\end{equation}
where \( y_i \) represents the true values, \( \hat{y}_i \) represents the predicted values, and \( n \) is the number of data points. We optimized the network using Bayesian regularization~\cite{Hagan-1997, MacKay-1992} to enhance weight estimation. Our recent work has highlighted the effectiveness of invariance-embedded machine learning models for SGS stress predictions, with different optimization strategies yielding varying accuracy and computational efficiency~\cite{hasan2025asme}.

The regression network's performance was evaluated through mean squared error metrics across different configurations, including varying the number of neurons and layers. An ensemble approach, combining outputs from multiple networks, was employed to improve prediction accuracy and reduce error variance.

\section{Results and Discussions} \label{sec:Res_Dis}
\subsection{Results from Classification Models}\label{sec:Results_class}
\subsubsection{Pointwise Learning}\label{subsec:Results_p2p}
After data preparation, we proceeded to the training phase using a feed-forward NN architecture. In pointwise learning, the network was tasked with classifying points based on the magnitude of the Smagorinsky coefficient, $C_s$, using the five tensor invariances shown in Table~\ref{tab:tensor_invariants} as input features. The tensor invariances were divided into positive and negative groups based on a threshold of 0.001 for the magnitude of $C_s$. Labels were assigned accordingly: 1 for positive and 0 for negative samples. We employed a linear model structure, exploring various configurations to assess their influence on the model's predictive capabilities.

\begin{table}[h]
\centering
\caption{Classification Neural Network Model Configurations}
\label{tab:model_configurations}
\begin{tabular}{ccccc}
\toprule
Model Names & \quad No. of Neurons \quad & \quad No. of Layers \quad & \quad Activation Function \quad & \quad No. of Epochs \quad \\
\midrule
NN-1 & 10   & 2 & ReLU  & 50,000 \\
NN-2 & 50   & 2 & ReLU  & 50,000 \\
NN-3 & 200  & 2 & ReLU  & 50,000 \\
NN-4 & 10   & 2 & Tanh  & 50,000 \\
NN-5 & 50   & 2 & Tanh  & 50,000 \\
NN-6 & 200  & 2 & Tanh  & 50,000 \\
\bottomrule
\end{tabular}
\end{table}

We conducted a series of experiments to understand the effect of network complexity on learning efficiency and generalization. The configurations differed in the number of neurons per layer (10, 50, and 200) and the type of activation function used (ReLU and Tanh) as summarized in Table~\ref{tab:model_configurations}.

\begin{figure}[htbp]
    \centering
    \includegraphics[width=0.49\linewidth]{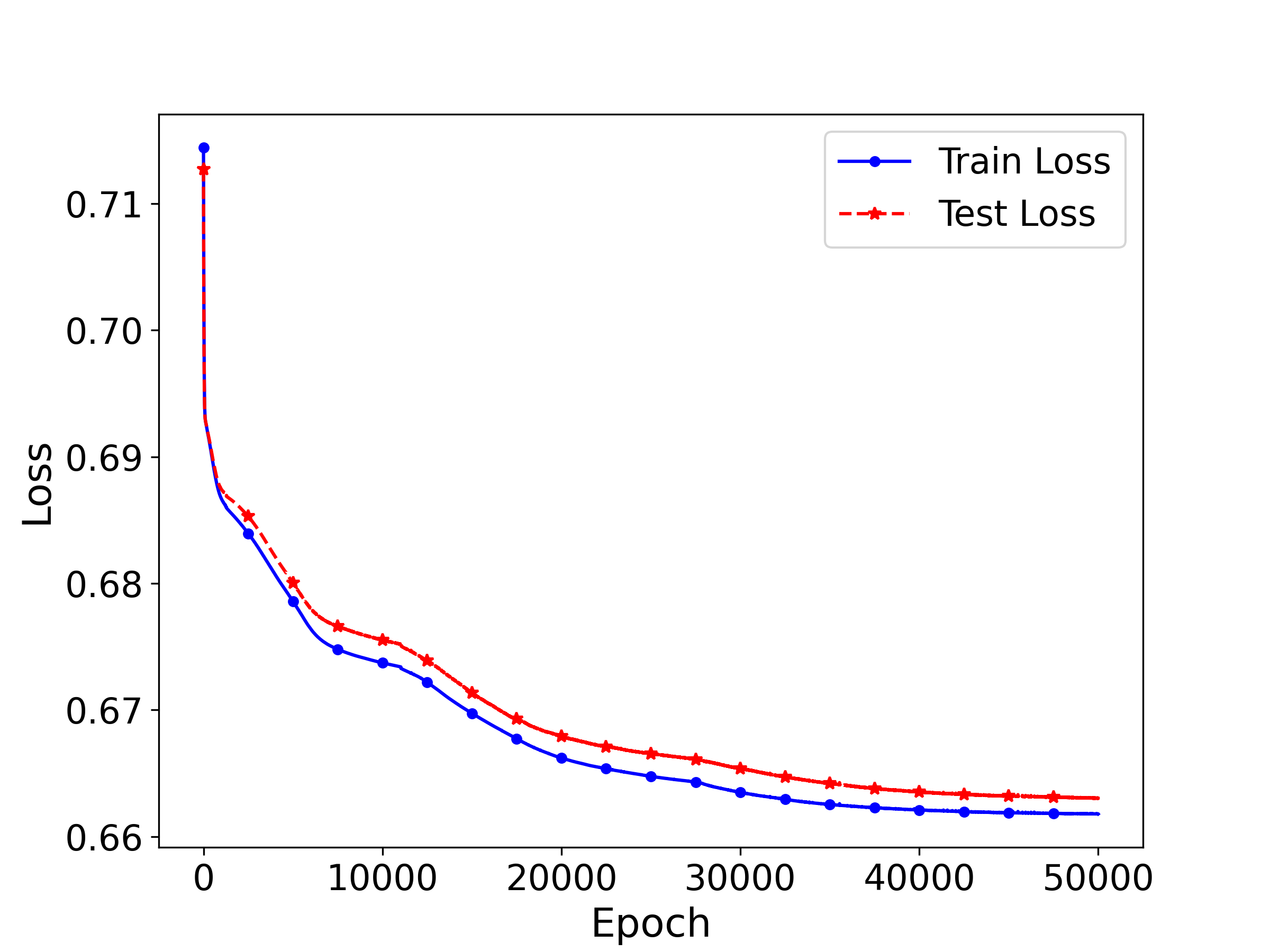}
    \includegraphics[width=0.49\linewidth]{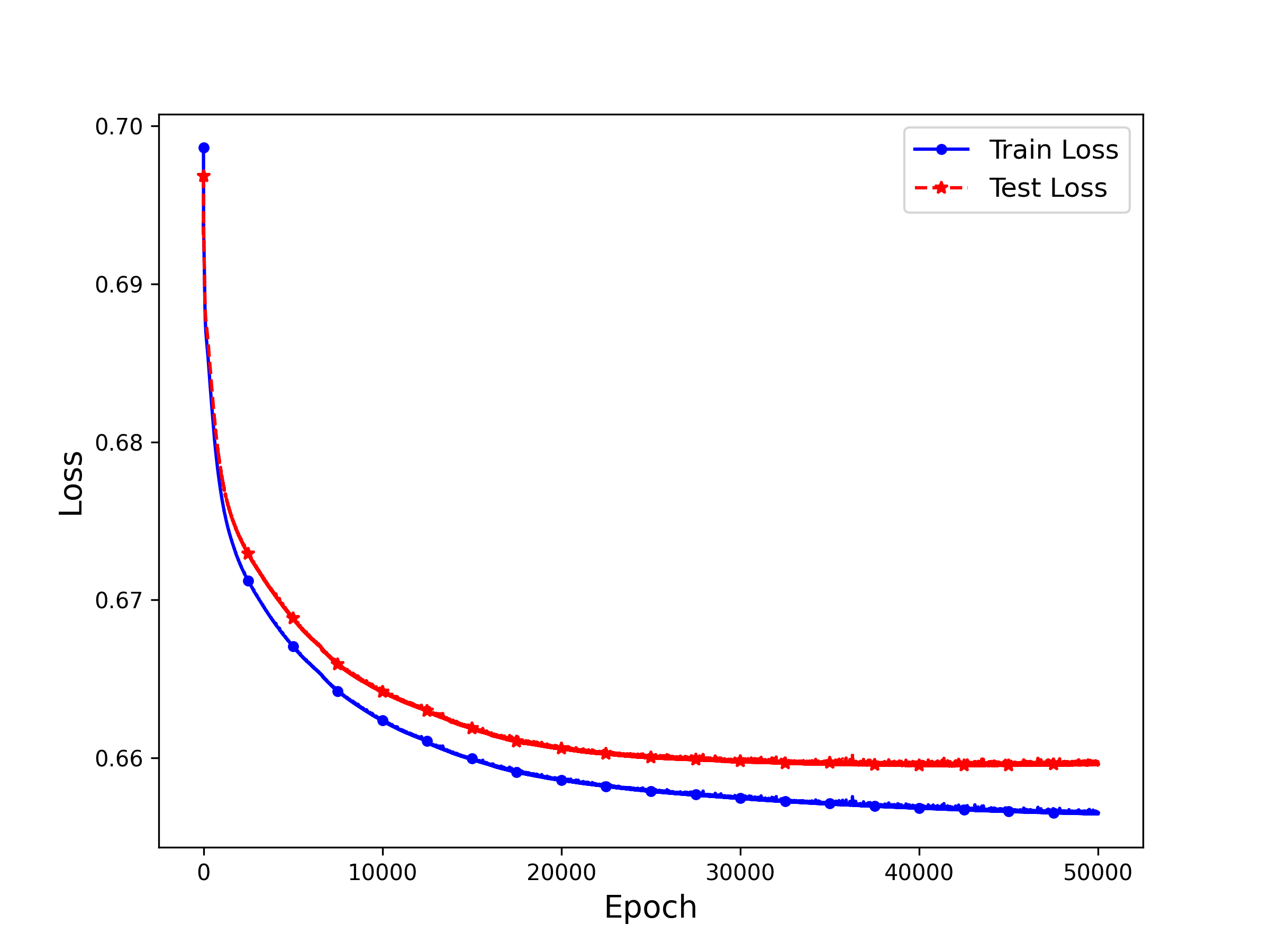}

    \vspace{1ex}
    \begin{minipage}[t]{0.49\linewidth}
        \centering \small (a) NN-1
    \end{minipage}%
    \begin{minipage}[t]{0.49\linewidth}
        \centering \small (b) NN-3
    \end{minipage}

    \vspace{1.5ex}
    \includegraphics[width=0.49\linewidth]{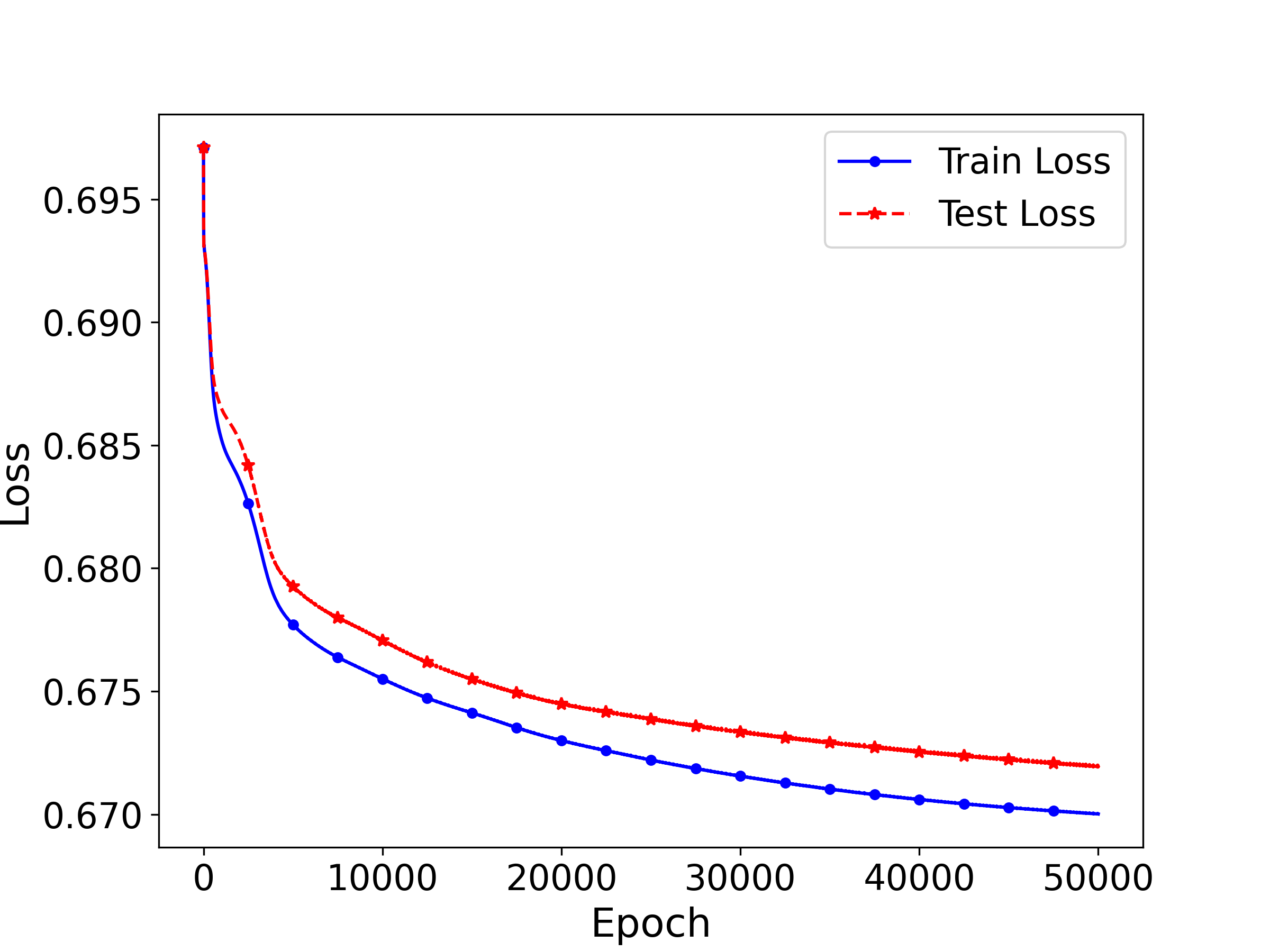}

    \vspace{1ex}
    \begin{minipage}[t]{0.49\linewidth}
        \centering \small (c) NN-4
    \end{minipage}

    \caption{Convergence performance of three typical NNs, i.e., NN-1, NN-3, and NN-4, when pointwise learning is used to train classification NNs.}
    \label{fig:nn_1}
\end{figure}

The loss curves in Figure~\ref{fig:nn_1} illustrate the training and test loss across epochs for typical configurations, NN-1, NN-3, and NN-4, highlighting the convergence behavior of each model. NN-1 and NN-4 configurations demonstrate a steady reduction in loss over the 50,000 epochs, with the training loss generally lower than the test loss, indicating effective learning but also a possible minor overfitting tendency, as seen in the consistent gap between the two losses. Interestingly, NN-3, despite having significantly more neurons (200 per layer), does not exhibit notably better convergence behavior than simpler models like NN-1 and NN-4. This suggests that increasing model complexity by adding neurons does not necessarily lead to enhanced model performance or improved generalization. The gradual decline in all configurations' losses suggests that the models effectively learn to classify more accurately as training progresses. The near-convergence of the training and test losses towards the end of training implies a stabilized learning process, where the models balance learning and generalization.

The quantitative performance of each model configuration is summarized in Table~\ref{tab:model_metrics_1}, presenting key classification metrics. The accuracy reflects the proportion of true results (both true positives and true negatives) among the total number of cases examined. Precision assesses the model's ability to classify the positive class correctly, while recall measures the model's capability to identify all positive cases. The F1-Score combines precision and recall into a single metric, balancing the model's performance. Here, the formulas for calculating Accuracy, Precision, Recall, and F1-Score are given below:
\[
\text{Accuracy} = \frac{\text{True Positives (TP)} + \text{True Negatives (TN)}}{\text{TP} + \text{TN} + \text{False Positives (FP)} + \text{False Negatives (FN)}}
\]

\[
\text{Precision} = \frac{\text{True Positives (TP)}}{\text{True Positives (TP)} + \text{False Positives (FP)}}
\]

\[
\text{Recall} = \frac{\text{True Positives (TP)}}{\text{True Positives (TP)} + \text{False Negatives (FN)}}
\]

\[
\text{F1 Score} = 2 \times \frac{\text{Precision} \times \text{Recall}}{\text{Precision} + \text{Recall}}
\]
where True Positives (TP): Instances correctly predicted as 1,  False Positives (FP): Instances incorrectly predicted as 1, False Negatives (FN): Instances incorrectly predicted as 0, and True Negatives (TN): Instances correctly predicted as 0.

\begin{table}[h]
\centering
\caption{NN Model Performance Metrics for Pointwise Training}
\label{tab:model_metrics_1}
\begin{tabular}{ccccc}
\toprule
          Model Configuration &  \quad Accuracy \quad &  \quad Precision \quad &  \quad Recall \quad &  \quad F1-Score \\
\midrule
 NN-1 &      0.59 &       0.60 &   0.55 &     0.57 \\
 NN-2 &      0.59 &       0.59 &   0.61 &     0.60 \\
 NN-3 &      0.60 &       0.59 &   0.61 &     0.60 \\
 NN-4 &      0.59 &       0.60 &   0.51 &     0.55 \\
 NN-5 &      0.58 &       0.59 &   0.52 &     0.55 \\
 NN-6 &      0.58 &       0.59 &   0.54 &     0.56 \\
\bottomrule
\end{tabular}
\end{table}

The metrics in Table~\ref{tab:model_metrics_1} demonstrate a relatively uniform accuracy across all configurations, with most models achieving scores around 0.58 to 0.60. This moderate accuracy, paired with precision values close to 0.59 and recall values ranging from 0.51 to 0.61, suggests that while the models can identify positive cases reasonably well, they may struggle with balancing precision and recall consistently across configurations. The F1-Scores, which combine precision and recall, show limited variation across models, indicating that the number of neurons and activation functions did not substantially impact the model’s ability to balance precision and recall. This consistency in F1-Scores and overall performance implies that increasing the model complexity in terms of neurons and activation function did not significantly improve the classification performance of pointwise learning solely with physical invariance embedded into the NN model.

\subsubsection{Group Learning without Geometric Invariance}
In this section, we developed a custom function to capture local spatial relationships by correlating each data point with its neighbors (Figure~\ref{fig:data_1}(b)). Tensor invariants were grouped based on previously described threshold criteria. Balanced sampling ensured representative data coverage and the resulting dataset was randomized and converted to PyTorch tensors for neural network training.

\begin{figure}[htbp]
    \centering
    \includegraphics[width=0.49\linewidth]{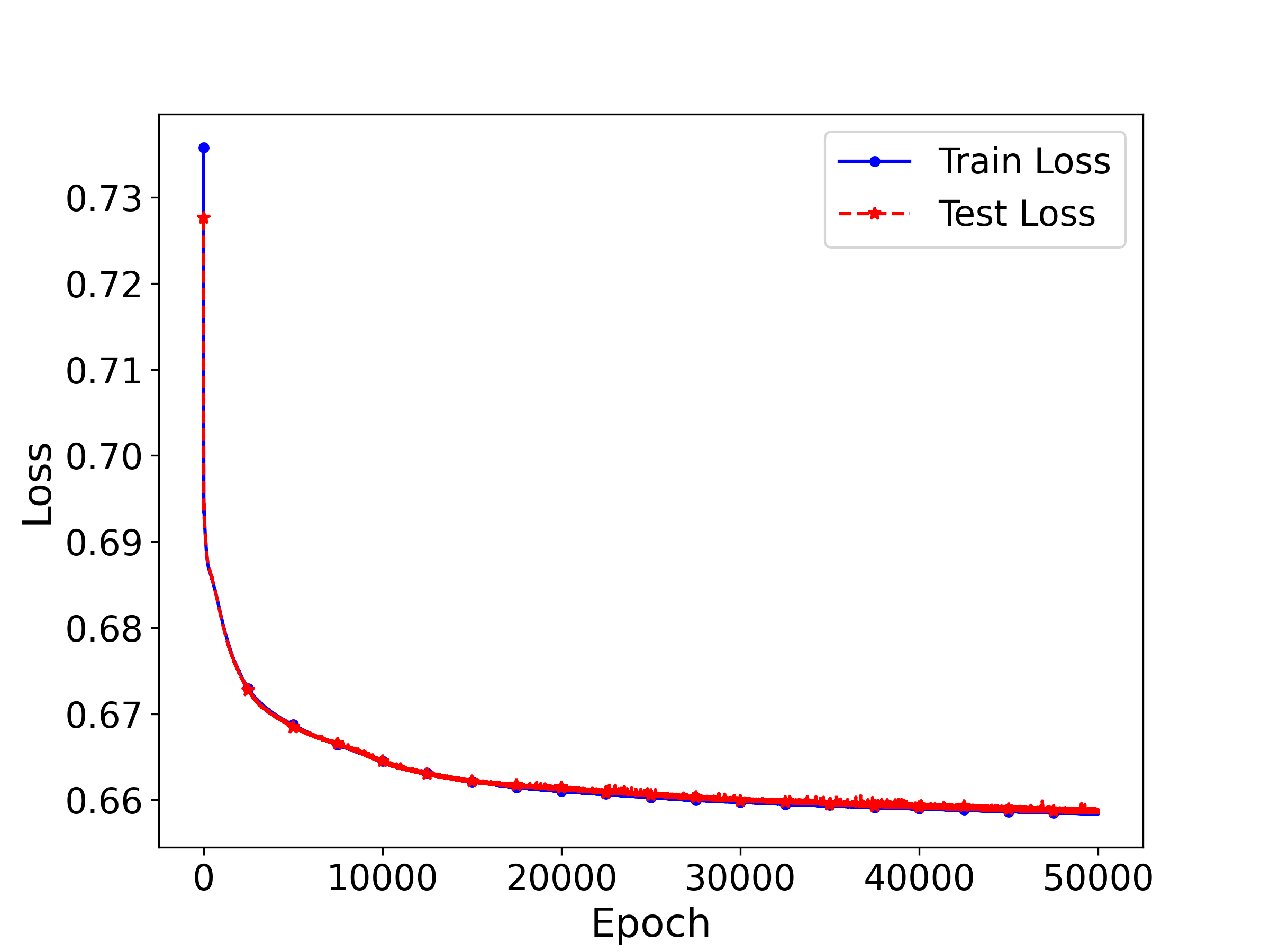}
    \includegraphics[width=0.49\linewidth]{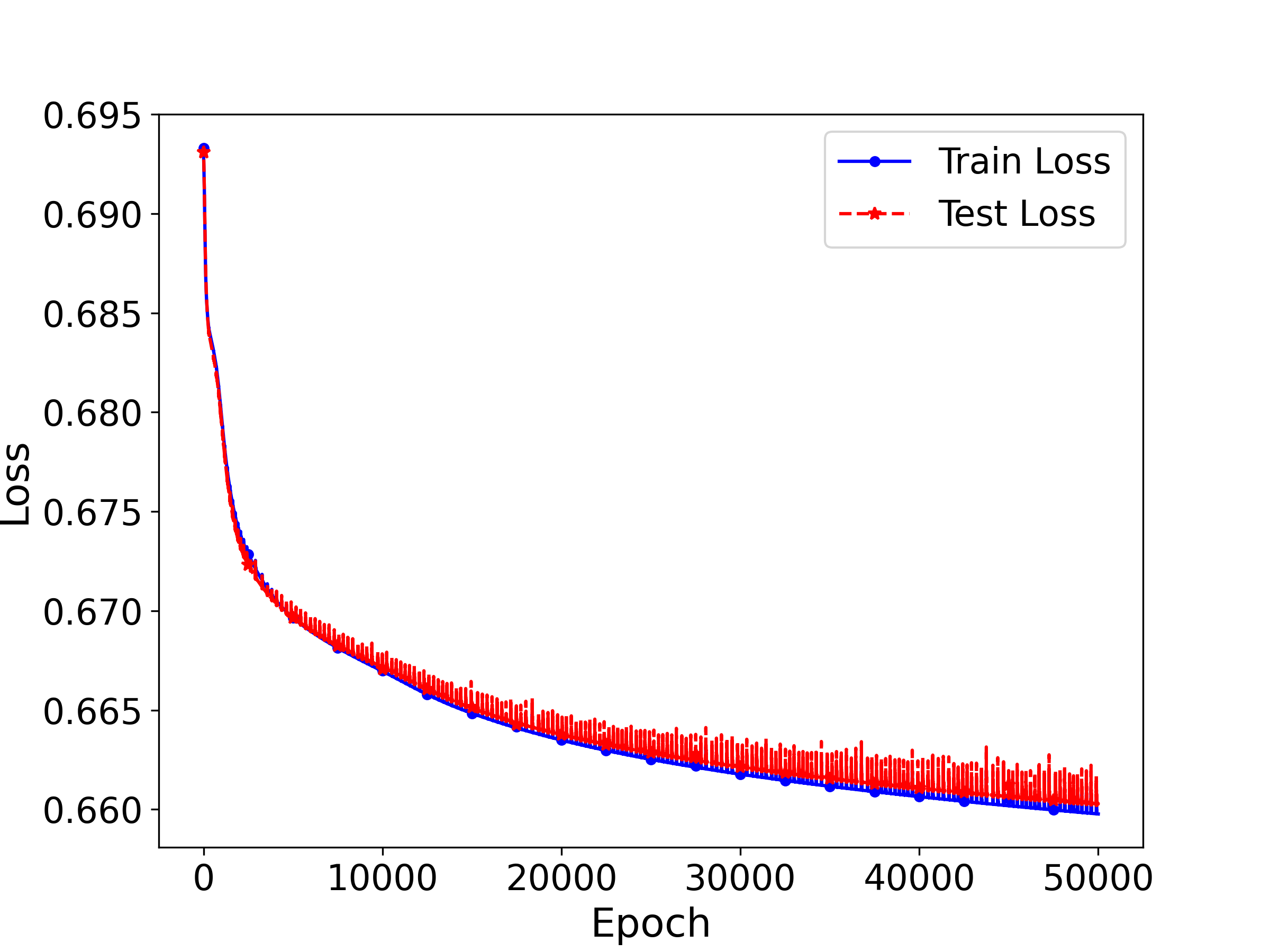}

    \vspace{1ex}
    \begin{minipage}[t]{0.49\linewidth}
        \centering \small (a) NN-1
    \end{minipage}%
    \begin{minipage}[t]{0.49\linewidth}
        \centering \small (b) NN-4
    \end{minipage}

    \caption{Convergence performance of two typical NNs, i.e., NN-1 and NN-4, when group learning without geometric invariance is used to train classification NNs.}
    \label{fig:nn_2}
\end{figure}

Similar NN from Sub-section~\ref{subsec:Results_p2p} were tested with the new dataset under a group learning approach without incorporating geometric invariance. As shown in Figure~\ref{fig:nn_2}, the training and test loss over epochs reveal the learning progression and provide insights into the model’s generalization capabilities. The models with more neurons show a faster convergence, indicating an enhanced capacity to capture complex data patterns. However, the training and test loss curves remain closely aligned across configurations, suggesting that overfitting was effectively controlled, even with higher neuron counts.

\begin{table}[h]
\centering
\caption{NN Model Performance Metrics for Group Learning without Geometric Invariance}
\label{tab:model_metrics_2}
\begin{tabular}{ccccc}
\hline
Model Configuration & \quad Accuracy \quad & \quad Precision \quad & \quad Recall \quad & \quad F1-Score \\ \hline
NN-1 & 0.60 & 0.62 & 0.51 & 0.56 \\
NN-2 & 0.61 & 0.60 & 0.67 & 0.63 \\
NN-3 & 0.62 & 0.61 & 0.67 & 0.64 \\
NN-4 & 0.60 & 0.61 & 0.54 & 0.57 \\
NN-5 & 0.60 & 0.61 & 0.54 & 0.57 \\
NN-6 & 0.60 & 0.61 & 0.55 & 0.58 \\ \hline
\end{tabular}
\end{table}

Table~\ref{tab:model_metrics_2} summarizes the model's performance, which presents accuracy, precision, recall, and F1-score for each configuration. The accuracy across configurations remains relatively uniform, around 0.60 to 0.62, with precision generally around 0.61 to 0.62. Recall values vary more widely, ranging from 0.51 to 0.67, indicating differences in the model's ability to identify positive cases across configurations. Higher recall values are observed in configurations with more neurons, reflecting their enhanced feature-capturing ability. Overall, the model with 200 neurons and ReLU activation (NN-3) achieved the highest scores across most metrics, suggesting that this configuration was most effective in capturing complex patterns in the data while balancing precision and recall. The consistent patterns across configurations indicate that increasing model complexity, specifically regarding neurons and activation functions, positively impacted classification performance under group learning without geometric invariance.

\subsubsection{Group Learning with Geometric Invariance}
In contrast to pointwise learning, group learning leverages spatial relationships by considering each data point and its neighboring points. Specifically, we focused on points within a defined annular region of the hurricane boundary layer due to their significance in capturing turbulence dynamics. Neural networks were trained using tensor invariants as inputs to classify the magnitude of the \( C_s \). SVD was employed to incorporate geometric invariances into the dataset, ensuring robustness against coordinate variations. Balanced sampling was implemented to maintain unbiased and representative data for effective neural network training.

\begin{figure}[htbp]
    \centering
    \includegraphics[width=0.49\linewidth]{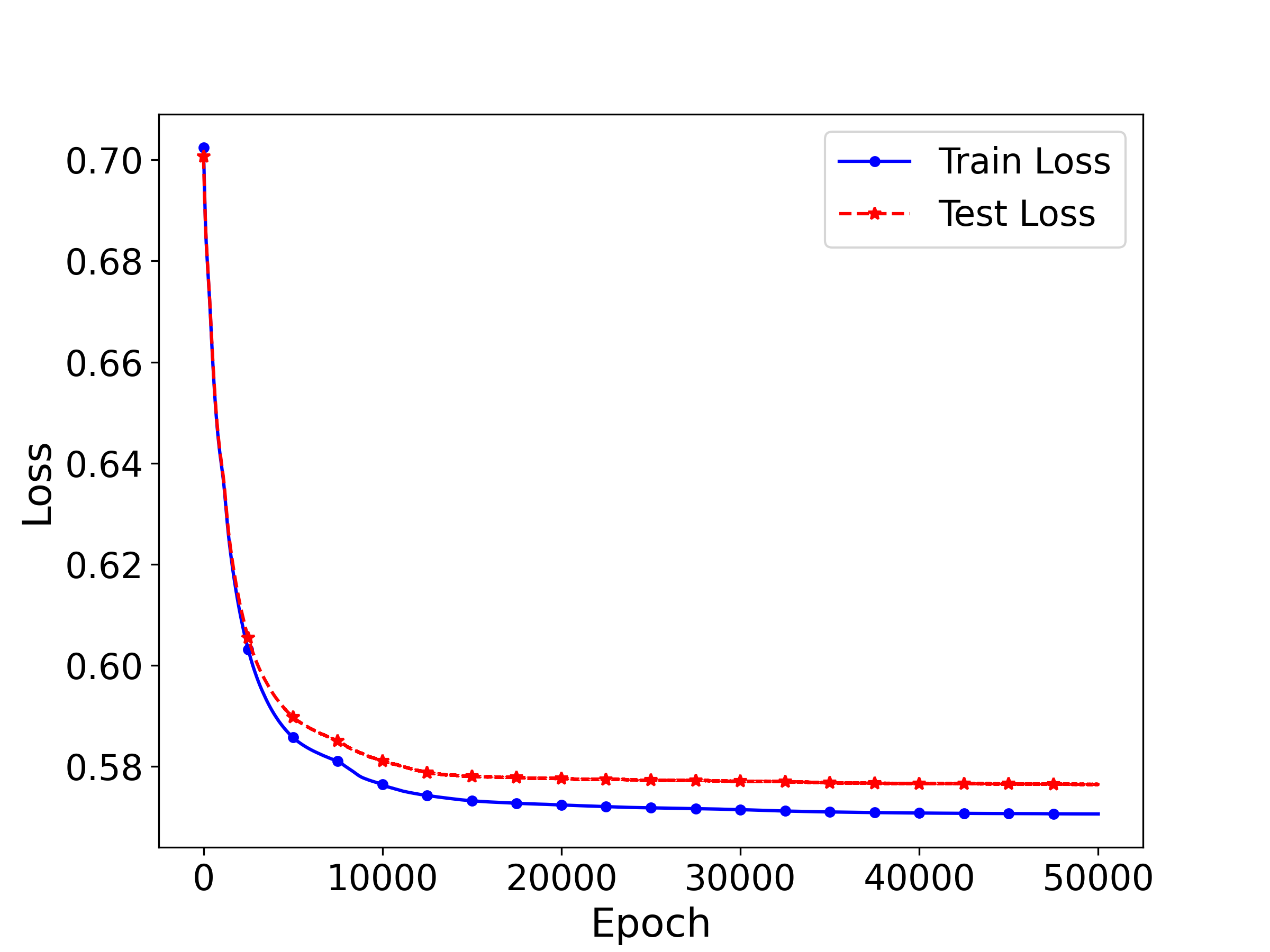}
    \includegraphics[width=0.49\linewidth]{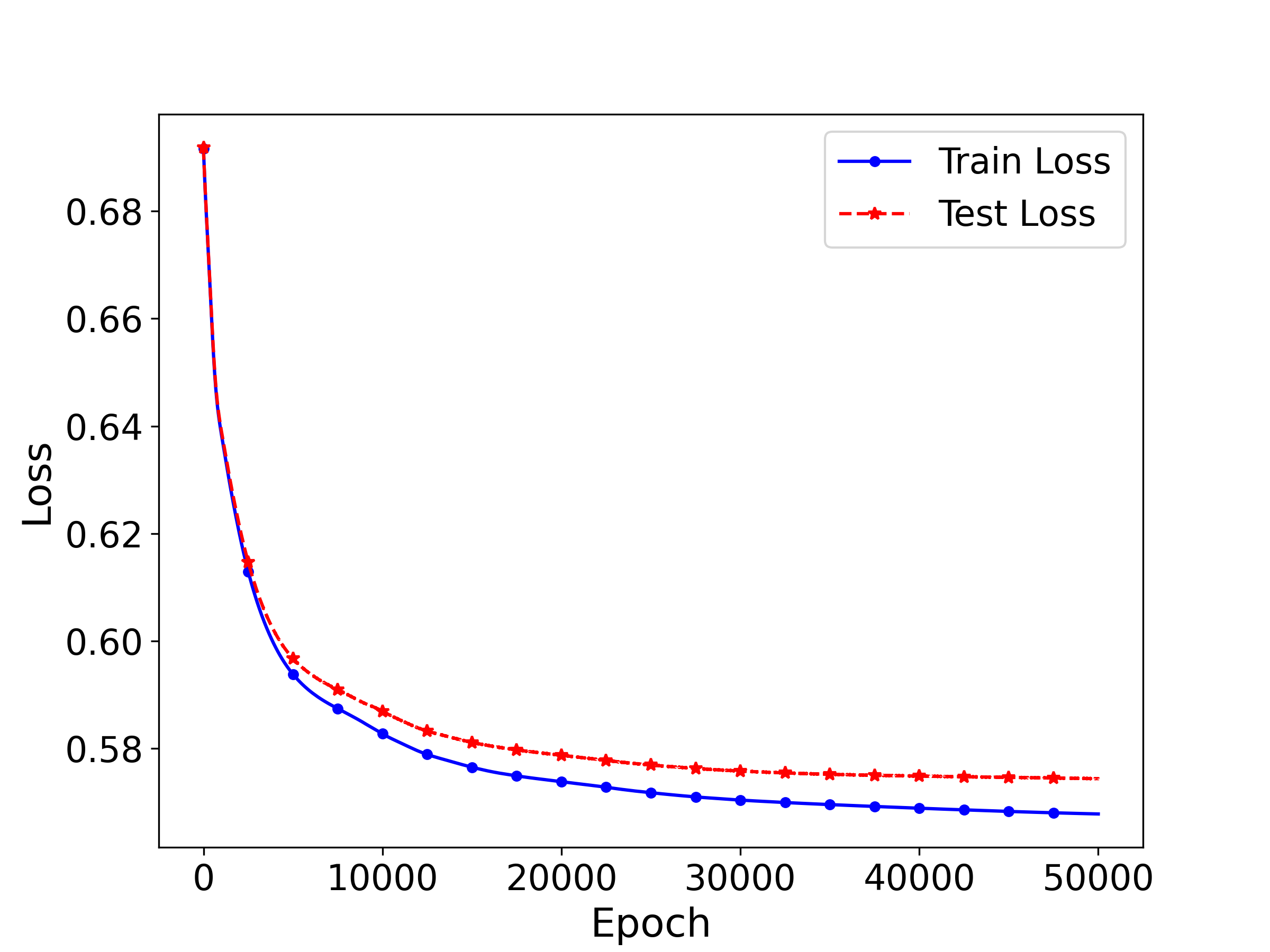}

    \vspace{1ex}
    \begin{minipage}[t]{0.49\linewidth}
        \centering \small (a) NN-1
    \end{minipage}%
    \begin{minipage}[t]{0.49\linewidth}
        \centering \small (b) NN-4
    \end{minipage}

    \caption{Convergence performance of NNs, i.e., NN-1 and NN-4, when group learning with geometric invariance on a $3 \times 3$ grid ($9$ points) is used to train classification NNs.}
    \label{fig:nn_3}
\end{figure}

Figure~\ref{fig:nn_3} presents the training and test loss curves over epochs for two representative configurations: NN-1 and NN-4. NN-1 exhibits a steady decline in training and test loss, reflecting a consistent learning process with minimal overfitting. NN-4, while following a similar trajectory, shows a slightly slower convergence, likely due to the smoother gradients provided by the Tanh activation function compared to ReLU.

\begin{table}
\centering
\caption{NN Model Performance Metrics for Group Learning with Geometric Invariance on a $3\times3$ grid ($9$ points)}
\label{tab:model_metrics_3}
\begin{tabular}{ccccc}
\toprule
Model Configuration \quad & \quad Accuracy \quad & \quad Precision \quad & \quad Recall \quad & \quad F1-Score \\
\midrule
NN-1 & 0.70 & 0.69 & 0.73 & 0.71 \\
NN-2 & 0.70 & 0.69 & 0.74 & 0.71 \\
NN-3 & 0.71 & 0.69 & 0.75 & 0.72 \\
NN-4 & 0.70 & 0.69 & 0.73 & 0.71 \\
NN-5 & 0.70 & 0.69 & 0.74 & 0.71 \\
NN-6 & 0.70 & 0.69 & 0.74 & 0.71 \\
\bottomrule
\end{tabular}
\end{table}

As presented in Table~\ref{tab:model_metrics_3}, the performance metrics reveal the classification results for all configurations. Networks with more neurons, such as NN-3, consistently outperform those with fewer neurons regarding recall (0.75) and F1-Score (0.72). This indicates that larger networks have a greater capacity to model complex patterns within the data, particularly when using ReLU activation. Conversely, models with Tanh activation, such as NN-6, show comparable performance but slightly slower convergence, as observed in the loss curves.

The study enhanced its data preprocessing method to optimize the classification of turbulence model parameters by incorporating an expanded neighboring context for singular value decomposition. The initial approach focused on a $3 \times 3$ grid (9 points) around a central point, while this enhanced approach considers a larger, $5 \times 5$ grid (25 points). Including a broader neighboring grid likely provides a more comprehensive view of the local turbulence structures. This wider perspective could capture variations in the turbulence characteristics that smaller neighborhoods might miss, allowing for a more nuanced understanding of the local interactions and dependencies. The enhanced feature set obtained via SVD from these 25-point neighborhoods then informs the training of the NN models.

\begin{figure}[htbp]
    \centering
    \includegraphics[width=0.49\linewidth]{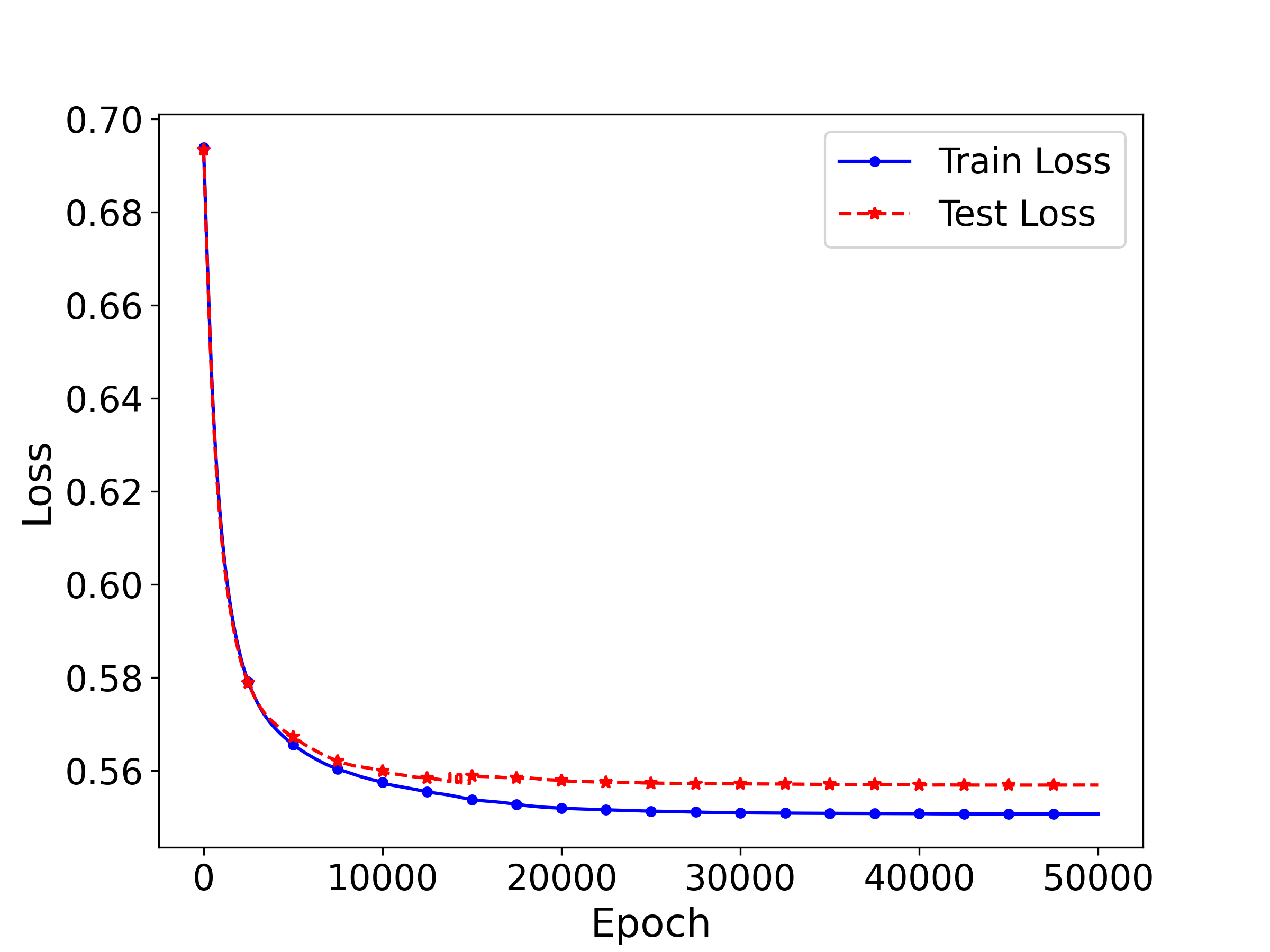}
    \includegraphics[width=0.49\linewidth]{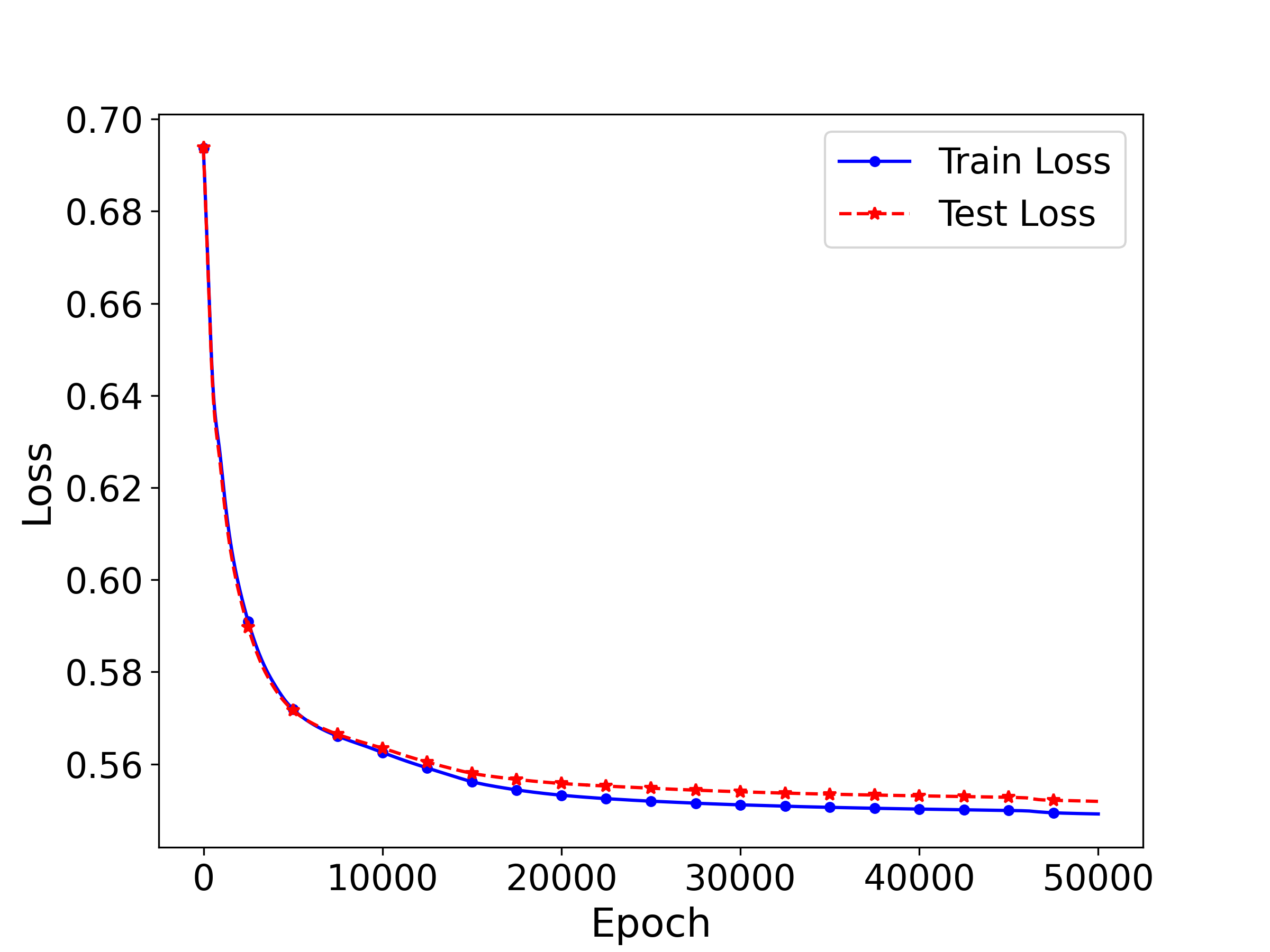}

    \vspace{1ex}
    \begin{minipage}[t]{0.49\linewidth}
        \centering \small (a) NN-1
    \end{minipage}%
    \begin{minipage}[t]{0.49\linewidth}
        \centering \small (b) NN-4
    \end{minipage}

    \caption{Convergence performance of two typical NNs, i.e., NN-1 and NN-4, when group learning with geometric invariance on a $5 \times 5$ grid ($25$ points) is used to train classification NNs.}
    \label{fig:nn_4}
\end{figure}

In Figure~\ref{fig:nn_4}, models trained on a $5 \times 5$ grid exhibit similar trends. The training and test loss curves for NN-1 and NN-4 show a rapid decline in the initial epochs, followed by a plateau, indicating that the expanded spatial context aids in capturing relevant features early in the training process. The consistent alignment between training and test loss curves demonstrates effective generalization.

Table~\ref{tab:model_metrics_4} provides the performance metrics for the $5 \times 5$ grid. Across all configurations, accuracy improves to 0.72, and recall increases to 0.76 for models with 200 neurons (NN-3, NN-6). This improvement in recall reflects the model’s enhanced ability to identify true positives, likely attributable to the larger spatial context in the $5\times5$ grid, which provides richer information for classification.

\begin{table}
\centering
\caption{NN Model Performance Metrics for Group Learning with Geometric Invariance on a $5 \times 5$ grid ($25$ points)}
\label{tab:model_metrics_4}
\begin{tabular}{ccccc}
\toprule
Model Configuration \quad & \quad Accuracy & \quad Precision \quad & \quad Recall \quad & \quad F1-Score \\
\midrule
NN-1 & 0.72 & 0.70 & 0.75 & 0.72 \\
NN-2 & 0.72 & 0.70 & 0.76 & 0.73 \\
NN-3 & 0.72 & 0.70 & 0.76 & 0.73 \\
NN-4 & 0.72 & 0.70 & 0.75 & 0.73 \\
NN-5 & 0.72 & 0.70 & 0.76 & 0.73 \\
NN-6 & 0.72 & 0.70 & 0.76 & 0.73 \\
\bottomrule
\end{tabular}
\end{table}

The results indicate that increasing the neighborhood size for SVD significantly impacts performance metrics, particularly recall and F1-score. The broader spatial context in the 5x5 grid enables the models to capture better local turbulence structures, which are crucial for accurately classifying turbulence dynamics. The improved metrics suggest that spatial relationships in the input features play a more critical role in enhancing classification performance than merely increasing the NN’s complexity. The consistent architecture across configurations highlights the importance of balancing network complexity and activation function to achieve optimal learning. ReLU-based configurations generally converge faster and achieve better classification metrics due to their ability to handle non-linearities and avoid vanishing gradients. Meanwhile, Tanh-based configurations demonstrate a more gradual learning process, which may lead to better generalization under certain conditions.

In summary, transitioning from a $3 \times 3$ grid to a $5 \times 5$ grid demonstrates the importance of incorporating broader spatial contexts in turbulence modeling. The enhanced classification performance across all metrics reinforces the potential of machine learning models, combined with geometric invariance, to provide more accurate and reliable predictions in complex fluid dynamics. These findings underscore the critical role of spatial feature representation in advancing machine-learning-based approaches for turbulence modeling.

\subsection{Results from Regression Models}\label{sec:Results_regression}

This section evaluates the performance of NN models in predicting the signed Smagorinsky coefficient ($C_s$), a critical parameter for turbulence modeling. Both single NNs and ensemble methods are assessed for their effectiveness and reliability in capturing complex patterns in the data.

\subsubsection{Single Neural Network Performance}

\begin{figure}[htbp]
    \centering
    \includegraphics[width=0.9\linewidth]{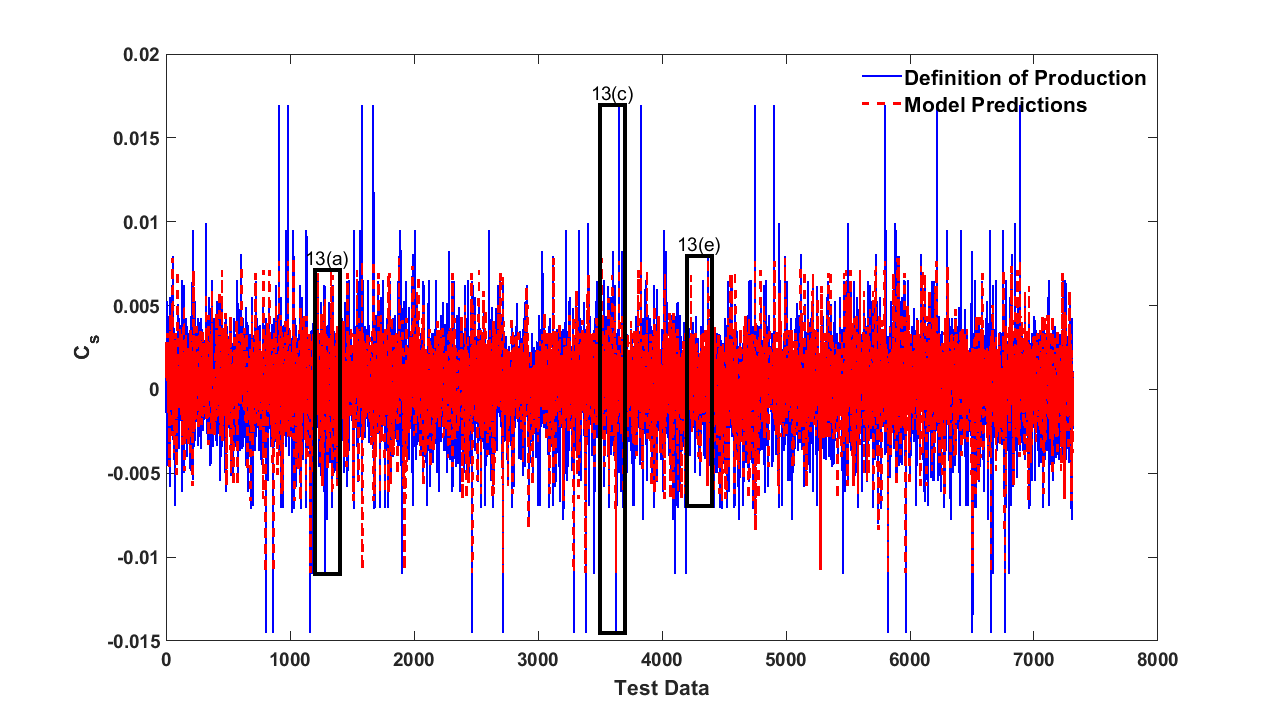}

    \vspace{0.5ex}
    \begin{minipage}{0.9\linewidth}
        \centering \small (a) Single Network: \(n = 8\), \(l = 2\)
    \end{minipage}

    \vspace{2ex}
    \includegraphics[width=0.9\linewidth]{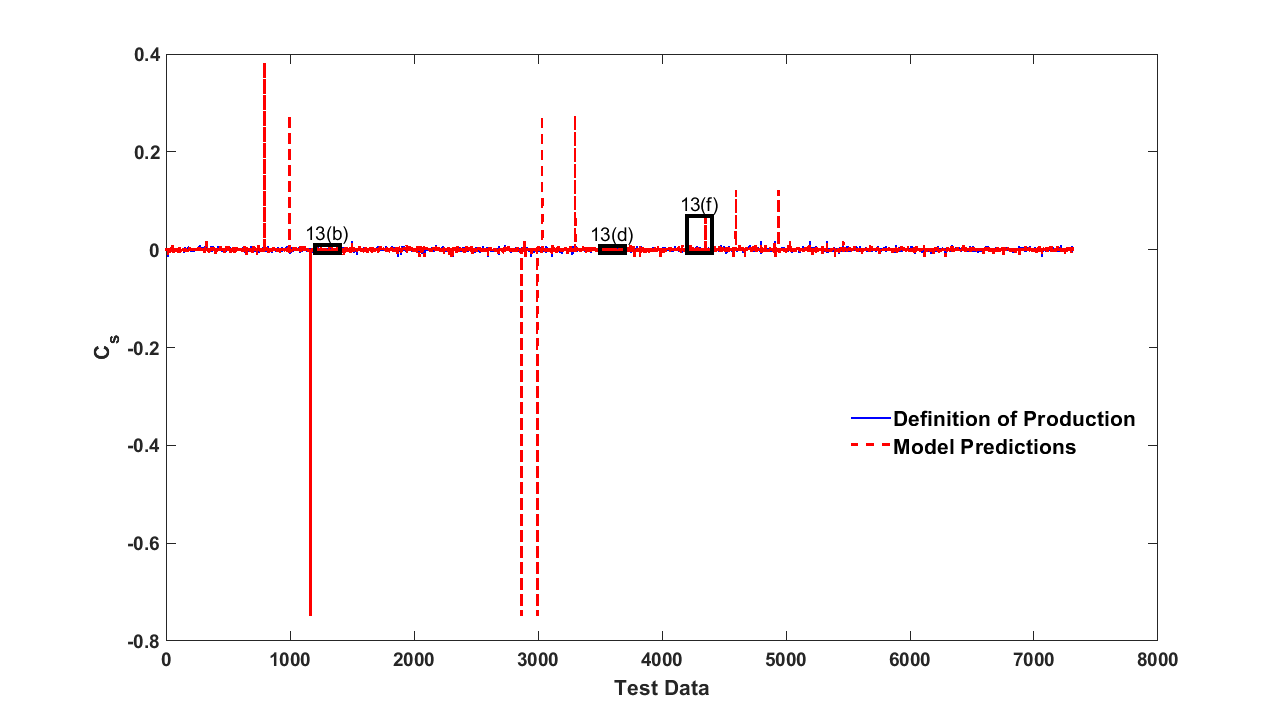}

    \vspace{0.5ex}
    \begin{minipage}{0.9\linewidth}
        \centering \small (b) Single Network: \(n = 32\), \(l = 2\)
    \end{minipage}

    \caption{Comparison of Definition of Production (blue solid line) and single NN model predictions (red dashed line) for two different network configurations. The highlighted boxes indicate regions selected for detailed examination in Figure~\ref{fig:single_NN_comp_zoom}.}
    \label{fig:single_nn_results}
\end{figure}

Figure~\ref{fig:single_nn_results} presents a detailed comparison of single NN predictions against the true values of the signed Smagorinsky coefficient. Two representative network configurations are analyzed: one relatively simple with fewer neurons (\(n=8, l=2\)) and a slightly more complex configuration (\(n=32, l=2\)). Herein, $n$ is the number of neurons within each hidden layer, and $l$ is the number of hidden layers in NN.

Both network configurations effectively capture the overall trend of \(C_s\) yet display variations in predictive fidelity, particularly in regions characterized by high variability. While the simpler network (Figure~\ref{fig:single_nn_results}(a)) demonstrates strong overall agreement with minor deviations, the more complex network (Figure~\ref{fig:single_nn_results}(b)) marginally improves alignment, suggesting enhanced capability to represent nuanced variations in the flow data.

\begin{figure}[htbp]
    \centering

    % Row 1
    \includegraphics[width=0.49\linewidth]{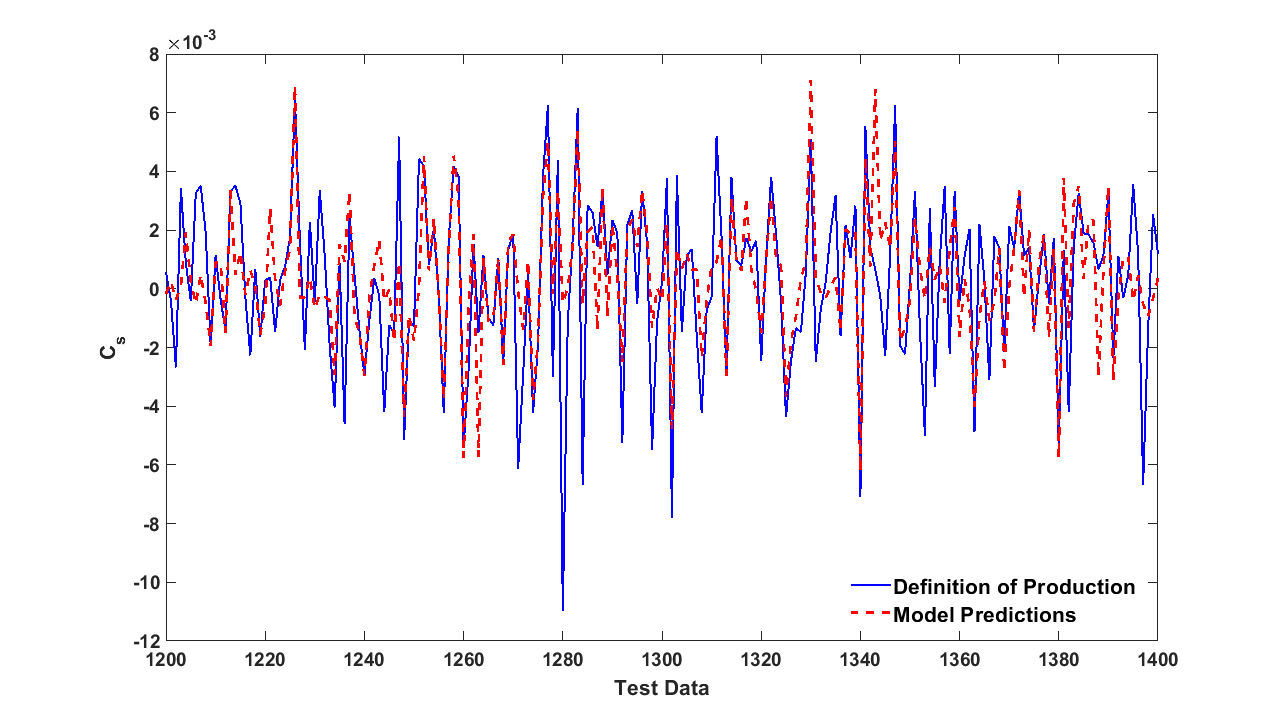}
    \includegraphics[width=0.49\linewidth]{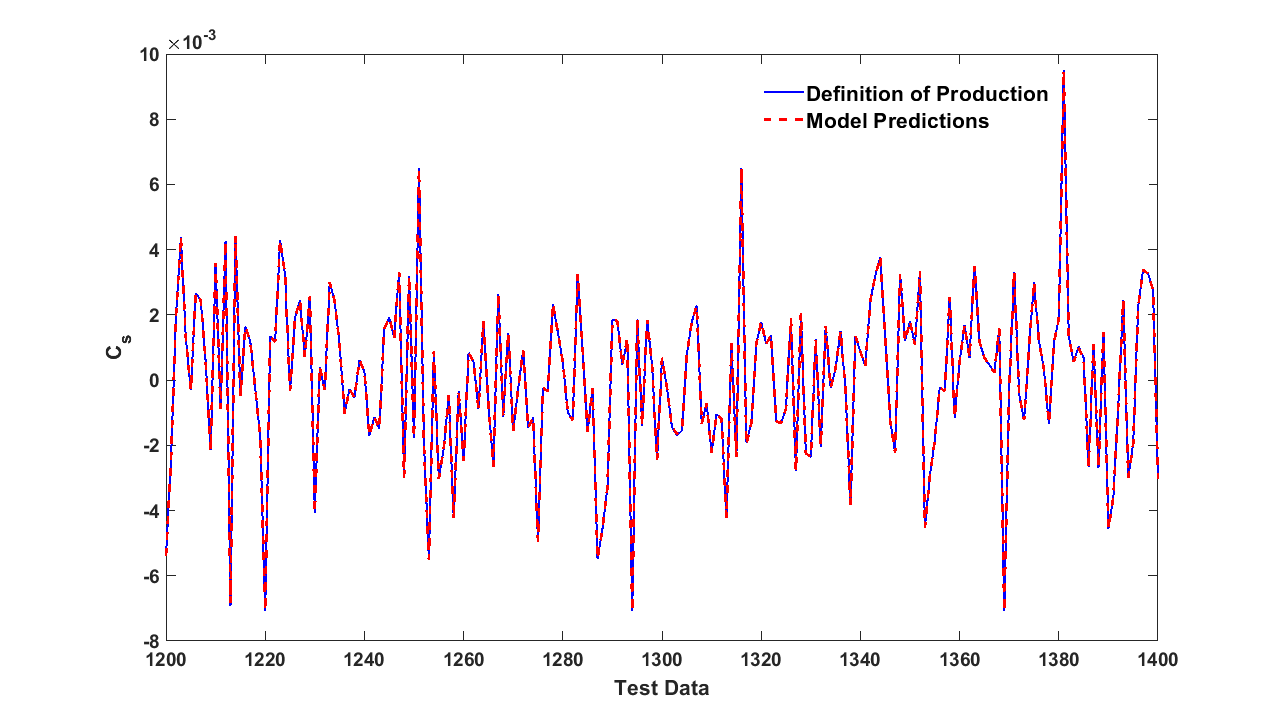}

    \vspace{0.5ex}
    \begin{minipage}[t]{0.49\linewidth}
        \centering \small (a) \(n = 8, l = 2\), Zoomed Area 1
    \end{minipage}%
    \begin{minipage}[t]{0.49\linewidth}
        \centering \small (b) \(n = 32, l = 2\), Zoomed Area 1
    \end{minipage}

    \vspace{2ex}

    % Row 2
    \includegraphics[width=0.49\linewidth]{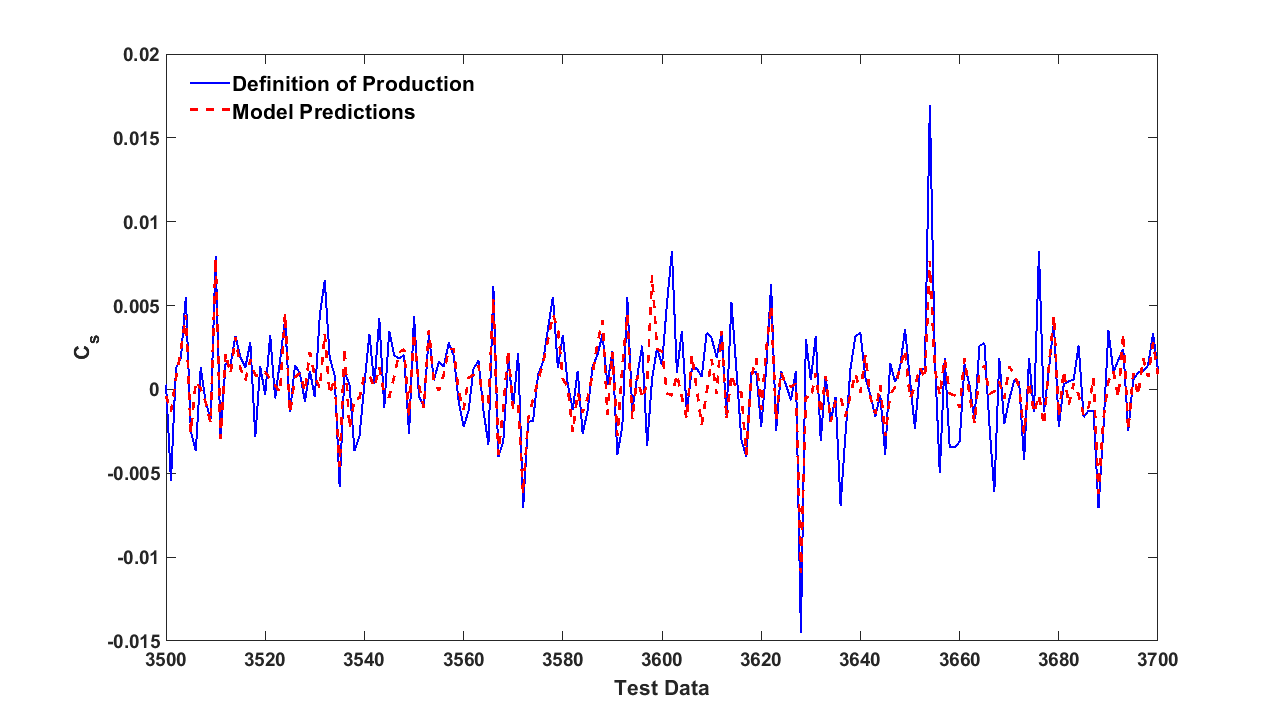}
    \includegraphics[width=0.49\linewidth]{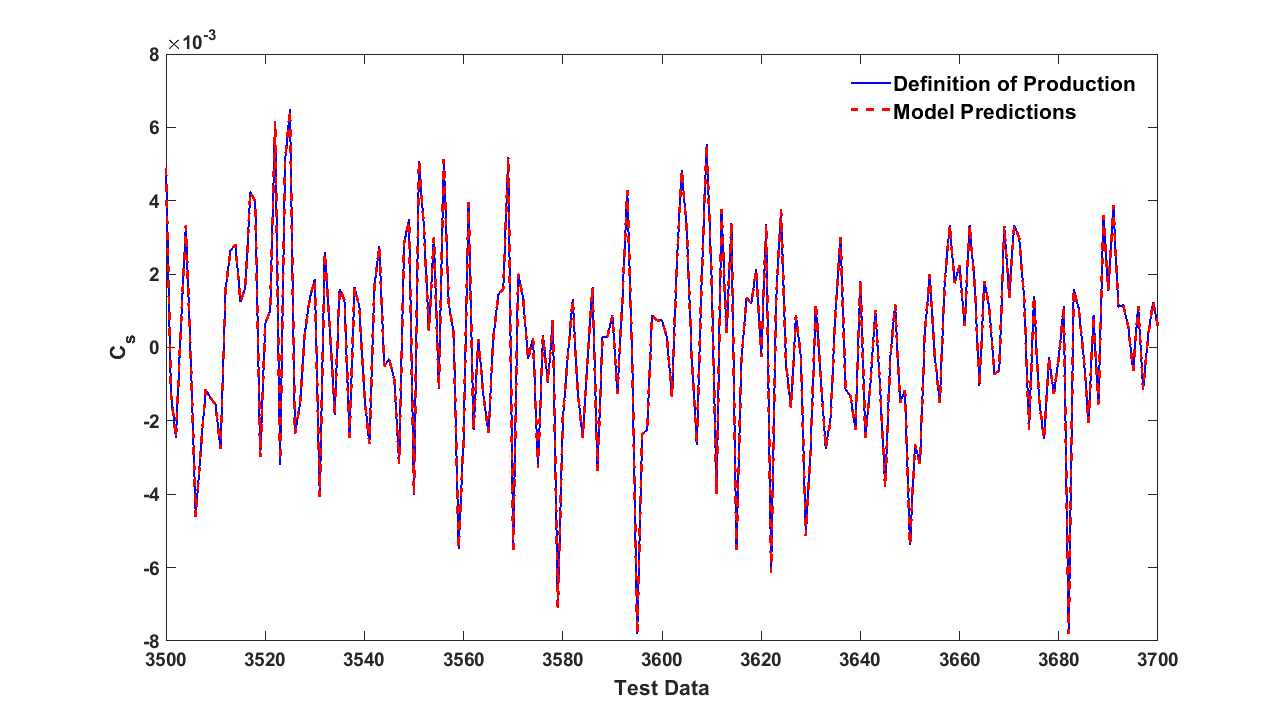}

    \vspace{0.5ex}
    \begin{minipage}[t]{0.49\linewidth}
        \centering \small (c) \(n = 8, l = 2\), Zoomed Area 2
    \end{minipage}%
    \begin{minipage}[t]{0.49\linewidth}
        \centering \small (d) \(n = 32, l = 2\), Zoomed Area 2
    \end{minipage}

    \vspace{2ex}

    % Row 3
    \includegraphics[width=0.49\linewidth]{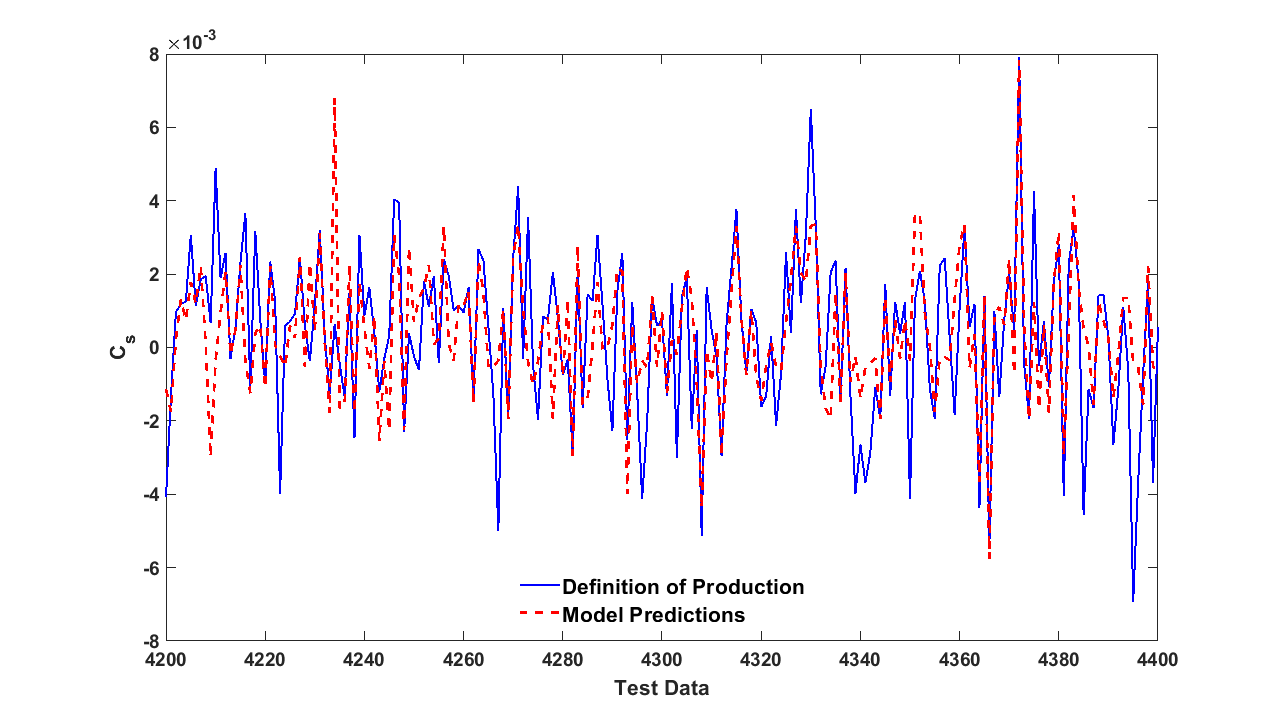}
    \includegraphics[width=0.49\linewidth]{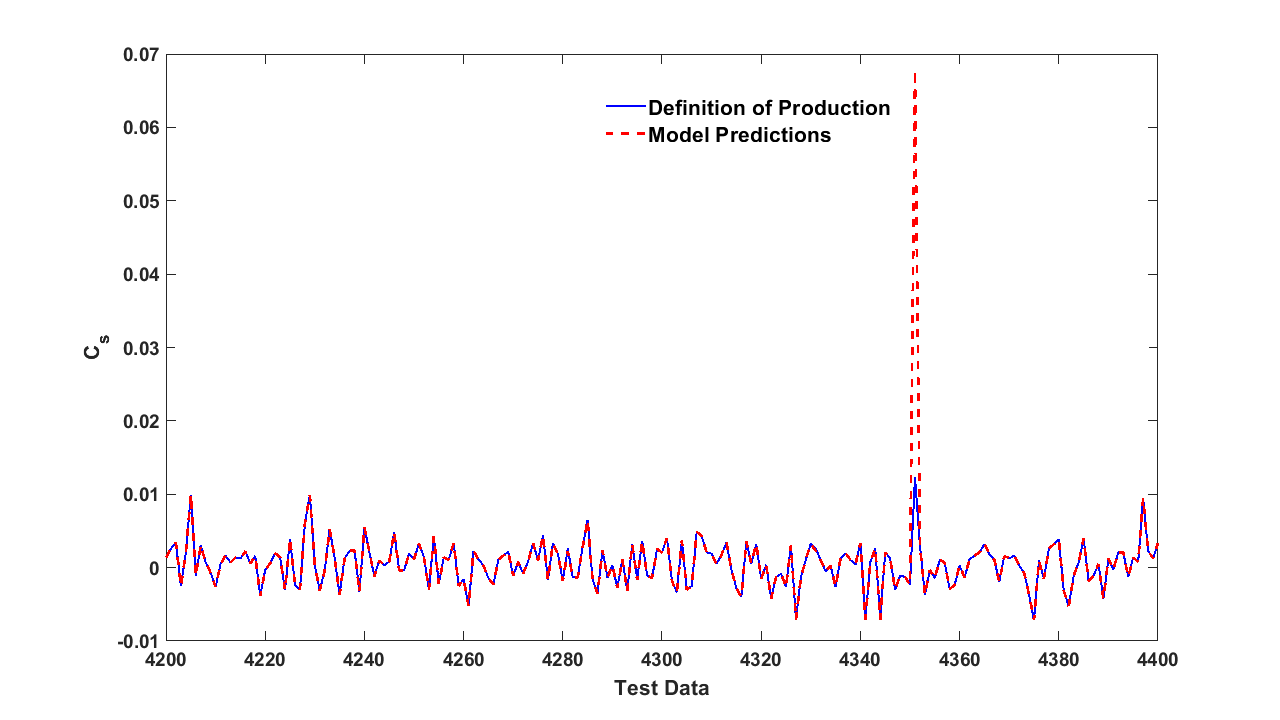}

    \vspace{0.5ex}
    \begin{minipage}[t]{0.49\linewidth}
        \centering \small (e) \(n = 8, l = 2\), Zoomed Area 3
    \end{minipage}%
    \begin{minipage}[t]{0.49\linewidth}
        \centering \small (f) \(n = 32, l = 2\), Zoomed Area 3
    \end{minipage}

    \caption{Detailed zoomed-in views of selected regions from Figure~\ref{fig:single_nn_results}, highlighting the specific prediction strengths and deviations of the two NN configurations.}
    \label{fig:single_NN_comp_zoom}
\end{figure}

To better illustrate specific model behaviors, zoomed-in views of highlighted regions from the main predictions are shown in Figure~\ref{fig:single_NN_comp_zoom}. These provide insights into the network’s local performance and areas of notable discrepancies.

These detailed views clarify how different network architectures respond to local fluctuations in turbulent flow data. The simpler network occasionally struggles with highly variable regions, resulting in visible prediction errors (Figures~\ref{fig:single_NN_comp_zoom}(a), \ref{fig:single_NN_comp_zoom}(c), \ref{fig:single_NN_comp_zoom}(e)). Conversely, the more complex network generally improves local predictions but occasionally shows larger errors, highlighting a trade-off between network complexity and predictive robustness.

Quantitative assessments of the model performance for various configurations are summarized in Table~\ref{tab:single_nn}, where networks with fewer neurons and moderate depth exhibit favorable MSE values, indicating their potential for better generalization across the complex hurricane boundary layer flows.
However, this better generalization feature is not favored in negative Smagorinsky coefficient $C_s$ prediction, which is the key to model energy backscatter. Note that in the work by Guan et al.~\cite{guan_stable_2022} with CNNs directly modeling SGS stresses, it has been demonstrated that an accurate evaluation of SGS stresses that contribute to energy backscatter is of prominent importance in reliable \textit{a posteriori} model tests.
We also mention that the runs presented in Table~\ref{tab:single_nn} represent instantaneous model evaluations without explicitly setting target MSE values; further improvements in predictive performance can be achieved by specifying desired target MSE thresholds and employing ensemble techniques. 

\begin{table}[!htb]
    \centering
    \caption{Test Data MSE for All Single Network Configurations}
    \label{tab:single_nn}
    \begin{tabular}{|c|c|c|}
        \hline
        Neurons (\(n\)) & Hidden Layers (\(l\)) & Test MSE \\
        \hline
        8 & 2 & \(3.57 \times 10^{-6}\)\\
        8 & 3 & \(4.42 \times 10^{-6}\)\\
        8 & 4 & \(1.82 \times 10^{-6}\)\\
        16 & 2 & \(1.13 \times 10^{-5}\)\\
        16 & 3 & \(3.51 \times 10^{-5}\)\\
        16 & 4 & \(2.82 \times 10^{-5}\)\\
        32 & 2 & \(2.87 \times 10^{-4}\)\\
        32 & 3 & \(5.59 \times 10^{-6}\)\\
        \hline
    \end{tabular}
\end{table}

\subsubsection{Ensemble Neural Network Performance}

\begin{figure}[htbp]
    \centering

    \includegraphics[width=0.8\linewidth]{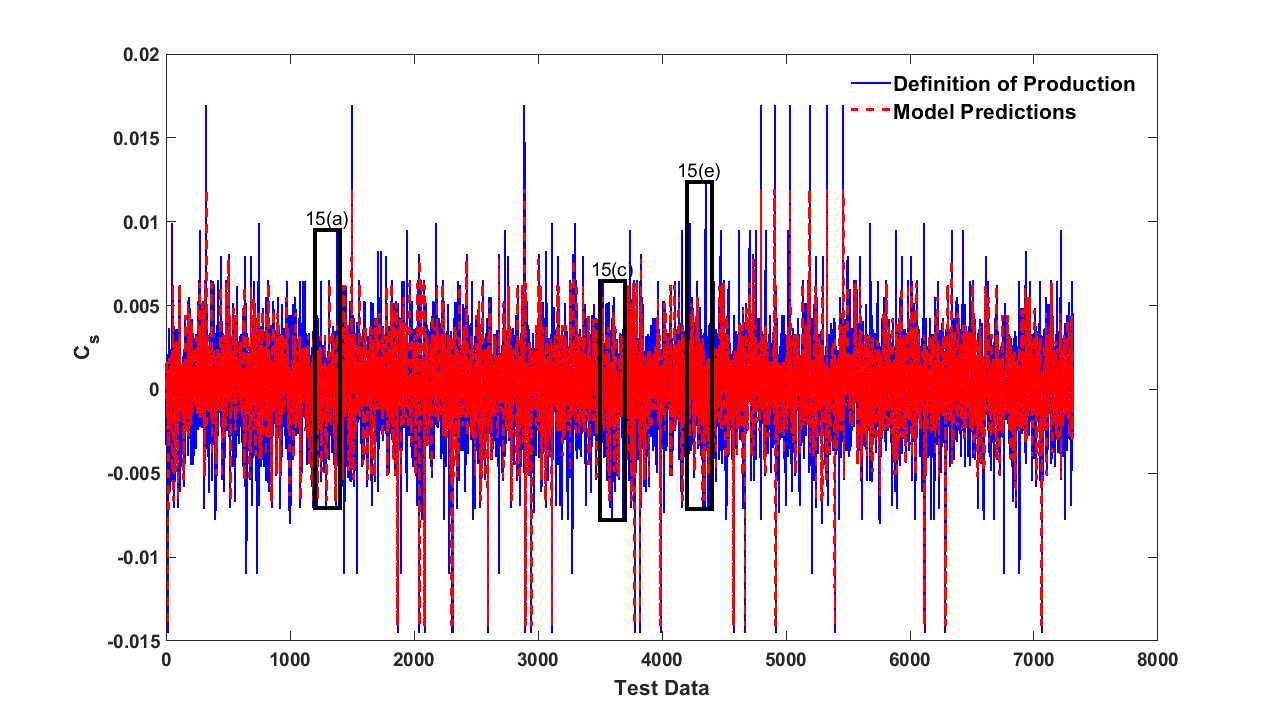}

    \vspace{0.5ex}
    \begin{minipage}{0.8\linewidth}
        \centering \small (a) Ensemble Network: $n = 8$, $l = 2$
    \end{minipage}

    \vspace{2ex}

    \includegraphics[width=0.8\linewidth]{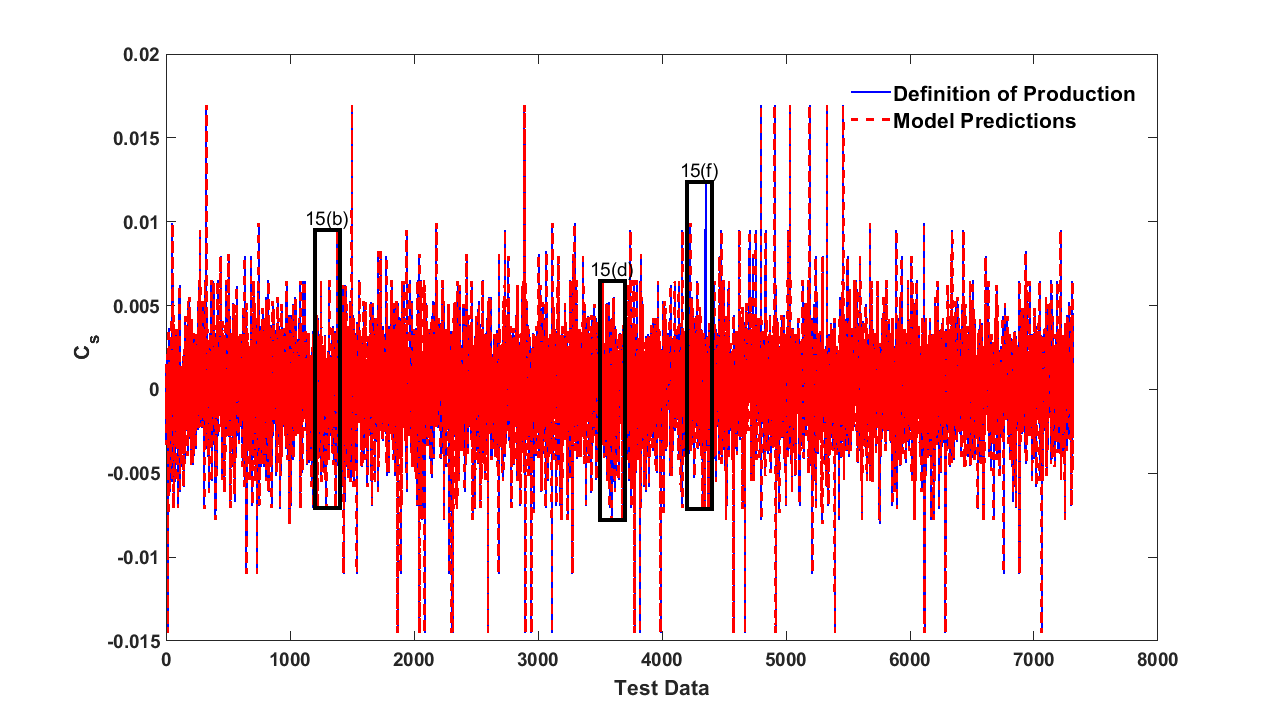}

    \vspace{0.5ex}
    \begin{minipage}{0.8\linewidth}
        \centering \small (b) Ensemble Network: $n = 32$, $l = 3$
    \end{minipage}

    \caption{Comparison of true values (blue) and ensemble network predictions (red) for two representative configurations. Highlighted boxes indicate zoomed regions explored in Figure~\ref{fig:ensemble_NN_comp_zoom}.}
    \label{fig:ensemble_nn_results}
\end{figure}

An ensemble NN approach was adopted to enhance accuracy and stability in predicting the Smagorinsky constant ($C_s$). Figure~\ref{fig:ensemble_nn_results} compares the true $C_s$ values (solid blue line) with predictions from ensemble NNs (red dashed line) for two representative configurations: $n=8, l=2$ and $n=32, l=3$. The ensemble method significantly reduces prediction variability and aligns closely with the actual data throughout the test dataset. This improved alignment emphasizes the ensemble's robustness in capturing both general trends and localized fluctuations of $C_s$.

\begin{figure}[htbp]
    \centering

    % Row 1
    \includegraphics[width=0.49\linewidth]{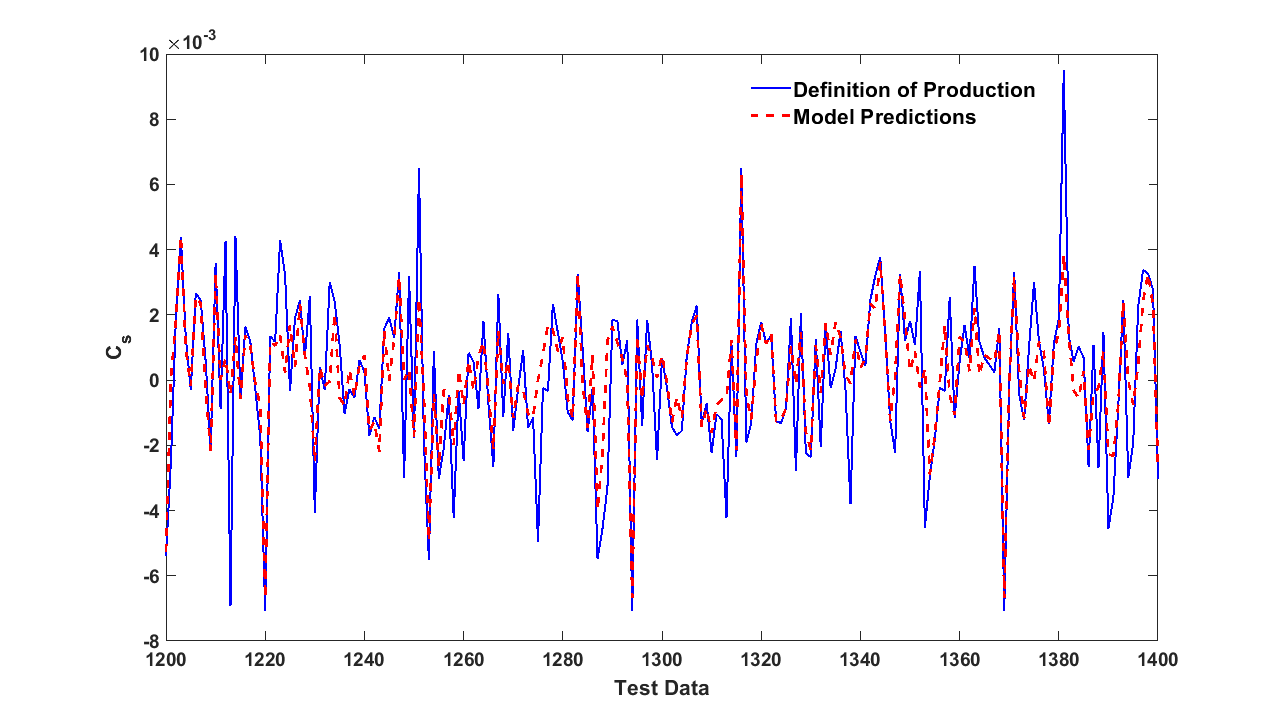}
    \includegraphics[width=0.49\linewidth]{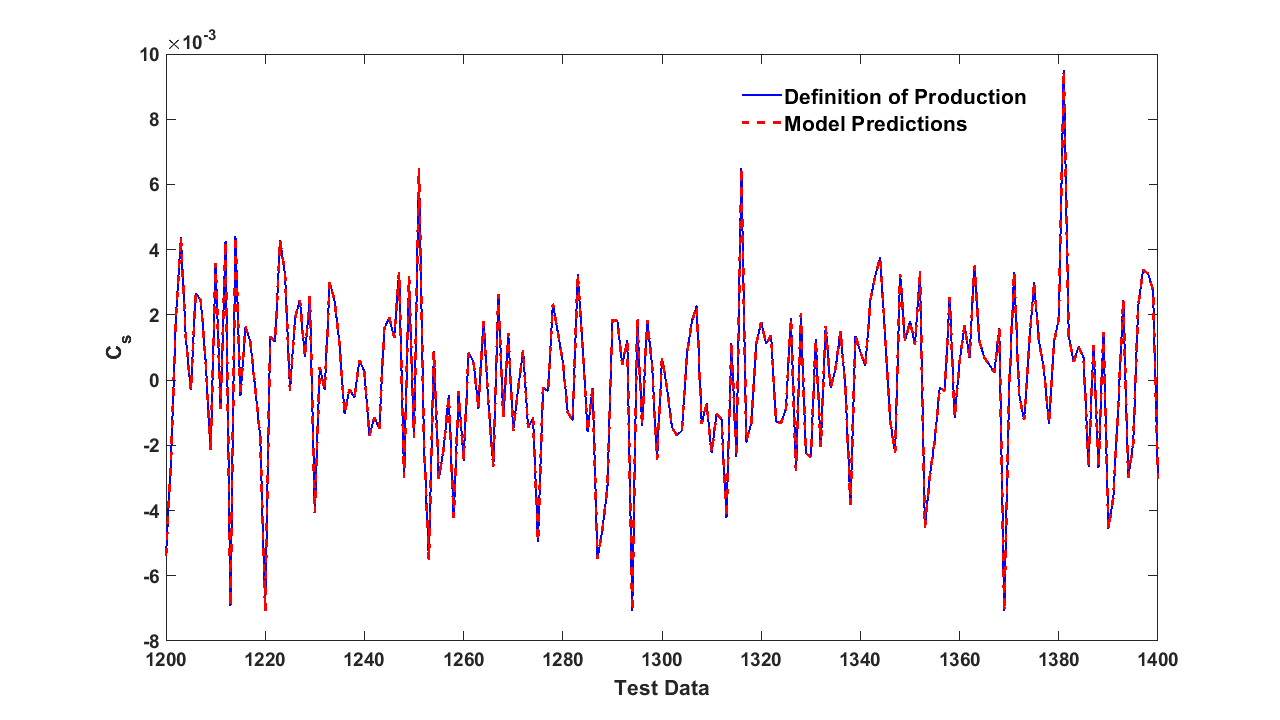}

    \vspace{0.5ex}
    \begin{minipage}[t]{0.49\linewidth}
        \centering \small (a) $n = 8$, $l = 2$, Zoomed Area 1
    \end{minipage}%
    \begin{minipage}[t]{0.49\linewidth}
        \centering \small (b) $n = 32$, $l = 3$, Zoomed Area 1
    \end{minipage}

    \vspace{2ex}

    % Row 2
    \includegraphics[width=0.49\linewidth]{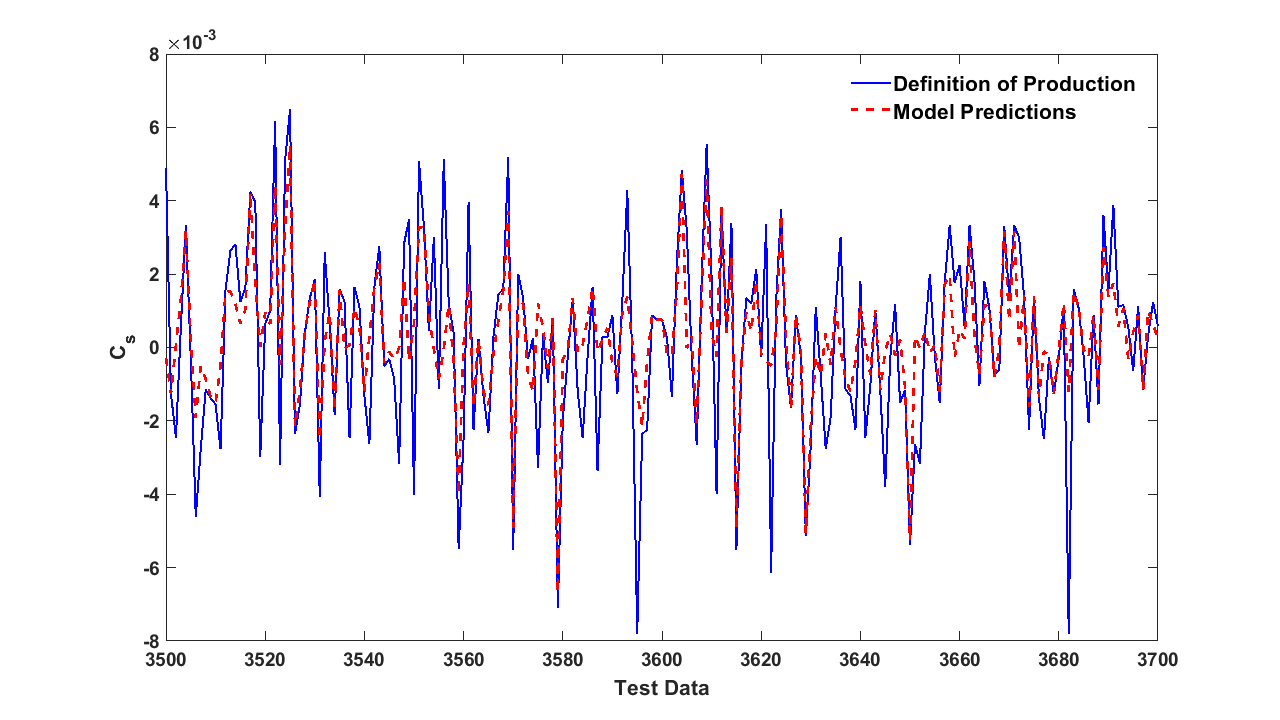}
    \includegraphics[width=0.49\linewidth]{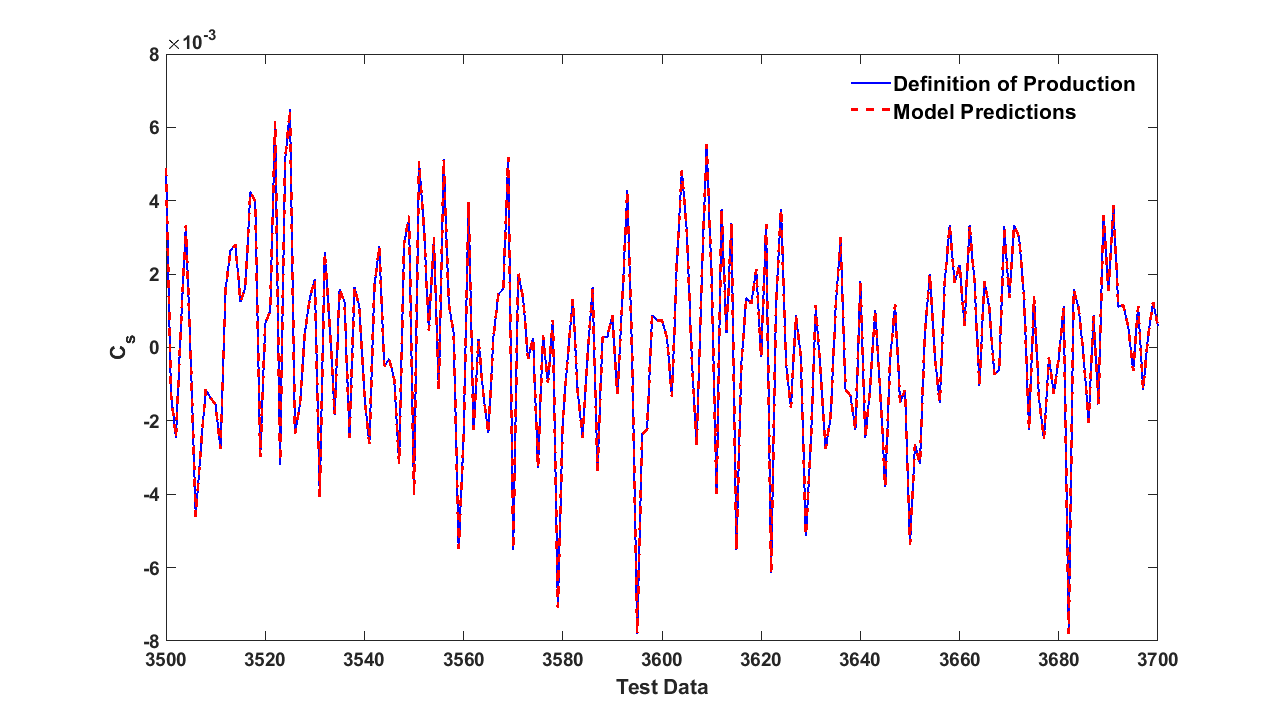}

    \vspace{0.5ex}
    \begin{minipage}[t]{0.49\linewidth}
        \centering \small (c) $n = 8$, $l = 2$, Zoomed Area 2
    \end{minipage}%
    \begin{minipage}[t]{0.49\linewidth}
        \centering \small (d) $n = 32$, $l = 3$, Zoomed Area 2
    \end{minipage}

    \vspace{2ex}

    % Row 3
    \includegraphics[width=0.49\linewidth]{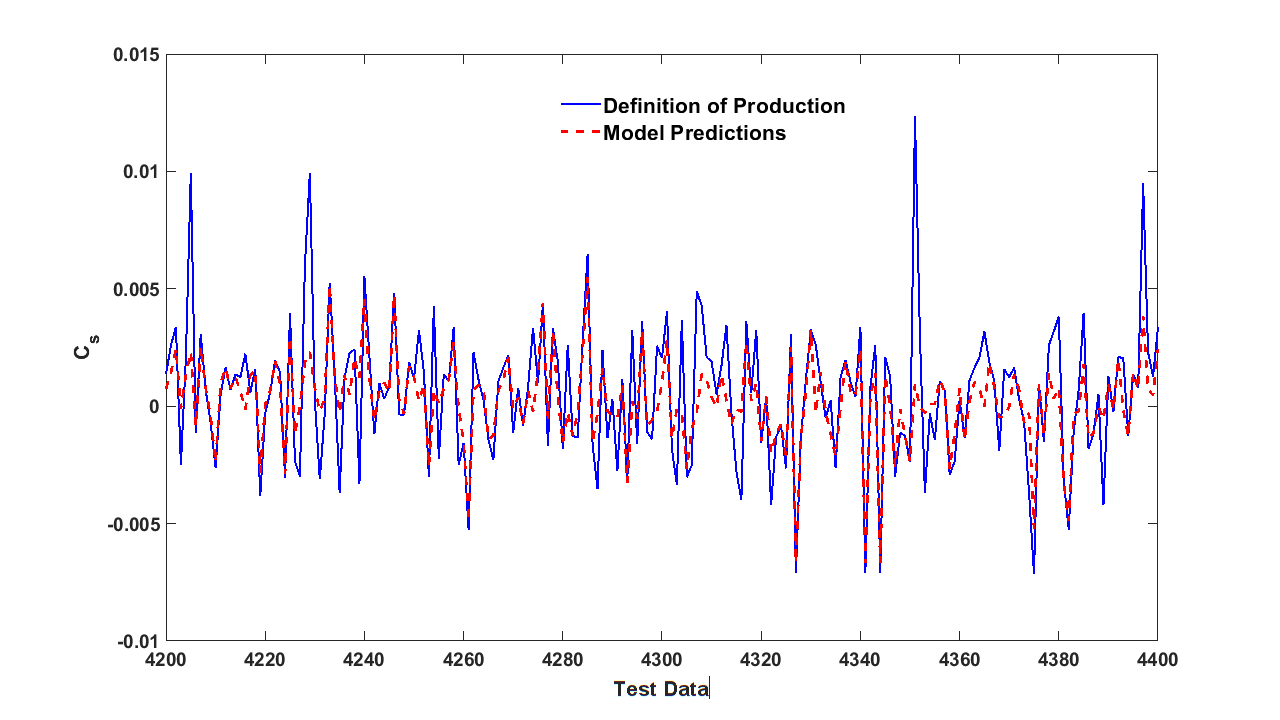}
    \includegraphics[width=0.49\linewidth]{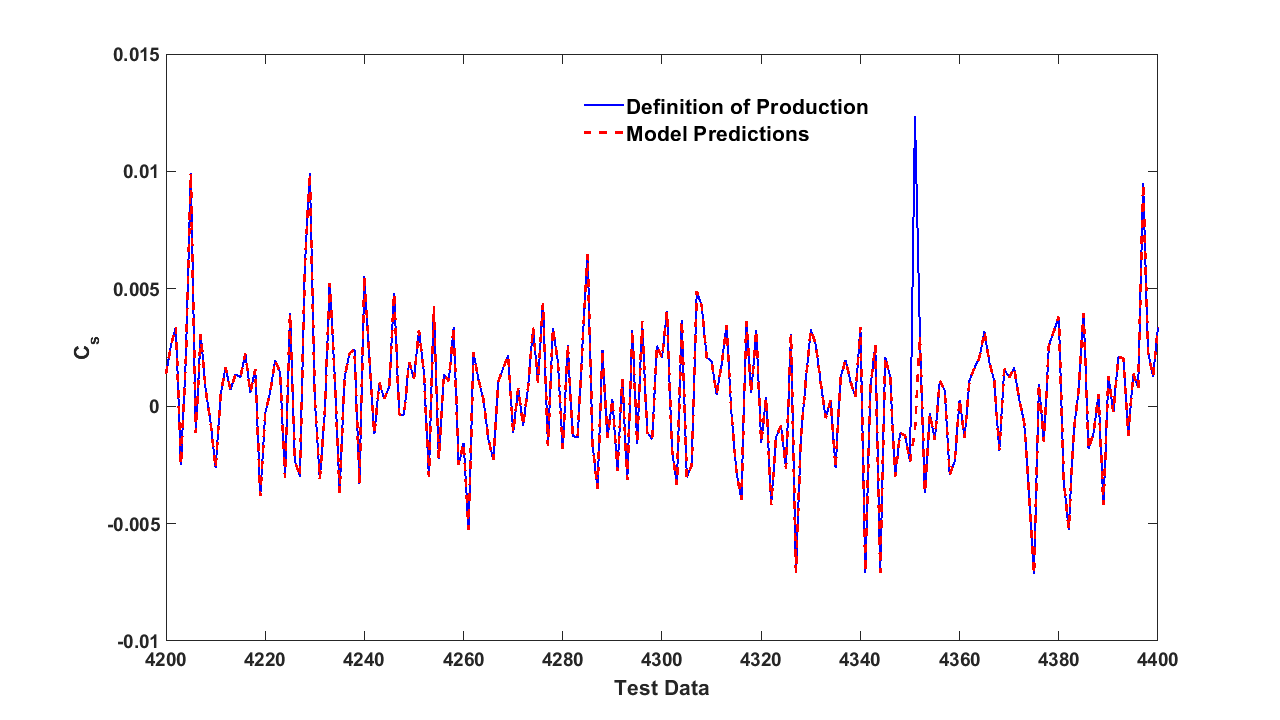}

    \vspace{0.5ex}
    \begin{minipage}[t]{0.49\linewidth}
        \centering \small (e) $n = 8$, $l = 2$, Zoomed Area 3
    \end{minipage}%
    \begin{minipage}[t]{0.49\linewidth}
        \centering \small (f) $n = 32$, $l = 3$, Zoomed Area 3
    \end{minipage}

    \caption{Detailed view of selected regions highlighting improved accuracy and reliability of ensemble NN predictions compared to single NN models.}
    \label{fig:ensemble_NN_comp_zoom}
\end{figure}

The effectiveness of ensemble networks becomes clearer through detailed inspections presented in Figure~\ref{fig:ensemble_NN_comp_zoom}, where zoomed-in views of selected regions illustrate significant improvements in local predictions. For example, Figures~\ref{fig:ensemble_NN_comp_zoom}(a) and \ref{fig:ensemble_NN_comp_zoom}(b) highlight enhanced predictive consistency at a challenging local peak around test data indices 1200–1400. Similarly, Figures~\ref{fig:ensemble_NN_comp_zoom}(c) and \ref{fig:ensemble_NN_comp_zoom}(d), depicting the data indices around 3500–3700, demonstrate the ensemble's capability to capture complex local variations with reduced deviations compared to single network configurations. Finally, Figures~\ref{fig:ensemble_NN_comp_zoom}(e) and \ref{fig:ensemble_NN_comp_zoom}(f) illustrate improved accuracy in ensemble predictions within highly fluctuating regions (indices around 4200–4400).

The quantitative analysis presented in Table~\ref{tab:ensemble_nn} further emphasizes the advantages of ensemble NNs. Ensembles consistently achieved significantly lower MSE values across all tested configurations than single networks. The configuration with 32 neurons and three hidden layers achieved the lowest MSE ($2.49 \times 10^{-8}$). The ensemble predictions presented here were conducted using MATLAB, which implements bootstrap aggregating (bagging). That trains multiple NNs on different bootstrap subsets of the dataset, each optimized using Bayesian regularization (`trainbr'), and records individual network MSE. Subsequently, the script aggregates predictions from networks meeting a specified MSE threshold ($1 \times 10^{-8}$), taking their median to enhance robustness. The resulting ensemble significantly reduces prediction errors compared to individual networks, notably achieving an MSE as low as \(2.49\times10^{-8}\). 

\begin{table}[h]
    \centering
    \caption{Test Data MSE for All Ensemble Network Configurations (10 Networks)}
    \label{tab:ensemble_nn}
    \begin{tabular}{|c|c|c|}
        \hline
        Neurons (\(n\)) & Hidden Layers (\(l\)) & Ensemble Test MSE \\
        \hline
        8 & 2 & \(2.75 \times 10^{-6}\)\\
        8 & 3 & \(1.61 \times 10^{-6}\)\\
        8 & 4 & \(1.02 \times 10^{-6}\)\\
        16 & 2 & \(5.87 \times 10^{-7}\)\\
        16 & 3 & \(6.49 \times 10^{-7}\)\\
        16 & 4 & \(5.29 \times 10^{-7}\)\\
        32 & 2 & \(3.73 \times 10^{-7}\)\\
        32 & 3 & \(2.49 \times 10^{-8}\)\\
        \hline
    \end{tabular}
\end{table}

The results conclusively demonstrate that ensemble NNs significantly outperform single-network configurations, primarily through their ability to average out individual model biases and variances. This ability is especially critical in turbulence modeling, where capturing detailed and nuanced variations in data is vital for accurate predictions. Therefore, ensemble methods represent a powerful tool in operational hurricane boundary-layer simulations, providing enhanced prediction reliability and accuracy.

\subsection{Comparison of Machine Learning Models and Dynamic Smagorinsky Model}\label{Sec:Compare_pro_dyn}

The coefficient \( C_s \) in the standard Smagorinsky model is considered constant in the computational domain. To overcome these limitations, dynamic models change their parameters in real time and space to reflect the flow condition. Because of this idea, the dynamic Smagorinsky model (DSM) emerged. Germano et al.~\cite{germano1991dynamic} improved the basic Smagorinsky model by dynamically adjusting the Smagorinsky constant \( C_s \) based on the flow's current condition, rather than establishing a fixed value in advance. This dynamic approach involves applying a test filter $\{\Tilde{.}\}$ to the LES data. The LES data is then utilized to compute the resolved stress tensor. 

\begin{equation}
\mathscr{L}_{i j} = T_{ij} - \widetilde{\tau_{ij}} = \widetilde{\overline{u_i}} \hspace{1mm} \widetilde{\overline{u_j}} - \widetilde{\overline{u_i} \hspace{1mm} \overline{u_j}}
\end{equation}

Assuming that the same functional form can be used to parametrize both $T_{i j}$ and $\tau_{i j}$ (the Smagorinsky model, for example), let $M_{i j}$ and $m_{i j}$ be the models for the anisotropic parts of $T_{i j}$ and $\tau_{i j}$ :
\begin{equation}
    \tau_{i j}-\left(\delta_{i j} / 3\right) \tau_{k k} \simeq m_{i j}= 2 C \bar{\Delta}^2|\overline{S}^{*} | \overline{S}_{ij}^{*}, 
    \label{eq:tau}
\end{equation}
\begin{equation}
    T_{i j}-\left(\delta_{i j} / 3\right) T_{k k} \simeq M_{i j}= 2 C \tilde{\bar{\Delta}}^2|\tilde{\overline{S}}^{*}| \tilde{\overline{S}}^{*}_{ij},
    \label{eq:T}
\end{equation}
where
$$
|\overline{S}_{ij}^{*}| = \frac{1}{2} (\frac{\partial \overline{u_i}}{\partial x_j} + \frac{\partial \overline{u_j}}{\partial x_i} - \frac{2}{3}\frac{\partial \overline{u_k}}{\partial x_k} \delta_{ij}), \quad|\tilde{\overline{S}}^{*}|=\sqrt{2 \tilde{\overline{S}}^{*}_{mn} \tilde{\overline{S}}^{*}_{mn}},
$$
$\bar{\Delta}$ is the characteristic filter width associated with $\bar{G}$, and $\widetilde{\bar{\Delta}}$ is the filter width associated with $\widetilde{\widetilde{G}}$.  Now, at first subtracting filtered Eq.~\eqref{eq:tau} from Eq.~\eqref{eq:T}, and contracting  by $\bar{S}_{i j}$ gives
$$
[T_{i j}-\left(\delta_{i j} / 3\right) T_{k k} - [\widetilde{\tau_{i j}} - \left(\delta_{i j} / 3\right) \widetilde{\tau_{k k}}]] \overline{S}_{ij}^{*} = 2 C\{ \tilde{\bar{\Delta}}^2|\tilde{\overline{S}}^{*}| \tilde{\overline{S}}^{*}_{ij} \overline{S}_{ij}^{*} - \bar{\Delta}^2|\widetilde{\overline{S}^{*}|  \overline{S}_{ij}^{*}} \overline{S}_{ij}^{*} \}
$$
$$
 [T_{ij} - \widetilde{\tau_{ij}}] \overline{S}_{ij}^{*} - \frac{\delta_{i j}}{3} [T_{k k} - \widetilde{\tau_{k k}} ] \overline{S}_{ij}^{*}
= 2 C\{ \tilde{\bar{\Delta}}^2|\tilde{\overline{S}}^{*}| \tilde{\overline{S}}^{*}_{ij} \overline{S}_{ij}^{*} - \bar{\Delta}^2|\widetilde{\overline{S}^{*}|  \overline{S}_{ij}^{*}} \overline{S}_{ij}^{*} \}.
$$
For incompressible flow $\overline{S}_{ii}^{*} = 0$ so , $\frac{\delta_{i j}}{3} [T_{k k} - \widetilde{\tau_{k k}}] \overline{S}_{ij}^{*} = \frac{\overline{S}_{ii}^{*}}{3} [T_{k k} - \widetilde{\tau_{k k}}] = 0$. Therefore, $C_{s}$ can be defined as 

\begin{equation}
    C_{s} =  \frac{1}{2} \frac{\mathscr{L}_{i j}\overline{S}_{ij}^{*}}{\{ \tilde{\bar{\Delta}}^2|\tilde{\overline{S}}^{*}| \tilde{\overline{S}}^{*}_{ij} \overline{S}_{ij}^{*} - \bar{\Delta}^2|\widetilde{\overline{S}^{*}|  \overline{S}_{ij}^{*}} \overline{S}_{ij}^{*} \}}
    \label{eq:dyn_Cs}
\end{equation}

On the other hand, the production-based formulation defines the true $C_s$ from the TKE production. The $C_s$ evaluation details are given in Sect.~\ref{subsec:Calc}. This production-based $C_s$ serves as the ground truth in our comparative analysis.

\begin{figure}[htbp]
\centering
\includegraphics[width=0.9\linewidth]{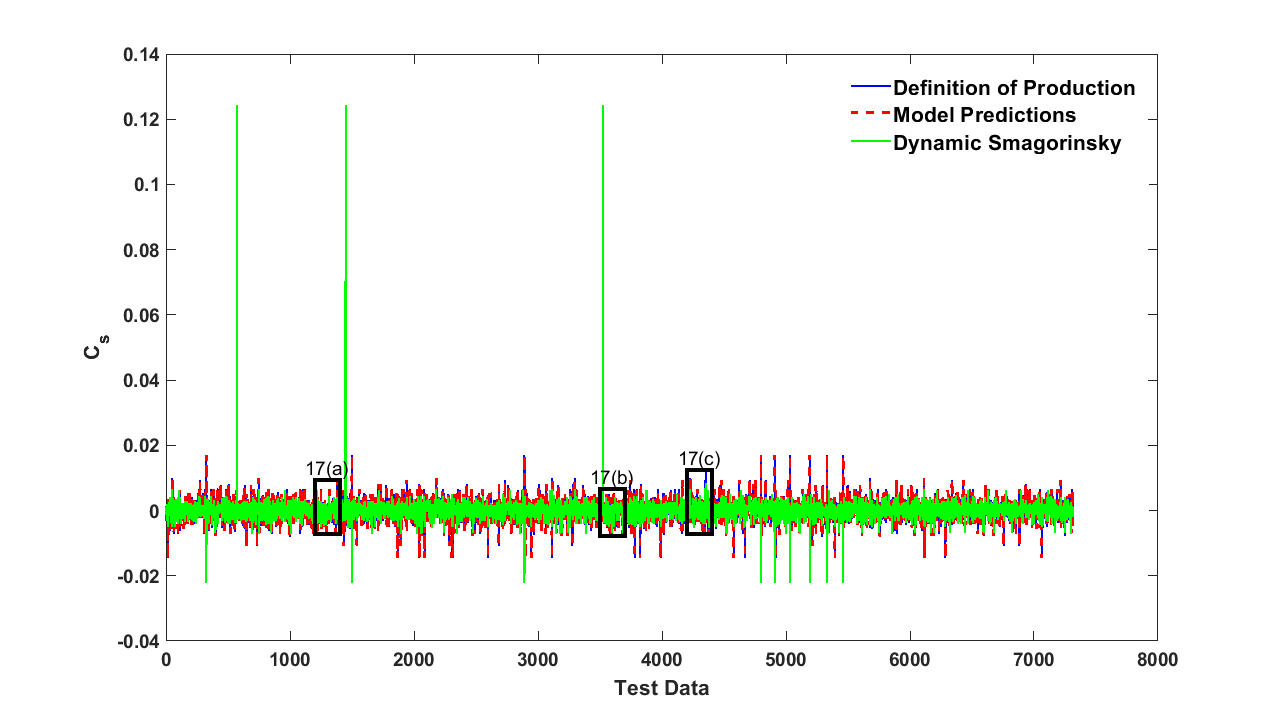}
\caption{Ensemble NN predictions compared with the definition of production formulation and dynamic Smagorinsky model.}
\label{fig:Dyn_NN_comp}
\end{figure}

Next, we compare the predictions of $C_s$ using two methods, the dynamic Smagorinsky method, and ensemble NN-based predictions, against the definition of production formulation as true values. The comparison is illustrated through figures highlighting the trends, variations, and zoomed regions for better visualization and interpretation.

\begin{figure}[htbp]
    \centering

    \includegraphics[width=0.7\linewidth]{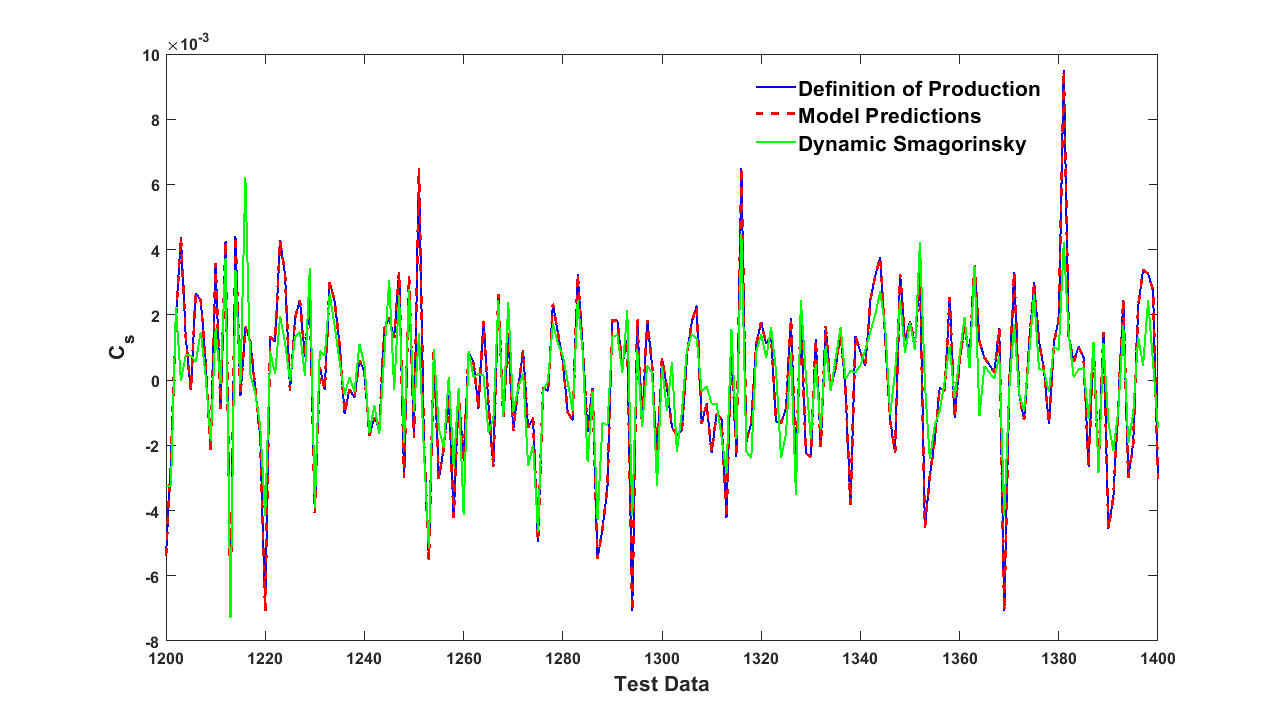}
    \vspace{0.5ex}
    \begin{minipage}{0.7\linewidth}
        \centering \small (a) Zoomed area 1
    \end{minipage}

    \vspace{2ex}

    \includegraphics[width=0.7\linewidth]{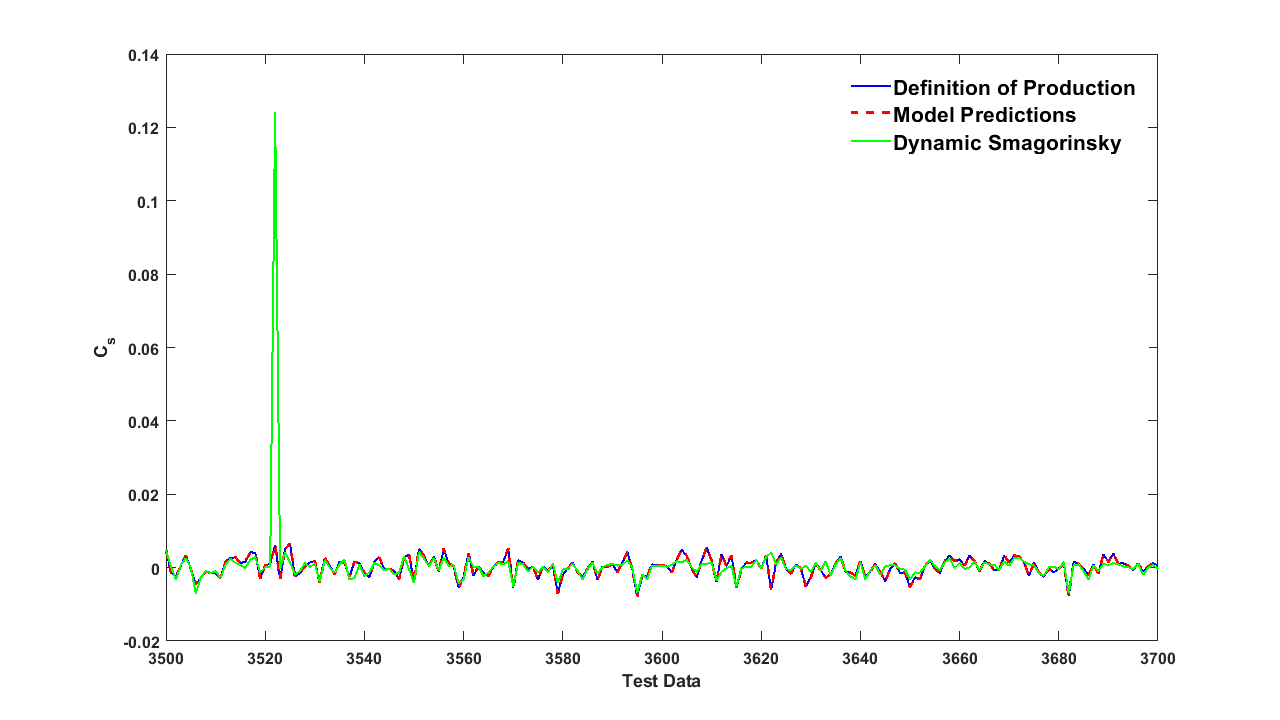}
    \vspace{0.5ex}
    \begin{minipage}{0.7\linewidth}
        \centering \small (b) Zoomed area 2
    \end{minipage}

    \vspace{2ex}

    \includegraphics[width=0.7\linewidth]{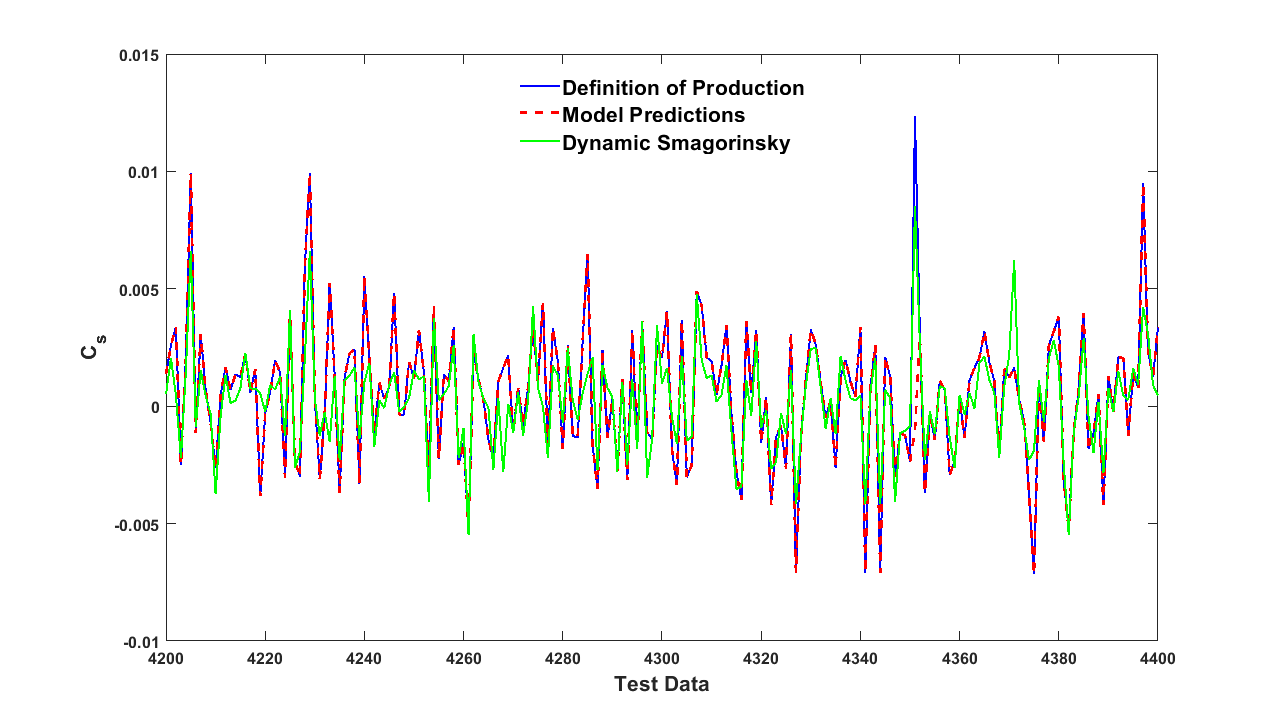}
    \vspace{0.5ex}
    \begin{minipage}{0.7\linewidth}
        \centering \small (c) Zoomed area 3
    \end{minipage}

    \caption{Zoomed areas showing NN predictions alongside other methods for $C_s$.}
    \label{fig:Dyn_NN_comp_zoom}
\end{figure}

Figures~\ref{fig:Dyn_NN_comp} and \ref{fig:Dyn_NN_comp_zoom} present a side-by-side comparison of the $C_s$ predicted by three different approaches: the definition of production formulation (considered as the ground truth, shown in solid blue), the DSM (shown in solid green), and the ensemble NN model (shown in dashed red). NNs in an ensemble configuration offer a promising signed $C_s$ prediction. By leveraging training on large datasets, they effectively interpolate between the smoothness of the production formulation and the adaptability of the dynamic Smagorinsky method. Their predictions closely follow the dynamic Smagorinsky model while mitigating its extreme fluctuations, making them a more robust alternative for SGS modeling. The ability of NNs to generalize well across different turbulence regimes ensures their viability as a scalable and computationally efficient solution for atmospheric simulations, thus bridging the gap between traditional turbulence models and modern data-driven approaches. 

To better understand these differences, Figure~\ref{fig:Dyn_NN_comp_zoom} zooms into three specific regions of the test data. These close-up views show that while the dynamic model may overshoot or undershoot in certain areas, the NN model stays tightly aligned with the true curve. The ensemble NN effectively filters out noise while preserving important physical features, providing a more stable and accurate estimate of $C_s$ across the domain.

In summary, the ensemble NN is a middle ground between the production formulation’s physical accuracy and the dynamic model’s adaptivity. It offers reliable predictions that avoid the instability of DSM while preserving the variability needed to represent turbulence. These qualities make ensemble NNs a strong candidate for replacing traditional SGS models in large-scale atmospheric simulations.

\section{Conclusions}\label{Sec:Conclusion}
This study presents a novel framework for SGS turbulence modeling in mesoscale atmospheric simulations by integrating machine learning with physical and geometric invariances. The primary goal of this research is to improve the prediction of the signed Smagorinsky coefficient ($C_s$) by embedding domain-specific knowledge into ML models through tensor invariants and singular value decomposition of local flow field features. These techniques enhance the models' ability to capture essential turbulence structures while maintaining interpretability and physical consistency. By utilizing high-fidelity LES data, the proposed ML-based approach improves the accuracy of SGS stress modeling, demonstrating significant advantages over traditional turbulence closure methods.

A key contribution of this study is the interpretability of the ML framework. Unlike many existing ML-based SGS models that function as ``black-box'' predictors, our approach explicitly incorporates physical constraints and geometric invariances. Using tensor invariants ensures that the predictions remain consistent with fundamental turbulence principles, while SVD of local flow features allows for an effective representation of anisotropic turbulence structures. This incorporation of physical knowledge makes the model more robust and transparent, addressing a major challenge in applying ML to scientific modeling.

The comparative analysis of the invariance-embedded ML predictions of signed $C_s$ and those with the dynamic Smagorinsky method demonstrate that ensemble NN predictions successfully capture localized turbulence variations similar to the dynamic Smagorinsky model while avoiding the extreme fluctuations that often lead to numerical instabilities.

The study also explores different NN configurations, evaluating their effectiveness in predicting $C_s$. Classification models are shown to effectively distinguish between different turbulence regimes, while regression models accurately predict signed $C_s$ values pointwise. The implementation of ensemble learning further enhances prediction stability and accuracy, reducing the variability associated with individual models due to the statistical learning nature of ML. This demonstrates that combining multiple NNs can improve the reliability and generalizability of ML-based turbulence modeling.

This research establishes that ML-based approaches can significantly enhance the accuracy and efficiency of SGS modeling in mesoscale atmospheric flows when designed with physical and geometric considerations. By improving the representation of turbulence in numerical simulations, this work contributes to the ongoing advancement of computational fluid dynamics and atmospheric modeling, with the potential to improve predictive capabilities in hurricane forecasting and other high-impact weather phenomena.

\begin{acknowledgments}
The hardware used in the computational studies is part of the UMBC High Performance Computing Facility (HPCF). The facility is supported by the U.S. National Science Foundation through the MRI program (grant nos. CNS-0821258, CNS-1228778, and OAC-1726023) and the SCREMS program (grant no. DMS-0821311), with additional substantial support from the University of Maryland, Baltimore County (UMBC). Hasan and Yu are grateful to Dr. Heng Xiao for the constructive discussions.
\end{acknowledgments}

\section*{Data Availability Statement}
The data that support the findings of this study are available from the corresponding author upon reasonable request. WRF simulation~\cite{Ren-2020} data used in this study is available on the Research Data Archive of NCAR (https://rda.ucar.edu/datasets/d301000/).

\appendix
\section*{Appendix: Model Performance Metrics for Different Classifier Models}\label{Sec:Appendix}

This appendix provides a comparison of performance metrics for various machine learning classifier models, including Logistic Regression (LR), Support Vector Machines (SVMs), Random Forest (RF), Gradient Boosting (GB), and Neural Network (NN), and different configurations tested in this study. The metrics include Accuracy, Precision, Recall, and F1-Score (see the introduction in Section~\ref{subsec:Results_p2p}), with a particular emphasis on the F1-Score as it balances precision and recall, which is crucial for datasets with potentially imbalanced classes. Different machine learning classifier models are briefly explained in the sections below for completeness. For a detailed description of these models, configurations, and their implementation with data from fine-scale hurricane boundary layer flow simulations, readers are referred to Hasan et al.~\cite{hasan2025aiaa}.

\subsubsection{Logistic Regression}
Logistic Regression~\cite{hosmer2013applied} is a linear model used for binary classification tasks. It models the probability that a given input \(x\) belongs to a particular class. The logistic function, or sigmoid, is used to map predicted values to probabilities between 0 and 1:
\[
P(y=1|x) = \frac{1}{1 + e^{- (w^T x + b)}},
\]
where \(w\) is the weight vector, \(x\) is the input feature vector, and \(b\) is the bias term. In the context of LR, \(y\) represents the binary outcome or target variable for classification. It can take on one of two possible values, typically labeled as \(0\) or \(1\). For our study, in a binary classification problem, \(y = 1\) indicates the large absolute $C_s$ values, while \(y = 0\) represents small values. The model aims to predict the probability that \(y = 1\), given the input features \(x\), using the logistic function to map the predicted value to a range between 0 and 1. The model is trained by minimizing the binary cross-entropy loss, which measures the difference between the predicted probabilities and the true labels:
\[
L(w, b) = - \frac{1}{N} \sum_{i=1}^{N} \left( y_i \log(P(y_i|x_i)) + (1 - y_i) \log(1 - P(y_i|x_i)) \right) + \frac{1}{2C} \|w\|^2,
\]
where \(N\) is the number of samples, \(y_i\) represents the true label for sample \(i\), and \( \| \cdot \|^2 \) denotes the $L_2$ norm operator. The regularization term \( \frac{1}{2C} \|w\|^2 \) penalizes large weights, with the parameter \( C \) controlling the strength of the regularization. A smaller \( C \) value applies stronger regularization, penalizing large weights more heavily and encouraging the model to generalize better by simplifying its decision boundary. This is particularly useful for preventing overfitting in cases with noisy data. Conversely, a larger \( C \) reduces the regularization effect, allowing the model to fit the training data more closely and potentially capture more detailed patterns, albeit at the risk of overfitting.

The weights are optimized using gradient descent to best fit the training data, considering both the data fitting and regularization terms. For LR, the decision line is strongly set by setting the probability threshold, which is usually 0.5, to classify the input effectively:

\[
f(x) = \begin{cases} 
1 & \text{if } P(y=1|x) \geq 0.5 \\
0 & \text{otherwise}
\end{cases}.
\]

LR is simple and effective for linearly separable data, making it a popular baseline for classification tasks. By tuning the \( C \) parameter, the model can find a balance between bias and variance, optimizing its performance for specific datasets.
\subsubsection{Support Vector Machines}\label{subsubsec:SVM}

Support Vector Machines~\cite{cortes1995support} were also employed to classify the magnitude of the Smagorinsky coefficient, $C_s$. We tested different kernel functions, specifically the Radial Basis Function (RBF) and linear kernel, varying the regularization parameter ($C$) and the kernel coefficient ($\gamma$), to determine their impact on model accuracy. The SVM decision function is given by:

\[ f(x) = \text{sign}\left( \sum_{i=1}^{N} \alpha_i y_i K(x_i, x) + b \right), \]
where \(\alpha_i\) are the Lagrange multipliers, \(y_i\) are the target labels, \(K(x_i, x)\) is the kernel function, and \(b\) is the bias term. The SVM optimization problem, which includes the regularization parameter \(C\), is formulated as:

\[ \min_{\alpha} \frac{1}{2} \sum_{i=1}^{N} \sum_{j=1}^{N} \alpha_i \alpha_j y_i y_j K(x_i, x_j) - \sum_{i=1}^{N} \alpha_i, \]
subject to
\[ 0 \leq \alpha_i \leq C, \]
\[ \sum_{i=1}^{N} \alpha_i y_i = 0. \]

The optimal parameters were identified through cross-validation, balancing precision, and recall to achieve robust classification performance. The RBF kernel is particularly useful for handling nonlinear relationships within the data. The RBF kernel function is defined as:
\[ K(x_i, x) = \exp\left(-\gamma \|x_i - x\|^2\right), \]
where \(\gamma\) is a parameter that determines the spread of the kernel, and ${ \|\mathbf {x} -\mathbf {x'} \|^{2}}$ is the squared Euclidean distance between the two feature vectors.  
An equivalent definition involves a parameter  $\gamma ={\frac {1}{2\sigma ^{2}}}$, where $\sigma$ is a free parameter that controls the width of the Gaussian function. A smaller $\sigma$ value results in a narrower Gaussian, focusing the kernel on close points. A bigger $\sigma$ value results in a wider Gaussian, which helps the kernel catch connections between distant points. This kernel transforms the data into a higher-dimensional space where it becomes easier to separate classes that are not linearly separable in the original feature space. The RBF kernel can find the complex patterns in the dataset, which helps the model to classify data points more accurately.

\subsubsection{Random Forest}\label{subsubsec:RF}

Random Forest~\cite{breiman2001random} is an ensemble learning method for classification and regression tasks. It creates multiple decision trees during training and outputs the mode of the classes for classification or the average prediction for regression. This approach reduces the risk of overfitting compared to individual decision trees and increases model robustness.

The prediction from an RF is given by aggregating the predictions from all the trees in the forest, ensuring a reliable and robust outcome:
\[ f(x) = \frac{1}{M} \sum_{m=1}^{M} T_m(x), \]
where \(M\) is the number of trees, and \(T_m(x)\) represents the prediction from the \(m\)-th tree. Each tree is constructed using a bootstrap sample from the original dataset, and random subsets of features are selected for splitting at each node, thus introducing diversity and reducing the correlation between the trees.

The optimization problem for RF involves minimizing the impurity at each node split. The Gini impurity or entropy is often used as the criterion:
\[ G = 1 - \sum_{k=1}^{K} p_k^2, \]
where \(p_k\) is the proportion of samples belonging to class \(k\) at a particular node, and \(K\) is the number of classes. For classification tasks, \(K\) is determined by the number of unique class labels in the target variable. For example, if the target variable represents three categories (say, `A,' `B,' and `C'), then \(K = 3\). The impurity measure evaluates the ``purity'' of a node, where a lower value indicates that the node predominantly contains samples from a single class. When splitting the nodes, RF chooses the feature and threshold that result in the greatest reduction in impurity.

The hyperparameters tuned in this study included the number of trees (\(M\)) and the maximum depth of each tree. These hyperparameters were chosen using cross-validation, a technique used to assess how well a model will generalize to an independent data set. This process helps optimize the performance of the classification model.

\subsubsection{Gradient Boosting}\label{subsubsec:GB}

Gradient Boosting~\cite{friedman2001greedy} is an ensemble learning technique that builds a model sequentially, combining the strengths of weak learners (typically decision trees) to improve overall performance. Each new model is designed to correct the errors made by the previous models, ensuring a reliable and robust predictive model.

The GB algorithm minimizes a loss function by adding new models that approximate the negative gradient of the loss function concerning the current model predictions. The prediction at iteration \(m\) is given by:
\[ F_m(x) = F_{m-1}(x) + \eta \cdot h_m(x) \]
where \(F_{m-1}(x)\) is the prediction from the previous iteration, \(h_m(x)\) is the new weak learner (typically a decision tree), and \(\eta\) is the learning rate that controls the contribution of each tree to the final model.

The loss function minimized in GB is typically the mean squared error for regression tasks or the binary cross-entropy for classification tasks:
\[ L_m = \sum_{i=1}^{N} l(y_i, F_{m}(x_i) ) \]
where \(l(y_i, F_m(x_i))\) represents the loss for each data point \(i\).

The hyperparameters tuned in this study included the number of boosting stages, the learning rate (\(\eta\)), and the maximum depth of each tree. These parameters were optimized through cross-validation to achieve the best performance in classifying the signed Smagorinsky coefficient, \(C_s\).

\subsubsection{Neural Networks}\label{subsubsec:NN_class}

Neural networks~\cite{lecun2015deep, Goodfellow-et-al-2016, Schmidhuber2015} are computational models inspired by biological NNs. They consist of interconnected units or nodes, called neurons, organized in layers that transform input data \(x\) into output \(f(x)\) through a series of weighted connections and activation functions.

The output of a single-layer neuron can be represented as:
\[ f(x) = \sigma(w^T x + b) \]
where \(w\) is the weight vector, \(x\) is the input feature vector, \(b\) is the bias term, and \(\sigma\) is the activation function, such as ReLU or sigmoid. Deep neural networks extend this structure with multiple hidden layers, enabling the model to learn complex, hierarchical data representations.

Training an NN involves optimizing the weights and biases to minimize a loss function, such as binary cross-entropy for classification. The backpropagation algorithm~\cite{rumelhart1986} calculates the gradient of the loss concerning each parameter, and gradient descent is used to update the parameters to reduce the loss.

We designed and implemented a feedforward NN using the PyTorch~\cite{pytorch} library, exploring a range of fully connected architectures with diverse configurations to evaluate their impact on the model's performance. We meticulously adjusted key hyperparameters to optimize the model, including the number of hidden layers, nodes per layer, and activation functions. During training, our main focus is minimizing binary cross-entropy loss using backpropagation and gradient descent algorithms. The Binary Cross-Entropy Loss~\cite{Good-1952} is a commonly used loss function for binary classification tasks. The formula for the Binary Cross-Entropy (BCE) Loss is:
$$
\text{BCE Loss} = -\frac{1}{N} \sum_{i=1}^{N} \left[ y_i \log(\hat{y}_i) + (1 - y_i) \log(1 - \hat{y}_i) \right]
$$
where \( N \) is the number of samples, \( y_i \) is the true label for the \(i\)-th sample (0 or 1), and \( \hat{y}_i \) is the predicted probability for the \(i\)-th sample, which is the output of the sigmoid function applied to the model's output. The model's output is passed through a sigmoid function to produce probabilities, which are then compared to the true labels using the BCE Loss function.

\begin{figure}[htbp]
    \centering
    \includegraphics[width=0.7\textwidth]{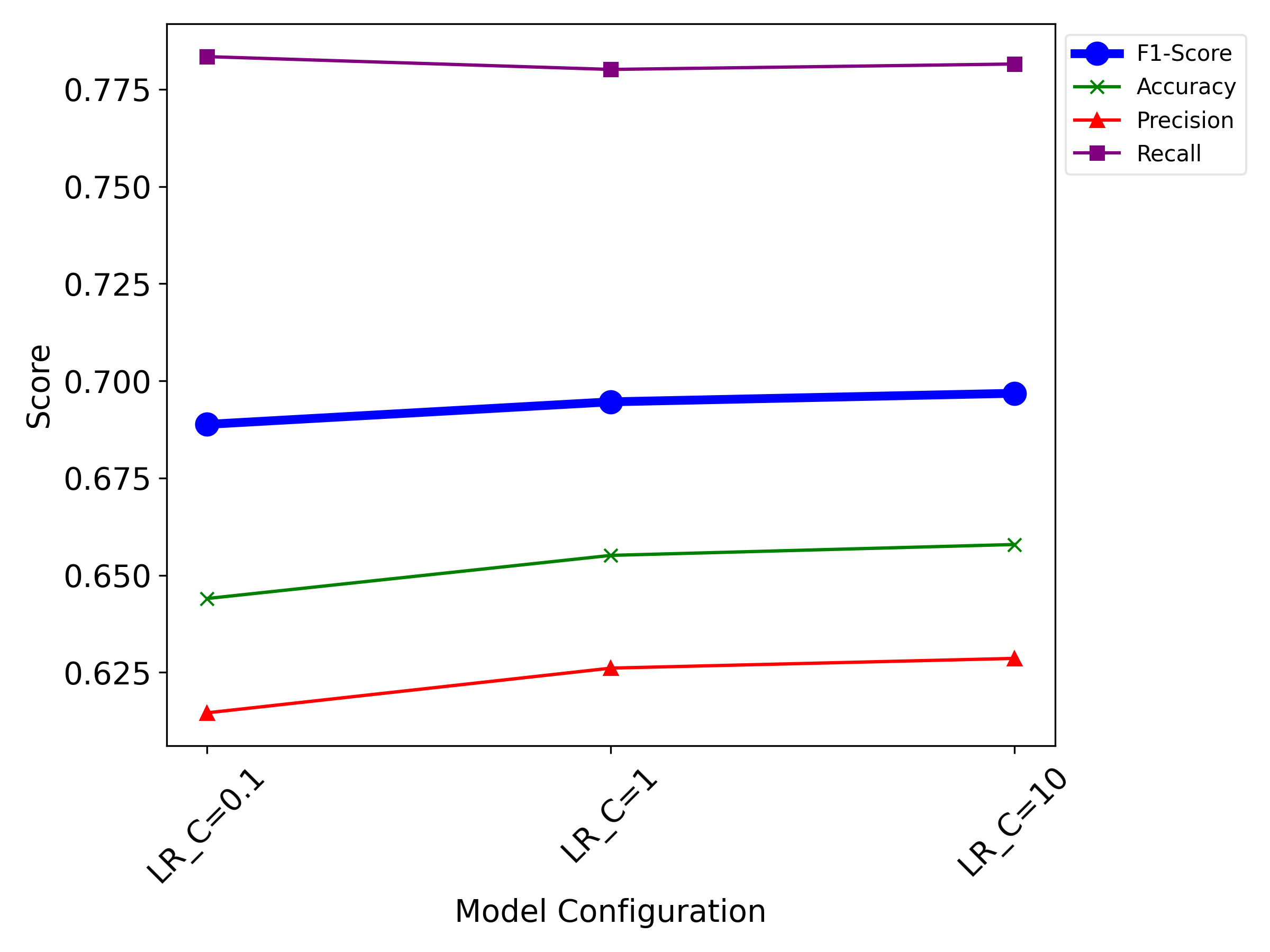}
    \caption{Performance metrics for Logistic Regression classifier configurations. Different values of the regularization parameter (\texttt{C}) are tested to assess the impact on the model's ability to classify accurately and generalize effectively.}
    \label{fig:LR}
\end{figure}

\begin{figure}[htbp]
    \centering
    \includegraphics[width=0.7\textwidth]{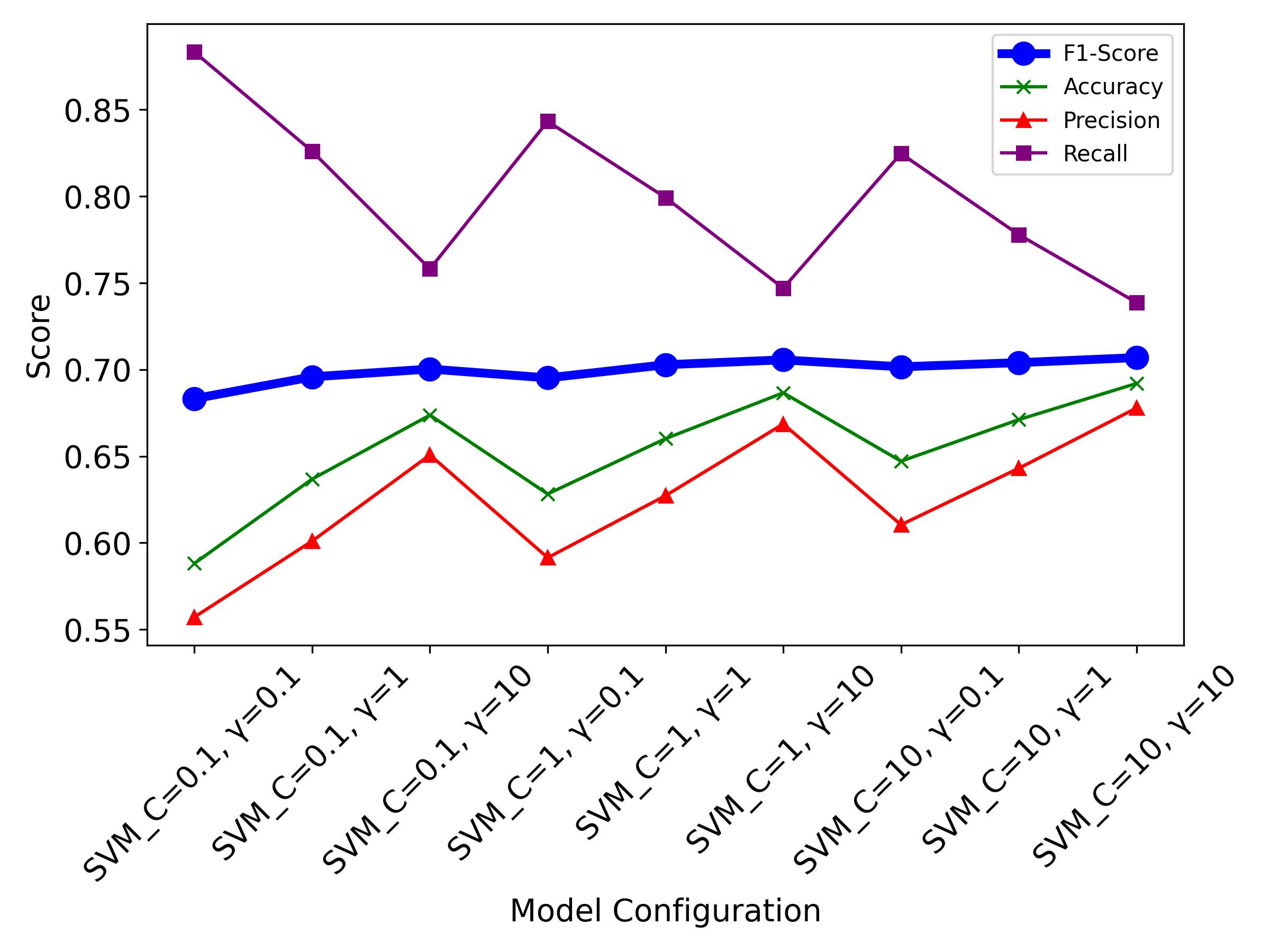}
    \caption{Performance metrics for Support Vector Machine classifier configurations. The regularization parameter (\texttt{C}) and kernel coefficient (\texttt{$\gamma$}) are tuned to explore their influence on the model’s decision boundary and classification accuracy.}
    \label{fig:SVM}
\end{figure}

\begin{figure}[htbp]
    \centering
    \includegraphics[width=0.7\textwidth]{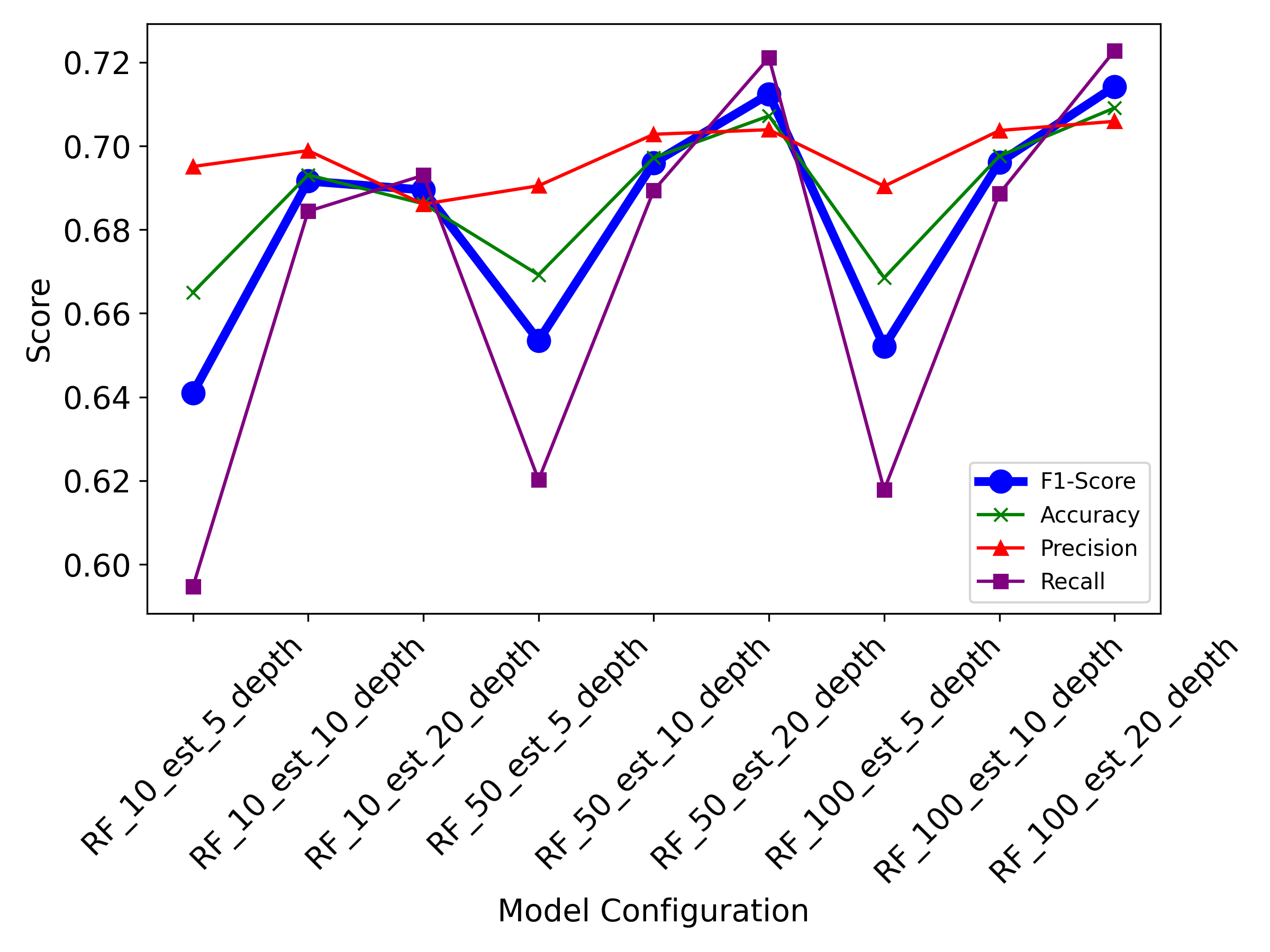}
    \caption{Performance metrics for Random Forest classifier configurations. Various combinations of estimators (\texttt{est}) and maximum depth (\texttt{depth}) are tested. The F1-Score and Recall values demonstrate the model’s sensitivity to positive cases across different configurations.}
    \label{fig:RF}
\end{figure}

\begin{figure}[htbp]
    \centering
    \includegraphics[width=0.7\textwidth]{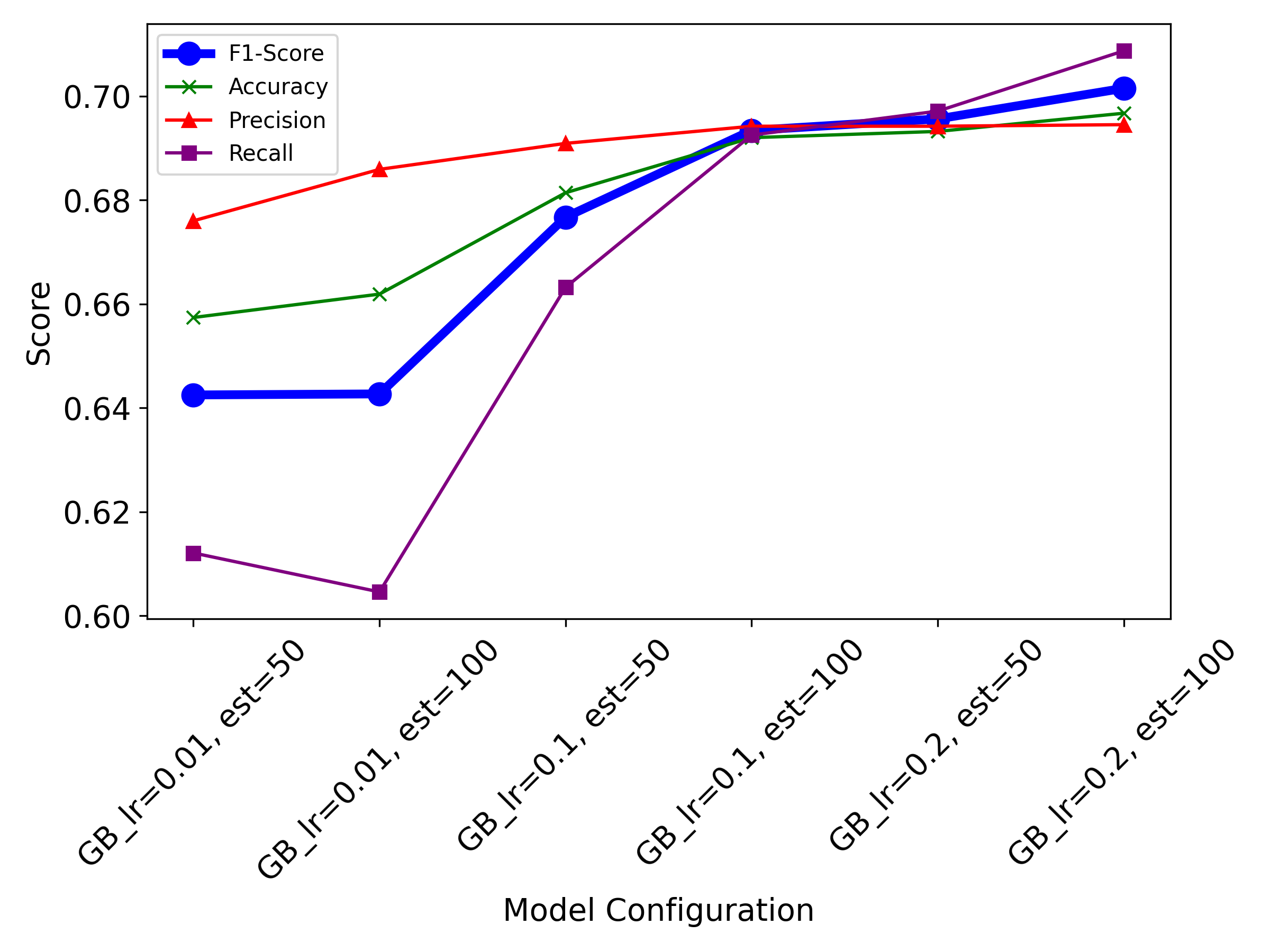}
    \caption{Performance metrics for Gradient Boosting classifier configurations. The configurations vary by learning rate (\texttt{lr}) and number of estimators (\texttt{est}). The F1-Score, indicated by the thicker line, highlights the model's overall balance in handling false positives and negatives across configurations.}
    \label{fig:GB}
\end{figure}

\begin{figure}[htbp]
    \centering
    \includegraphics[width=0.7\textwidth]{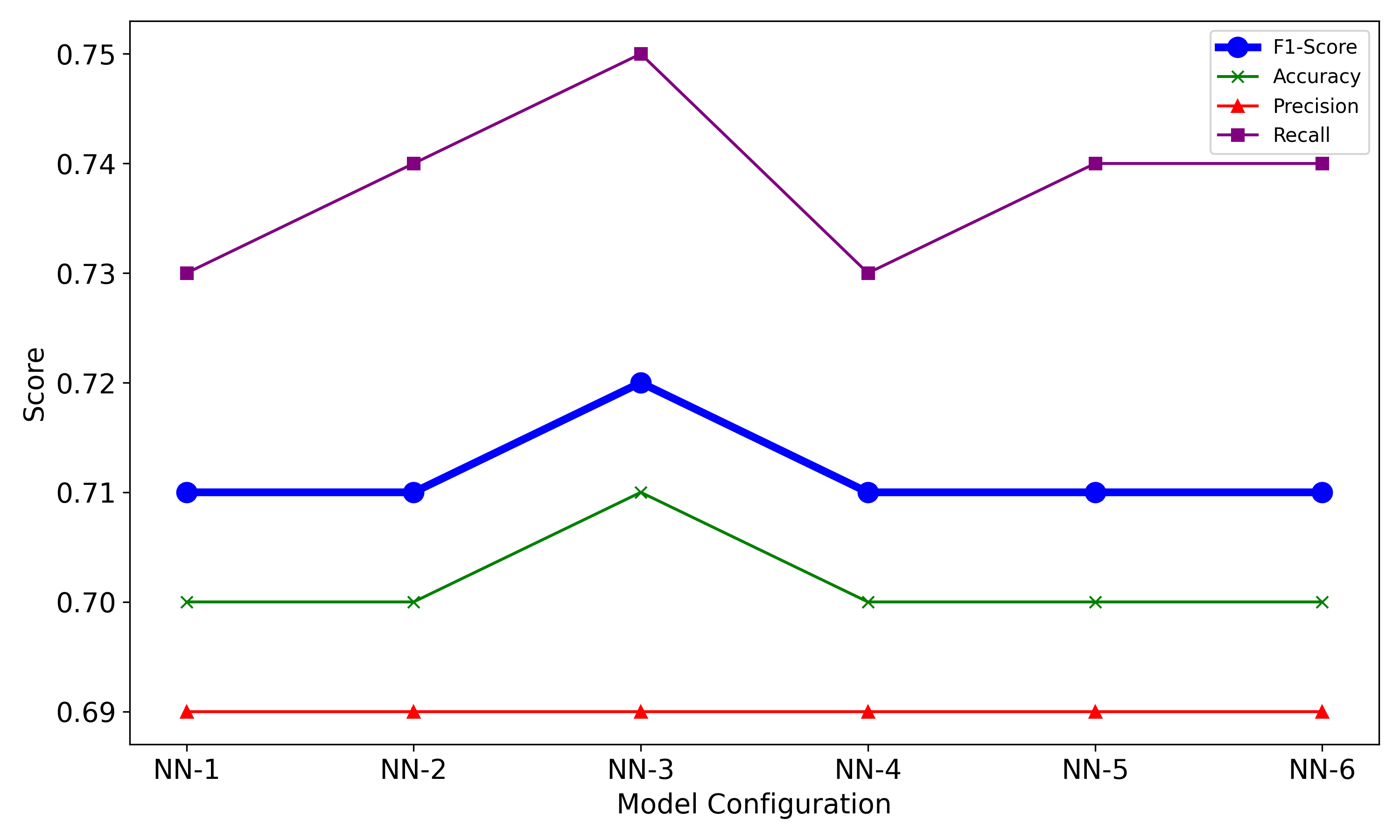}
    \caption{Performance metrics for NN classifier configurations. The model is tested with different activation functions and varying neuron counts (From Table~\ref{tab:model_configurations}) to examine their effect on model accuracy, precision, recall, and F1-Score.}
    \label{fig:NN_Classify}
\end{figure}

Each figure below shows the performance metrics for a specific classifier model type and configuration, helping to identify trends in how different parameters influence the effectiveness of each classifier, including that for LR in Figure~\ref{fig:LR}, SVM in Figure~\ref{fig:SVM}, RF in Figure~\ref{fig:RF}, GB in Figure~\ref{fig:GB}, and NN in Figure~\ref{fig:NN_Classify}.

%\section{Appendixes}

%\nocite{*}
%\bibliography{aipsamp}% Produces the bibliography via BibTeX.
\bibliography{badrul-ml_2}
%,xiao-refs
\end{document}